\documentclass[11pt,a4paper]{article}
\usepackage{jcappub}
\pdfoutput=1

\usepackage{epstopdf}
\usepackage{times, verbatim, xcolor, url, amsmath, amssymb, wasysym, graphicx, textcomp, array, tikz, subfigure,multirow}
\usetikzlibrary{shapes,arrows}
\usepackage{hyperref}
\usepackage[titletoc]{appendix}
\newcolumntype{P}[1]{>{\centering\arraybackslash}p{#1}}
\usepackage[utf8]{inputenc}
\usepackage{xspace}
\usepackage{ulem} 

\newcommand{\be}{\begin{equation}}
\newcommand{\ee}{\end{equation}}
\newcommand{\bea}{\begin{eqnarray}}
\newcommand{\eea}{\end{eqnarray}}

\newcommand{\Simbad}{\textsc{Simbad}\xspace}
\newcommand{\Planck}{$\rm Planck$\xspace}
\newcommand{\Riess}{$\mathcal{R}$\textsc{16}\xspace}
\newcommand{\Keenan}{$\mathcal{K}$\textsc{13}\xspace}
\newcommand{\Ned}{\textsc{Ned}\xspace}
\newcommand{\UTPO}{\textsc{Union 2.1}\xspace}
\newcommand{\TMPP}{\textsc{2M++}\xspace}
\newcommand{\Edd}{\textsc{Edd}\xspace}
\newcommand{\FN}[1]{{#1}}
\newcommand{\AER}[1]{{#1}}
\newcommand{\HWC}[1]{{#1}}

\newcommand{\COM}[1]{{}}

\newcommand{\HWCR}[1]{{#1}}
\newcommand{\COMT}[1]{{}}

\newcommand{\zmax}{z_{\rm max}} 
\newcommand{\zobs}{z_{\rm obs}} 
\newcommand{\zbar}{\bar{z}} 
\newcommand{\dd}{{\rm d}}
\newcommand{\Mpc}{{\rm Mpc}\xspace}
\newcommand{\Ocal}{\mathcal O}
\newcommand{\modelparam}{{\it Param.}\xspace}
\newcommand{\ksm}{${\rm km \, s^{-1} \, \Mpc^{-1}}$\xspace}
\newcommand{\ks}{${\rm km \, s^{-1}}$\xspace}

\begin{document}


\title{Probing homogeneity  with standard candles}

\author[a]{Hsu-Wen~Chiang
}
\author[b,g,h]{, Antonio~Enea~Romano
}
\author[a]{, Fabien~Nugier
}
\author[a,d,e,f]{and Pisin~Chen
}

\emailAdd{b98202036 [at] ntu.edu.tw}
\emailAdd{antonio.enea.romano [at] cern.ch}
\emailAdd{fnugier [at] ntu.edu.tw}
\emailAdd{pisinchen [at] phys.ntu.edu.tw}

\affiliation[a]{Leung~Center~for~Cosmology~and~Particle~Astrophysics, National~Taiwan~University, 
Taipei~10617, Taiwan, R.O.C.}
\affiliation[b]{Theoretical Physics Department, CERN, CH-1211 Geneva 23, Switzerland}
\affiliation[d]{Department~of~Physics, National~Taiwan~University, 
Taipei~10617, Taiwan, R.O.C.}
\affiliation[e]{Graduate~Institute~of~Astrophysics, National~Taiwan~University, 
Taipei~10617, Taiwan, R.O.C.}
\affiliation[f]{Kavli~Institute~for~Particle~Astrophysics~and~Cosmology, SLAC~National~Accelerator~Laboratory, Stanford~University, 
Stanford, CA~94305, U.S.A.}
\affiliation[g]{Department of Physics \& Astronomy, Bishop’s University,
2600 College Street, Sherbrooke, Qu´ebec, Canada J1M 1Z7
}

\affiliation[h]{Instituto de Fisica, Universidad de Antioquia, A.A.1226, Medellin, Colombia}


\abstract{
We show that standard candles can provide some valuable information about the density contrast, which could be particularly important at redshifts where other observations are not available.
We use an inversion method to reconstruct the local radial density profile from  luminosity distance observations assuming background cosmological parameters obtained from large scale observations. Using type Ia Supernovae
, Cepheids and the cosmological parameters from the Planck mission we reconstruct the radial density profiles along two different directions of the sky. We compare these profiles to other density maps obtained from luminosity density, in particular Keenan et al. 2013 and the \TMPP galaxy catalogue. The method independently confirms the existence of inhomogeneities, could be particularly useful to correctly normalize density maps from galaxy surveys with respect to the average density of the Universe, and could  clarify the apparent discrepancy between local and large scale estimations of the Hubble constant. When better observational supernovae data will be available, the accuracy of the reconstructed density profiles will improve and will allow to further investigate the existence of structures whose size is beyond the reach of galaxy surveys. 
}


\maketitle
\flushbottom


\section{Introduction}
\label{sec:intro}

The standard cosmological model 
has reached a high level of accuracy and self-consistency, accommodating in one unified theoretical framework different types of observations such as the cosmic microwave background (CMB), type Ia supernovae (SNe) and baryon acoustic oscillations (BAO). The population of SNe, historical probe of the cosmological constant $\Lambda$, is now comprising a large number of objects with 740 spectroscopically confirmed ones at redshifts $0.05<z<1.0$ (c.f. \cite{Betoule2014}). Thanks to many progresses, the precision on cosmological parameters is now approaching the percent level. Nevertheless, one parameter seems to create controversy in this consistent picture by showing a 2-3$\sigma$ tension between the different probes, and this is today's Hubble parameter $H_0$. Recently in Riess et al. (\cite{Riess2016}, abbreviated as \Riess) $H_0$ was re-evaluated with the best Cepheid calibration so far (details in \cite{Hoffmann2016,Cardona2016}), and it was found $H_0 = 73.24 \pm 1.74$ \ksm, hence raising the tension to $3.4\sigma$ against the $66.93 \pm 0.62$ \ksm value derived from the CMB observation of Planck \cite{PLANCK2016}, denoted in the rest of the paper as \Planck. It is thus important to explain this discrepancy.

Several ideas have been tested against data in order to resolve this problem.
One possibility is to modify our early-time picture of the Universe by changing our interpretation of CMB measurements \cite{Bernal2016}. It has also been proposed that the tension itself needs to be reinterpreted (c.f. \cite{Verde2013}).

A more conservative idea is to consider that different probes do not measure the same $H_0$.
In this spirit, inhomogeneous geometries have been tested with SNe data from the very start \cite{Zehavi1998,Giovanelli1999}, with perturbative \cite{Ichiki2015,Marra2013} and non-perturbative models like the Lema\^{i}tre-Tolman-Bondi (LTB) model  \cite{Cosmai2013,Fanizza2014,Romano:2013kua,Romano:2009mr,Romano:2009ej,Chung:2006xh} or the swiss-cheese model \cite{Fleury2013,Giblin16}. However, viable models must explain not only the magnitudes of SNe, but also pass constraints from other cosmological probes \cite{Alexander2007,Biswas2010,Wojtak2013,Nicola2016,Romano:2009mr}.
\par
It has been shown that \cite{Romano:2016utn} in the linear pertubative regime local inhomogeneities could resolve the apparent tension 
as the distance to the last scattering surface in negligibly affected by local structures, while at low redshift the local estimation of $H_0$ can be strongly affected. In the non perturbative regime the distance to the last scattering can be significantly affected by a local inhomogeneity, and in this case both the effects on local and large scale observations should be taken into account in order to determine if the tension can be partially resolved.   
In order to differentiate between the two measurements, from now on we denote the local value of the Hubble parameter as $H_0^{\rm loc}$, and the large scale value as $H_0^{\rm LS}$.

The statistical estimation of how much the $H_0^{\rm loc}$ value could vary among different regions of the Universe due to inhomogeneities was studied in \cite{Turner} and more recently in \cite{Marra2013,Ben-Dayan:2014swa}, where it was shown that there could be an uncertainty on $H_0^{\rm loc}$ of the same order of its current experimental errors. 
Nevertheless it should be noted that all these ``cosmic variance'' analyses involve an angular average which may underestimate the effects of anisotropy of local structure. Anisotropic effects could in fact play an important role \cite{Romano:2016utn,Romano:2014iea,Bengaly:2015nwa} because of the anisotropic distribution of SNe, even if the angular average of the effects on $H_0^{\rm loc}$ is not large enough to explain the tension.

This directional effect is mainly due to the assumption made in \Riess that the effect of inhomogeneities has been fully removed by the redshift correction based on \TMPP density map. This is not necessarily the case as the density maps used to perform the correction are limited by the observational depth. Notice that an \emph{isotropic inhomogeneity} extending in all directions beyond the depth of \TMPP density map is unlikely according to the $\Lambda$CDM structure formation predictions. However, an \emph{anisotropic inhomogeneity} extending only in some directions where most of SNe are located is a different case, and the probability estimation for the existence of such anisotropic inhomogeneities requires further investigations.

Local inhomogeneities explaining the tension between $H_0^{\rm loc}$ and $H_0^{\rm LS}$ could be tested independently using local density measurements and may also be related to the tensions in the estimation of the density parameter $\Omega_{m0}$ and the amplitude of the linear power spectrum $\sigma_8$ (c.f. \cite{Lee2013,Battye2014}). Considerable progress has been achieved in studying the bulk flow (e.g. \cite{Watkins2008,Turnbull2012,Hoffman2015,Bengaly2015}), with most studies converging on a velocity of several hundreds of \ks in a direction close to the CMB dipole of \cite{Aghanim2013}: $384 \pm 78 \, {\rm (stat)} \pm 115 \, {\rm (syst)}$ \ks pointing to the direction $(l,b) = (264^\circ,48^\circ)$ in galactic coordinates. 

Analyses are in general consistent with the $\Lambda$CDM model\cite{Park2012,Huterer2016} and it is well established observationally the existence of inhomogeneities extending in some directions up to a few hundred megaparsecs \cite{Frith2003,Alexander2007,Keenan}. \COMT{, which could produce an apparent directional dependence of cosmological parameters, as shown in some other cases \cite{Kalus2012,Bengaly2015}. Notice that as we stressed previously, whether these inhomogeneities themselves are consistent with $\Lambda$CDM is a separate issue.}
In particular, Keenan et al. (\cite{Keenan}, abbreviated as \Keenan) reports an under-density in some direction of the sky that extends up to $z \sim 0.07$ and suggests that a rescaling of the density map derived from the \TMPP catalogue \cite{Carrick15} is necessary. Direct measurements of $H(z)$ will probably help constraining $H_0$ in the future as well (c.f. \cite{Wang2016,Bonvin2016,Wei2016}).

In this paper we improve and apply the inversion method derived in \cite{Romano:2016utn} and \cite{inversion} to obtain the local density map from standard candles luminosity distance observations. The inversion method requires as input a smooth function for the luminosity distance $D_L(z)$ which we obtain with a model independent fit of a combination of Cepheids-hosting galaxies from \Riess and low redshift ($z<0.4$) SNe from the \UTPO catalogue of \cite{Union2.1}. The fit is based on radial basis functions (RBFs) and MCMC sampling. Because we are interested in a possible angular dependence of data and we want to establish a comparison with galaxy surveys, we consider 3 fields of view (F1,F2,F3) 
from \Keenan, to which we compare our results. All SNe in a given field give rise to a radial profile averaging over the corresponding window. Assuming that lensing effects can be neglected relative to peculiar velocities \cite{BenDayan:2012wi}, which is expected to be a valid approximation at small redshift \cite{Hilbert:2008kb}, we reconstruct the radial density profile for each field based on the inversion method described in \cite{inversion}, with some modifications necessary to take into account the corrections to the growth rate due to the cosmological constant. 

As the analyzed SNe are at small redshift, we also account for their galaxy plane motion by adding a constant 250 \ks velocity dispersion on SNe, and by comparing our results with \Keenan and \TMPP.
We also consider the velocity dispersion of Cepheids-hosting galaxies and show that its effect is negligible, 
and that the fit quality is acceptable
even without it, contrary to the assumption made in \Riess.

Our work reveals that in one field of view (F1) there is an under-density 
whose effect can partially account for the $H_0$ estimation discrepancy, due to the large number of low redshift SNe located along that direction.
The reconstructed density profile is in agreement with \TMPP rescaled according to \Keenan.
The reconstructed density profile along F3 also shows the presence of a large under-density in agreement with rescaled \TMPP. 

Our results show that SNe data can be a unique tool to probe structures whose size is larger than the depth of other observations such as the galaxy catalogues. Any deviation of cosmological parameters from their large scale estimation can consequently be interpreted as the evidence of structures whose size is beyond the reach of presently available astronomical data from which peculiar velocities and associated redshift corrections are derived. Once more SNe data will become available, they could be a valuable source of information about large scale structures and especially for the correct estimation of background density, complementary to the density maps from galaxy surveys.

The paper is organized as follows. In section~\ref{sec:data} we present the data employed in this study. This includes supernovae from the \UTPO catalogue, the latest Cepheids from \Riess, and the \TMPP catalogue. In section~\ref{sec:method} we explain the theoretical framework in which we interpret the data and describe the methods adopted in our statistical and numerical analysis. Section \ref{sec:setup}, \ref{sec:result} and \ref{sec:NOvcYESvd} are devoted to the presentation of our results, the physical interpretations and the comparison with the analyses of \Keenan, \Riess and \TMPP. In section~\ref{sec:discuss} we present some further discussions about the results, their limitations and possible improvements. In section~\ref{sec:conclusion} we summarize our conclusions. Complementary results and our dataset are presented in appendices \ref{sec:lambda}, \HWCR{\ref{sec:DirectionalLTB},} \ref{sec:NOvcNOvd} and \ref{app:U2p1plusRiessdata}.


\section{Data}
\label{sec:data}

We are interested in the effects of inhomogeneities on the luminosity distance of low-redshift SNe and Cepheids-hosting galaxies, referred together as standard candles.
At low redshift peculiar velocity can be important and produce an important contamination of the Hubble flow, and we will indeed devote a lot of attention to the problem of distinguishing appropriately between them.
Two different sources of the peculiar velocity are investigated. First for the velocity dispersion within the bounded structures, unlike the 250 \ks dispersion universal to all standard candles in \Riess, we differentiate between SNe and Cepheids-hosting galaxies. Second, for the bulk flow due to large scale inhomogeneities we reconstruct the density map from the luminosity distance of standard candles, and compare it with the map obtained from galaxy catalogues to test their consistency.

\subsection{Supernovae Ia and Cepheids-hosting galaxies}
\label{sec:sneia}

The supernovae dataset is extracted from the full \UTPO catalogue \footnote{\url{http://supernova.lbl.gov/union/}} of \cite{Union2.1} (Supernova Cosmology Project) with a redshift cut $z < \zmax$, where $\zmax$ is either 0.2 or 0.4 (for reasons explained in section~\ref{sec:2m++}), and the exclusion of ``bad" data points corresponding to types `p', `f' and `d' respectively: bad light curve fittings, long first phases after B-band maximum and SNe observed only less than 5 times (c.f. \cite{Union2}, table~13 therein). For other types among the remaining SNe, we give a more detailed account in appendix~\ref{app:U2p1plusRiessdata}. We additionally remove 3 SNe with unconventional names\footnote{Their names in the \UTPO dataset are 4064, 6968 and 10106.}, for which we cannot find the sky positions (see after), and rename two others \footnote{These are e020 and k429 that become respectively 2003kk and 2004hm thanks to \cite{Foley:2008qc}.}. Concerning the redshift $z$, distance modulus $\mu$ and its error $\Delta \mu$, we prefer the values from the short \UTPO list\footnote{\url{http://supernova.lbl.gov/union/figures/SCPUnion2.1_mu_vs_z.txt}} (more precise), otherwise we take them from the full catalogue (less precise).
To this dataset consisting of 288 ($\zmax = 0.2$) or 372 ($\zmax = 0.4$) SNe, we add 20 Cepheids-hosting galaxies from table~5 of \Riess. Except the masers-hosting anchor NGC 4258, each of Cepheids-hosting galaxies also hosts a type Ia supernova, among which 7 SNe can be found in \UTPO (though only 3 in the fields we considered, see after).
We use the redshift of the host galaxies given from the \Ned database.\footnote{\url{https://ned.ipac.caltech.edu/}}

\begin{figure}
\hspace{-.35cm}\includegraphics[width=1.04\textwidth]{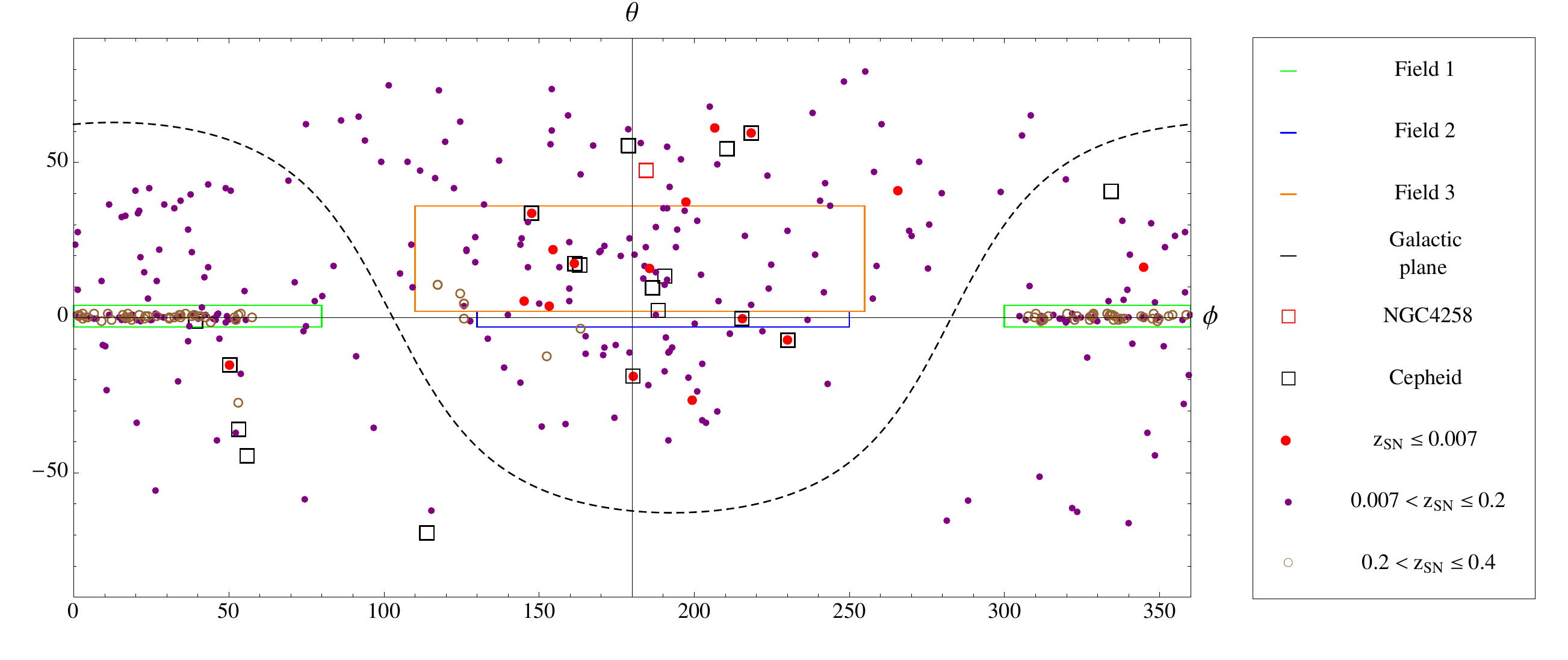}
\vspace{-1cm}
\caption{Sky map in ICRS coordinates of all SNe and Cepheids-hosting galaxies in our dataset. Three fields are specified according to \Keenan as the regions with luminosity density data. Our targets of interest are F1 and F3 which contain enough data points to fit the luminosity distance curve. For the sake of clarity we use (here and later on) the same colors as \Keenan, i.e.  green for F1 and orange for F3. \label{fig:skymap}}
\end{figure}

We are interested in the possible angular dependence of SNe due to local inhomogeneities. To study this dependence we consider 3 fields of view employed by \Keenan, to which we will compare our results. These fields are defined in table~\ref{tab:Fields}, shown in figure~\ref{fig:skymap}, and their total area of 6172 ${\rm deg}^2$ contains about half of the SNe. We extract the sky positions of all the \UTPO SNe and \Riess Cepheids-hosting galaxies automatically from the \Simbad database\footnote{\url{http://simbad.u-strasbg.fr/simbad/}} of \cite{Wenger:2000sw}. These positions are right ascension (R.A.) and declination (Dec.) expressed in the International Celestial Reference System (ICRS), written in decimal (degrees). We find more precisely for $\zmax = 0.2$ ($\zmax = 0.4$) that only 123 (203) SNe out of the 288 (372) \UTPO SNe are present in Fields 1, 2, and 3, as shown in table~\ref{tab:Fields} and illustrated in figure~\ref{fig:skymap}. As we can see only Fields 1 and 3 (abbreviated as ``F1" and ``F3") have enough data to be exploited, hence Field 2 (``F2") is excluded from our analysis. Among all these SNe, 1 in F2 and 2 in F3 belong to the Cepheids-hosting galaxies of \Riess (denoted as ``+1" and "+2"). These 3 SNe will be considered as independent data points from the host in the statistical analysis, for the reasons discussed at the end of this subsection and at the beginning of section~\ref{sec:2m++}. Cepheids-hosting galaxies associated to these 3 SNe are denoted through their type by `$\ast$' in appendix~\ref{app:U2p1plusRiessdata}. There are five other Cepheids-hosting galaxies in the fields we are interested in (1 in F1, 4 in F3) whose associated SNe are not included in \UTPO, and they are denoted by `$\dagger$' in appendix~\ref{app:U2p1plusRiessdata}.

\begin{table}[h]
\begin{center}
\begin{tabular}{|c|c|c|c|c|c|}
\hline
Field		& \multicolumn{2}{c|}{ICRS coordinates ($\rm deg$)}	& \multicolumn{2}{c|}{Number of SNe}& Cepheids-hosting	\\
\cline{2-5}
N$^\circ$	& R.A.					& Dec.						& $\zmax = 0.2$	& $\zmax = 0.4$		& galaxies of \Riess\\
\hline
1			& $[300,360]\cup[0,80]$	& $[-3,4]$					& 69			& 144				& 1					\\
2			& $[130,250]$			& $[-3,2]$					& 4+1			& 4+1				& 1					\\
3			& $[110,255]$			& $[2,36]$					& 47+2			& 52+2				& 6					\\
\hline
/			& \multicolumn{2}{c|}{Whole Sky}					& 288			& 372				& 20				\\
\hline
\end{tabular}
\caption{\label{tab:Fields} Fields of \Keenan and number of standard candles 
they contain from our dataset. $[A,B]$ stands for the angular range between A and B (including the boundaries).
}
\end{center}
\end{table}

\begin{figure}
\centering
\vspace{-.1cm}
\includegraphics[width=.7\textwidth]{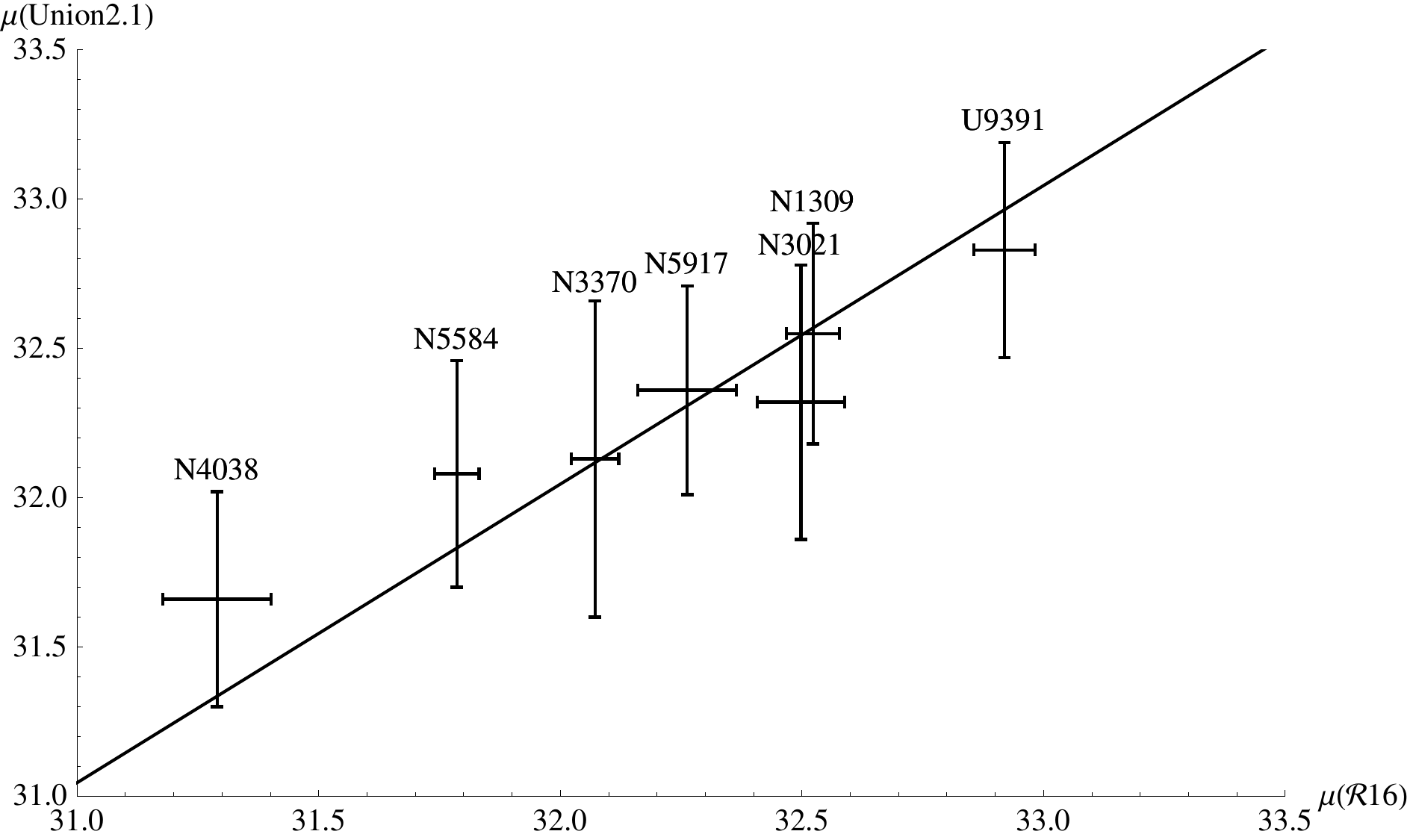}
\vspace{-.1cm}
\caption{$\mu(\mbox{\Riess})$ - $\mu(\rm{Union~2.1})$ plot with 7 host galaxies common to \Riess Cepheids and \UTPO SNe. The black line represents the shift of $5 \log_{10}\left(73.24/H_0^{\rm U2.1}\right)$ according to eq.~\eqref{eq:mucorr}.\label{fig:cal}}
\end{figure}

\COM{From the \UTPO SNe we have $(z,\mu,\Delta\mu)$, but we correct $\mu$ according to the calibrator difference between \UTPO\cite{Union2.1} and \Riess\cite{Riess2016}.} 
\AER{The \UTPO data contains no information about the absolute SNIa magnitudes. For this reason we convert the relative \UTPO magnitudes into absolute
magnitudes by calibrating with the 7 SNe common to the \Riess and \UTPO datasets.
It should be noted that the light curves fits of \UTPO are  obtained assuming a homogeneous model, and a full model independent analysis would require to also include the effects of inhomogeneities in the light curves fits or to consider different types of fitters \cite{Smale:2010vr,2011ApJ...740...72M,Bengochea:2014iha}, but this would go beyond the scope of this paper.  
} 
\COM{
\HWC{The calibration between these two datasets is expected to be redshift-independent. We also assume that Union2.1 magnitudes are independent from the $\Lambda$CDM model (though their light-curve fits involve that model). We believe the calibration of SNe Ia magnitudes and the focusing on small redshifts should limit the influence of the $\Lambda$CDM model on the light-curve parameters.}
}
\AER{ }
At low redshift the distance-modulus is given by
\be
\label{eq:mu}
\mu \equiv m-M = 25 - 5 \log _{10} H_0^{\rm loc} + 5 \log _{10} \left( H_0^{\rm loc} D_L\right) \approx 25 - 5 \log _{10} H_0^{\rm loc} + 5 \log _{10} c z \quad ,
\ee
where $H_0^{\rm loc}$ is the local Hubble parameter fitted with low-redshift standard candles. Keeping the redshift fixed, we expect
\be
\label{eq:mucorr}
\mu(\text{\Riess}) = \mu(\text{\UTPO}) - 5 \log_{10}\left(\frac{73.24}{70}\right) \;\, .
\ee
This is what we obtain by fitting $\mu(\text{\UTPO}) - \mu(\text{\Riess})$ with a \COM{shift}\HWC{difference} of $5 \log_{10}\left(73.24/H_0^{\rm U2.1}\right)$ (where $H_0^{\rm U2.1}$ is the only parameter of the fit), based on the 7 SNe common to \Riess and \UTPO, and weighted by \HWC{$\Delta\mu^2(\text{\UTPO}) + \Delta\mu^2(\text{\Riess})$}, as shown in figure~\ref{fig:cal}. More precisely, we find \HWC{$H_0^{\rm U2.1} = 70.05 \pm 2.20 \, \rm{km \, s^{-1} \, \Mpc^{-1}}$}, very close to the value 70 $\rm{km \, s^{-1} \, \Mpc^{-1}}$ assumed by \UTPO (and the value we keep). \HWC{The uncertainty of the calibration is propagated to the covariance matrix $\tilde{V}$ of the shifted distance modulus of \UTPO SNe with components given by
\be
\tilde{V}_{ij} = \Delta\mu_i^2 (\text{\UTPO}) \, \delta_{ij} + \text{Variance}\left( \mu(\text{\UTPO}) -  \mu(\text{\Riess}) \right) \,.
\ee
} In the remaining of this paper we will denote as $\mu$ the combination of shifted \UTPO SNe and \Riess Cepheids-hosting galaxies data\HWC{, as $V^{-1}$ the inverse covariance matrix of the combined data, and as $\Delta \mu^{-2}$ the diagonal part of $V^{-1}$.}.

Despite that the \Riess measurements of the distance modulus are almost the same as the corrected \UTPO moduli in eq.~\eqref{eq:mucorr} (difference always $<0.3$ mag among the 7 SNe), the values of the error in $\mu$ are quite different due to precise luminosity distance measurements of Cepheids-hosting galaxies. Hence we choose to treat these sources as different. In practice though, the large uncertainty on the \UTPO values makes them less relevant to the fit that will be described in section~\ref{sec:method}. On the other hand, we do not have $z$ for Cepheids-hosting galaxies, so we take $z$ from \Ned and combine it with $(\mu,\Delta\mu)$ from \Riess. More details about our SNe dataset are presented in appendix~\ref{app:U2p1plusRiessdata}.

\subsection{Velocity dispersion, galaxy surveys and density maps}
\label{sec:2m++}

SNe and Cepheids-hosting galaxies are not isolated objects as they are located within bounded structures, and consequently inherit rotational motions that should not be attributed to large scale inhomogeneities. These additional sources of noise are described as velocity dispersions that affect the distance modulus through
\begin{align}
\Delta \mu _{\rm v.d.} \approx \frac{5}{\log 10} \frac{\Delta v}{c z} \,,
\end{align}
where $\Delta \mu _{\rm v.d.}$ is the additional dispersion in the distance modulus, $\Delta v$ is the velocity dispersion, and $c z$ is the recession velocity at small redshifts.
Following \Riess we will add a velocity dispersion of $250$ \ks to SNe in order to account for their galactic plane motion, shown in the $\Delta \mu^{250}$ column of tables in appendix~\ref{app:U2p1plusRiessdata}.
In addition, since Cepheids-hosting galaxies are included in our dataset, we will discuss about a $\sim 40$ \ks intra-filament velocity dispersion observed in \cite{Tully:2007ue} and \cite{2015ApJ...805..144K}, and its effect in section~\ref{sec:FullSky}.

\begin{figure}[h]
\centering
\subfigure[Field 1]{
\includegraphics[height=.5\textwidth]{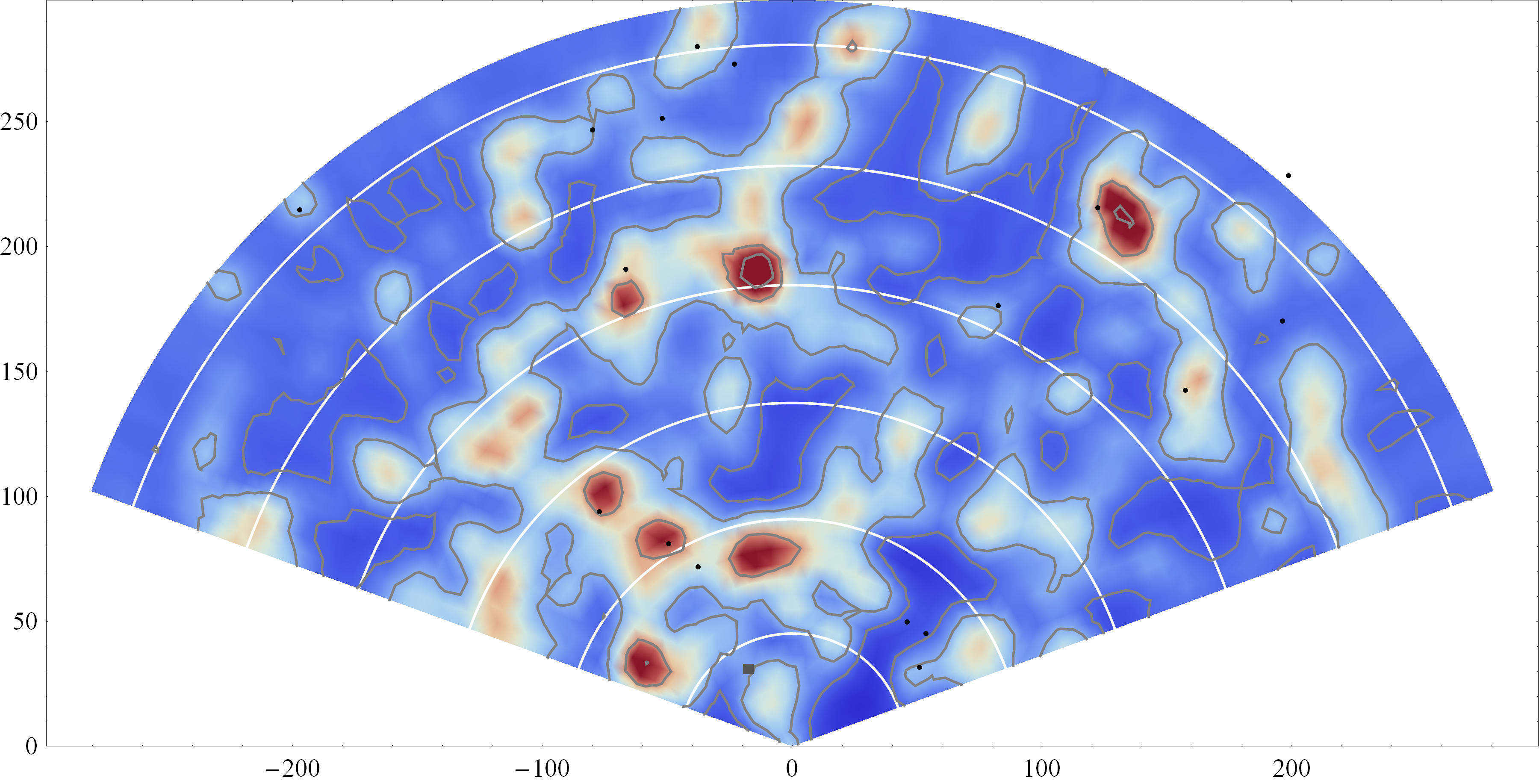}
}
\subfigure[Field 3]{
\includegraphics[height=.5\textwidth]{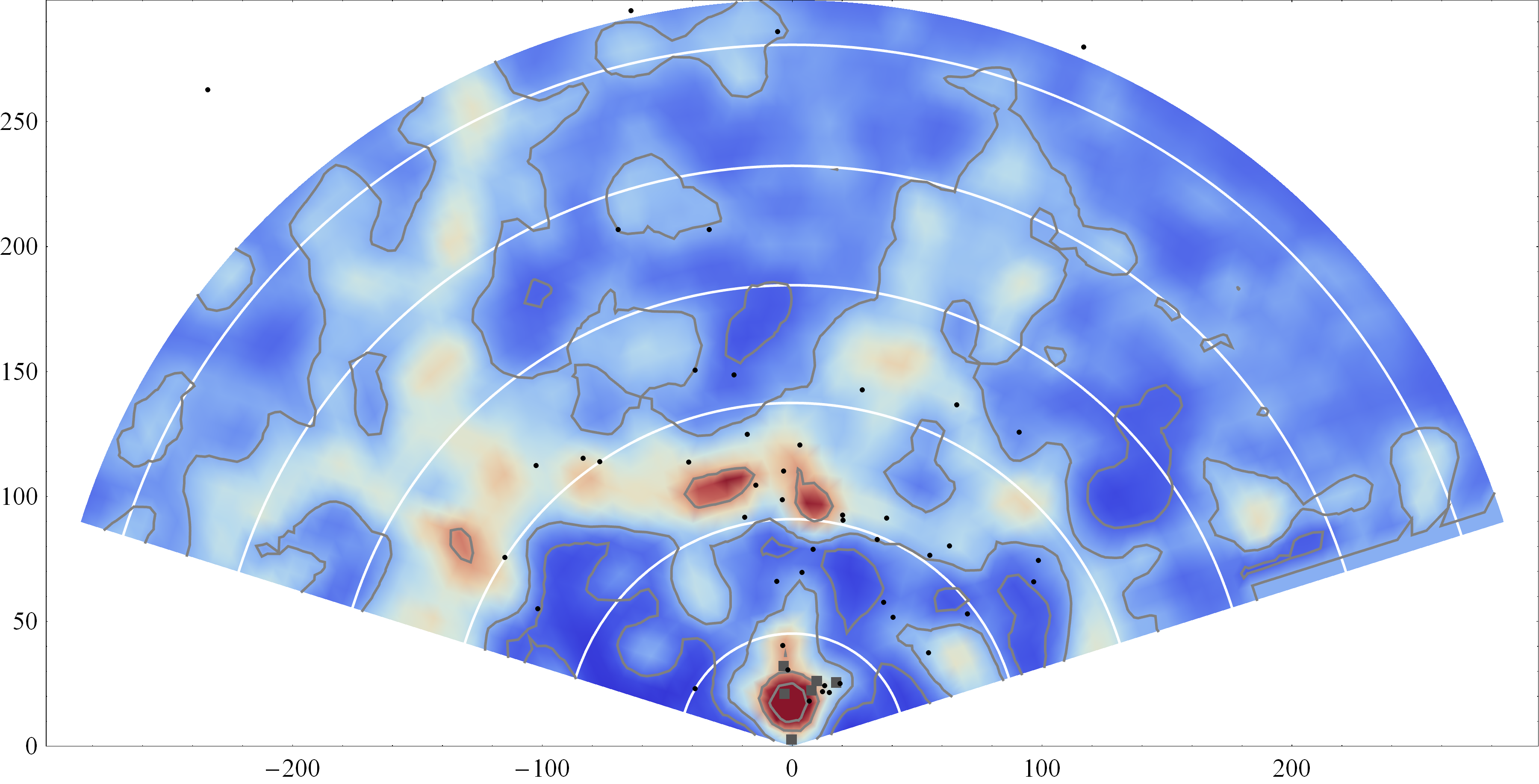}
}
\begin{tikzpicture}[overlay]
\node[rectangle,draw,thick,fill=white] (sc) at (-.8,7.85) {\includegraphics[height=.4\textwidth]{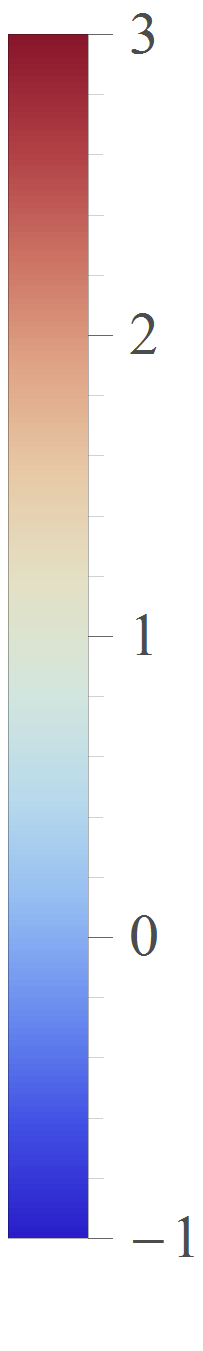}};
\end{tikzpicture}
\caption{Density map from \TMPP in Fields 1 and 3 of \Keenan, averaged along declination direction in ICRS coordinates. Gray squares correspond to Cepheids-hosting galaxies and black dots to SNe, using peculiar-velocity-corrected redshifts (see section~\ref{sec:NoVeloCorr}). White arcs correspond to $z=0.01,0.02,\ldots,0.06$ respectively, gray contours indicate iso-density lines of $\delta_C=-0.5,0,2,4$.}\label{fig:2mpp}
\end{figure}

We compare the results of our density field reconstruction to the luminous density data of the \TMPP galaxy redshift catalog \cite{Lavaux,Carrick15}. This catalogue extends the Two-Micron All-Sky Redshift Survey (2MRS), presenting photometry from 2MASS-XSC and redshifts of 2MRS, SDSS-DR7, and 6dFGRS (see references in these two papers). It covers almost the whole sky except for the zone of avoidance near the Milky-Way's galactic plane. Notice that in \TMPP data measured in redshift space are re-expressed in comoving coordinates and the (normalized) density contrast of observed galaxies $\delta_g^\ast(\vec{r})$ is transformed into matter density contrast $\delta(\vec{r})$ by $\delta_g^\ast (\vec{r}) = b^\ast \delta(\vec{r})$, with $b^\ast$ the linear bias factor. $\delta(\vec{r})$ is smoothed with a Gaussian filter of $4 \, h^{-1} \Mpc$. To visualize the distribution of the standard candles within the large scale inhomogeneities, the density maps of \TMPP averaged along the declination are shown for F1 and F3 in figure~\ref{fig:2mpp}.\par Peculiar velocities are obtained from the galaxy density through an equation of the form
\begin{equation}
\label{eq:velocity}
v(\vec{r}) = \frac{\beta^\ast}{4 \pi} \int_0^{R_{\rm max}} \dd^3 \vec{r}' \, \delta_g^\ast(\vec{r}') \frac{\vec{r}'}{r'^3} \quad ,
\end{equation}
where $\beta^\ast = 0.43$ is a best fit value and the upper limit of integration is the depth of the survey $R_{\rm max} = 200 \, h^{-1} \Mpc$, i.e. $z = 0.067$. 
Therefore \TMPP does not take into account the possibility of an inhomogeneity extending on scales larger than its depth, except for a dipole accounting for an external bulk flow. We remove this bulk flow so that peculiar velocity corrections are expressed in the CMB frame.
The lack of observations outside the depth of \TMPP could lead to a wrong estimation of the background density, the associated density contrast and consequently the peculiar velocity. It is therefore important to consider \Keenan as well since it is probing the density field on scales larger than \TMPP.


We compare our reconstructed density profiles to \Keenan analysis results (c.f. figure~11 of \cite{Keenan}), which are based on galaxies from the UKIDSS Large Area Survey and their spectroscopy taken from SDSS, 2DFGRS, and GAMA (c.f. references in \cite{Keenan}). For a better comparison with this study, we choose to do our fits up to a maximum redshift $z \leq \zmax = 0.2$. A second value of $\zmax = 0.4$ is considered to see how the fit stabilizes when considering the farthest \FN{farther} SNe, while remaining close enough for not having to consider weak lensing dispersion (cf. \cite{BenDayan:2013gc}). The under-density profiles presented in figure~11 of \Keenan extend up to $z \sim 0.07$ ($\sim 300 \, h_{70}^{-1} \Mpc$), with an over-dense shell surrounding at least F2 and to a lesser extent F3.
Since according to \Keenan there are evidences of inhomogeneities extending beyond the depth of \TMPP, we will not apply the \TMPP peculiar velocity correction to the data in section~\ref{sec:NOvcYESvd}, contrary to what was done in \Riess.
We will discuss more about the difference between \Keenan and \TMPP and especially the choice of the average density in section~\ref{sec:densityrescale}.


\section{Methodology}
\label{sec:method}

This section is devoted to the description of our statistical method employed to fit the distance modulus, and the description of the inversion method used to reconstruct the density profile along different directions. We also discuss the role of peculiar velocities in our study.

\subsection{Model independent distance modulus fitting}
\label{sec:distModulus}

We follow a systematic procedure to obtain model independent fits of the distance modulus data \HWC{$(z_i, \mu _i, V^{-1}_{ij})$ 
by minimizing the $\chi^2$ for the deviation from a homogeneous model prediction $\delta\mu_i = \mu_i-\mu^{\rm Planck} \left( z_i \right)$  
\be
\chi^2 = \sum _{ij} \Big{(} f \left( z_i \right) -\delta\mu_i \Big{)} V^{-1}_{ij}
\Big{(} f \left( z_j \right) - \delta\mu_j \Big{)} \,, \label{eq_chi2}
\ee
where $f$ is the fitting function that will be defined in eq. \ref{eq:RBF},  and} $\mu^{\rm Planck}(z)$ is the $\Lambda$CDM theoretical value of distance modulus at $z$ computed using \Planck cosmological parameters.

By model independent fit we mean that we do not make any a-priori assumption about the geometry of space-time and consequently about the functional form of the distance modulus. For example since we do not assume homogeneity, we consider fits more complicated than what was considered in \Riess, i.e. a simple shift. 
In this way the existence and the shape of inhomogeneities can be tested by comparing different fitting functions that we have no prior bias except for the degrees of freedom analysis presented at the end of this subsection.

Our model independent approach is based on decomposing the fitting function $f(z)$ with respect to a set of radial basis functions (RBFs) according to \footnote{It is known as the radial basis function network.}
\begin{align}
f (z) = w_0 + w_{-1} \, z + \sum_{m=1}^{N_{\rm NL}} w_m \, \Phi \left( \left| z - p_m \right| \right) \,, \label{eq:RBF}
\end{align}
where $\Phi$ is a very simple monotonic template function known as the radial basis function (RBF), here chosen to be $\Phi(r) = r^3$, $p_m$ are the non-linear parameters or ``centers'' of the RBFs, $w_m$ the linear parameters, and $N_{\rm NL}$ the number of RBFs. A fitting model is thus classified by a set of parameters $( N_0,N_{-1},N_{\rm NL} )$, where
\begin{itemize}
\item 
$N_0=1$ when the intercept parameter $w_0\neq 0$, and $N_0=0$ otherwise 
\item 
$N_{-1}=1$ when the slope parameter $w_{-1}\neq 0$, and $N_{-1}=0$ otherwise 
\item $N_{\rm NL}$ is the number of RBFs \,.
\end{itemize}
The total number of parameters for a $( N_0,N_{-1},N_{\rm NL} )$ model is thus $N_0 + N_{-1} + 2N_{\rm NL}$. 
In our analysis we will fit the function $f(z)= \left( \mu^{\rm obs}-\mu^{\rm Planck} \right) (z)$ as defined in eq. \eqref{eq_chi2}, where $\mu^{\rm obs}$ is the observed distance modulus. 
In the multi-dimensional case one could replace the redshift with the position vector, the centers $p_m$ with vectors $\vec{p}_m$, and $w_{-1}$ with a vector identifying a plane. The ``model parameters'' determining the fitting function, e.g. $\Phi(r)$ and $N_{\rm NL}$, are fixed for a given fitting model. But as explained later, we test and compare different models, with different numbers of RBFs, with/without the inclusion of $w_0$ (intercept) and/or $w_{-1}$ (slope) parameters. 

\begin{figure}[ht!]
\hspace{.08cm}\scalebox{0.8}{\input{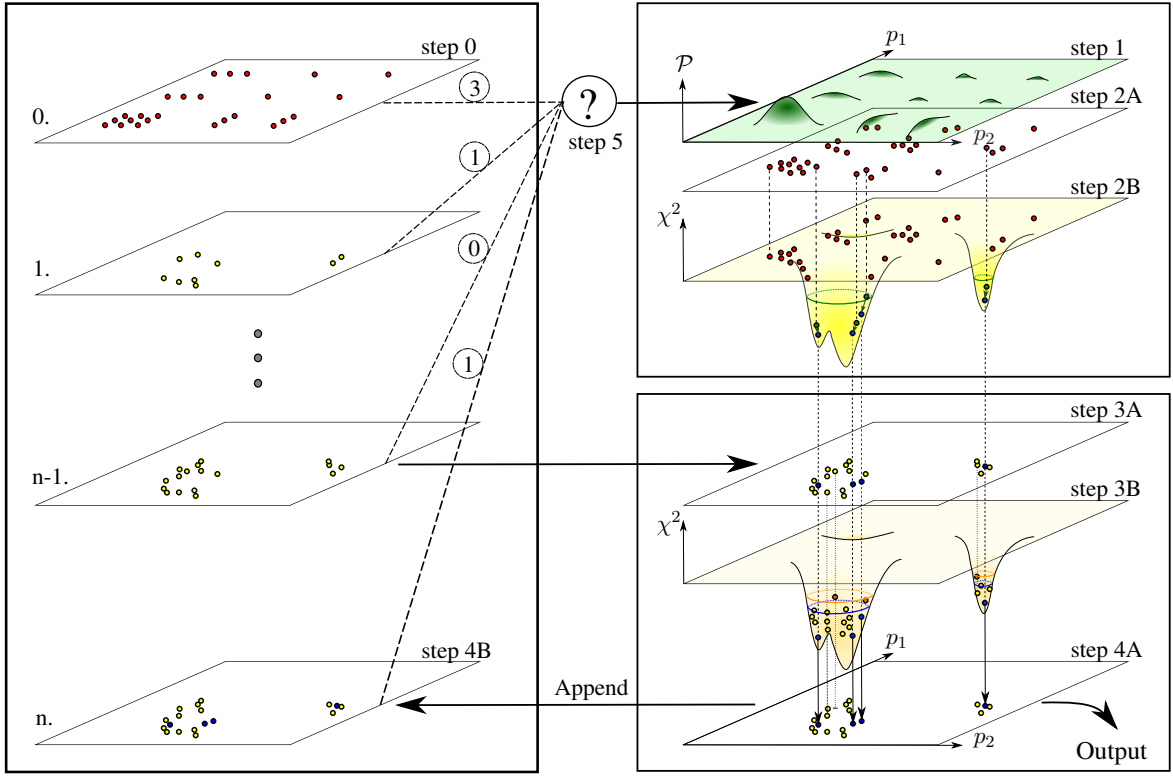}}
\caption{\label{fig:IllusMethod} Illustration of the different steps employed in our MCMC fitting method for non-linear parameters (centers) $\mathbf{p}$. Each dot in the figure corresponds to a specific $\mathbf{p}_a$, plus the associated $\mathbf{w}_a$ and $\chi^2$ given by Moore-Penrose pseudo-inverse. Red dots are points in $\mathbf{p}$ space, randomly sampled according to the probability density $\mathcal{P}$ from eq.~\eqref{eq:prob}, green dots are those selected for local optimization (LO), blue dots are points refined by LO algorithm, yellow dots are those from the previous loop, and orange dots are the points discarded according to the $\chi^2$ threshold. Green circles correspond to the threshold for local optimization. Orange and blue circles correspond to the selection threshold for the previous loop and the current one respectively. Numbers from 1 to n are labeling the sets $\{ \mathbf{p}_a \}_j$ inside the stack in the chronological order. Circled numbers illustrate the weights assigned to each of these sets. Although the number of sampling points is growing in the case represented in the diagram above, it can also decrease (in which case the time constraint plays an important role, as described in {\bf\textsc{Step 4A}}).}
\end{figure}

To find the best fit and the confidence band of a given model, we utilize Monte Carlo (MC), local optimization (LO) and linear regression (LR) for different types of parameters. For linear parameters $\mathbf{w} \equiv \left( w_{-1} \,,\ldots \,, w_{N_{\rm NL}} \right)$ we use the simple Moore-Penrose pseudo-inverse method, and for non-linear parameters $\mathbf{p} \equiv \left( p_1 \,,\ldots \,, p_{N_{\rm NL}} \right)$ we use a MC random sampling method and a LO algorithm, specifically Newton algorithm.
In order to speed up the MC process and make sure that the confidence band is fully exploited, we use a Monte Carlo Markov Chain (MCMC) algorithm to explore the non-linear parameter space. The different steps of the MCMC algorithm are illustrated in figure~\ref{fig:IllusMethod} and described in details in the following paragraphs.

\subsubsection{Details of the Monte Carlo Markov Chain method}
These are the steps of the MCMC method we employ for data fitting.

\paragraph{\textsc{Step 0 --- Initialization}:}

Due to the monotonic nature of RBFs, the initial distribution of the centers for the following MCMC analysis can be determined according to the positions of data points in $z$ space. We take redshifts of our SNe data points, using a Gaussian smoothing function specified in the next step to generate an identical distribution for each RBF center $p_m$. To generate an identical distribution for each RBF center $p_m$, we construct an initial input set $\left\{ \mathbf{p}_a \right\}_0$ for the Gaussian smoothing function by considering all possible combinations of redshifts from the data points
\begin{align}
\left\{ \mathbf{p}_a \right\}_0 &\equiv  \left\{ z_i \right\}_\mathcal{D}^{\otimes N_{\rm NL}} \;,
\end{align}
where $\left\{ \; \right\}_\mathcal{D}$ is the data set, $z_i$ is the redshift of the $i$-th data points, $\left\{ z_i \right\}_\mathcal{D}$ is the set of redshifts of all the data points, $N_{\rm NL}$ is the number of RBFs, $A^{\otimes n}$ is the tensor product of $A$ by $n$ times, and $\mathbf{p}_a$ in $\left\{ \mathbf{p}_a \right\}_0$ is the $a$-th combinatorial vector inside $\left\{ z_i \right\}_\mathcal{D}^{\otimes N_{\rm NL}}$, e.g. $\left( z_3\,, z_5\,, z_3\,,\ldots \right)$. $\left\{ \mathbf{p}_a \right\}_0$ is sent to the next step for distribution creation.

\paragraph{\textsc{Step 1 --- Distribution creation}:}

We create the distribution for MC by applying a Gaussian smoothing function acting on the input set $\{ \mathbf{p}_a \}$ with a bandwidth specified by the Silverman's rule. The Gaussian function is of the form
\begin{align}
\mathcal{P} \left( \mathbf{p} \right) &= \frac{1}{\mathcal{N}} \sum _{ \{ \mathbf{p}_a \} } e^{-\frac{1}{2} \left| \left( \mathbf{p} - \mathbf{p}_a \right) / \mathcal{B} \right| ^2} \;, \label{eq:prob}
\end{align}
where $\mathcal{P}$ is the probability in $\mathbf{p}$ space, $\mathcal{N}$ is the normalization factor, $\mathcal{B}$ is the bandwidth, and $\{ \mathbf{p}_a \}$ is a set of points in $\mathbf{p}$ space, being either $\{ \mathbf{p}_a \}_0$ (from {\bf\textsc{Step 0}}) or $\{ \mathbf{p}_a \}_j$ (from {\bf\textsc{Step 5}}).

\paragraph{\textsc{Step 2 --- MC sampling and local optimization}:}

(A) We use the distribution to generate random sampling points in $\mathbf{p}$ space, and (B) select only the best few percent according to the associated $\chi^2$ computed using linear parameters $\mathbf{w}$ given by Moore-Penrose pseudo-inverse. Then these selected sampling points are further refined slightly by the local optimization program using a multiple-step Newton algorithm.

\paragraph{\textsc{Step 3 --- Samples combining and selection}:}

(A) The sampling points produced in {\bf\textsc{Step 2B}} and the set of sampling points $\{ \mathbf{p}_a \}_{n-1}$ from the last ($(n-1)$-th) loop are combined together to form a new set $\{ \mathbf{p}_a \}_n$. Notice that the initial set $\{ \mathbf{p}_a \}_0$ 
should not be included in $\{ \mathbf{p}_a \}_1$. (B) We then keep sampling points with $\chi^2$ (computed like in {\bf\textsc{Step 2B}}) lower than the threshold defined as $\chi^2_{\rm thres} = t_{98\%} \min\limits_{\{ \mathbf{p}_a \}_n}(\chi^2)$, where $t_{98\%}$ is the 98th percentile of the student-t distribution with degrees of freedom of the system.

\paragraph{\textsc{Step 4 --- Output or storing in the stack}:}

(A) If the program takes too much time or the set of sampling points is large enough, the program terminates and the set $\{ \mathbf{p}_a \}_n$ will be the output of the whole MCMC program.
(B) If this is not the case, the set is stored in a stack consisting of all the previous sets, plus the initial one generated in {\bf\textsc{Step 0}}. For the newly stored list, we will assign a weight of 1 to it if the lowest $\chi^2$ in {\bf\textsc{Step 3B}} comes from the newly added samples, or 0 otherwise. The initial set $\{ \mathbf{p}_a \}_0$ has higher weight (3).

\paragraph{\textsc{Step 5 --- Selection of samples for distribution generation}:}

One of the sets in the stack ($\{ \mathbf{p}_a \}_j$) is selected as the input for {\bf\textsc{Step 1}}, with probability proportional to the weights set in {\bf\textsc{Step 4B}}, and we go back to {\bf\textsc{Step 1}} and start it all over again.

\subsubsection{Meta-fitting, F-test\HWC{ and the PRESS statistics}}

To determine the model parameters in eq.~\eqref{eq:RBF} we utilize the F-test
\begin{align}
\left( 1- \textit{Threshold} \right)^{~ \ell_1 - \ell_2} &\leq \mathcal{P}_\mathcal{F} \left( \ell_1 - \ell_2 ~,~ \ell_2 ~,~ \frac{\chi^2_1 - \chi^2_2} {\ell_1 - \ell_2} / \frac{\chi^2_2} {\ell_2} \right) \,, \\
\mathcal{P}_\mathcal{F} &\equiv 1 - \textit{CDF} \left( \mathcal{F} \left( \ell_1 - \ell_2 ~,~ \ell_2 \right) ~,~  \frac{\chi^2_1 - \chi^2_2} {\ell_1 - \ell_2} / \frac{\chi^2_2} {\ell_2}\right) \,,
\end{align}
where $\ell_A$ is the number of degrees of freedom corresponding to model $A$, {\it Threshold} is the likelihood of model $2$ having improvement over model $1$ (set to $\sim 95\%$ in our case), $\mathcal{F} \left( n_1 , n_2 \right)$ is the F-Ratio distribution of $\{ n_1 , n_2 \}$ parameters, $\textit{CDF} \left( \mathcal{D}, x \right)$ is the cumulative probability of the distribution $\mathcal{D}$ at $x$, and $\textit{CDF}$ stands for the cumulative distribution function.
We perform a step-wise regression by gradually increasing the number of fitting parameters and stop when the F-test fails.\footnote{Although it is known that step-wise regression underestimates the likelihood, we choose a very stringent threshold that compensates this effect.}

\HWC{We also consider the predicted residual error sum of squares (PRESS) statistics in conjunction with the F-test to make our model selection even more robust. The PRESS is defined as
\begin{align}
\text{PRESS} = \sum _{ij} \Big{(} f_i \left( z_i \right) - \delta\mu_i \Big{)} V^{-1}_{ij}
\Big{(} f_j \left( z_j \right) - \delta\mu_j \Big{)} \,, 
\end{align}
where $f_i \left( z_i \right)$ is the best fit function fitted using the data without the $i$-th entry, evaluated at that data point. The PRESS is a measure of the predictability of a model. Lower PRESS value indicates high predictive power and vice versa.}
In addition to different model parameters, we \COM{also} consider the effect of ``method parameters'' such as number of steps in local optimization, by taking results from different methods and combining them together (bagging). The bagging process stabilizes the outcome by further reducing the likelihood of falling into local minima.

\subsection{Inhomogeneity model and inverted density profile}
\label{sec:theory}

We will reconstruct the radial density profiles in different directions using an inversion method which was developed in \cite{inversion} for an observer at the center of an isotropic and inhomogeneous pure dust universe, modeled by a LTB metric. 
\AER{This is justified by the fact that
at low redshift the dominant linear order perturbative effect on the luminosity distance is the Doppler effect due to the component of the peculiar velocity along the line of sight, as shown for example in \cite{Romano:2016utn,Bolejko:2012uj}.
If we choose a spherical coordinates system centered at our position, the along the line of sight component of the peculiar velocity is simply the radial component, which at linear order is mainly affected by the radial distribution of matter. \HWCR{In appendix \ref{sec:DirectionalLTB} we show that up to the first order in the gravitational potential the luminosity distance can be approximately expressed in terms of the density along the line of sight, if lensing contribution are negligible, as supported by observations\cite{Hikage:2018qbn,Troxel:2017xyo,Kohlinger:2017sxk,Hildebrandt:2016wyi,Kilbinger:2014cea,VanWaerbeke:2013eya}, and that the relation between the density contrast along the line of sight and the gravitational potential in a generic space is  the same as for a central observer in a symmetric space.} This justifies mathematically  the use of a central observer in a LTB metric, which should be a good approximation as long as the non linear effects of lensing contributions are negligible.\HWCR{

}We do not assume in any way that the local structure is spherically symmetric respect with our position, and in fact we reconstruct different density profiles along different directions. 
In other words we are not assuming to be located at the center of spherical symmetry, since no center of symmetry is assumed, but at the center of a spherical coordinates system. The central position of the observer in a  LTB  metric is just a convenient choice of coordinates to compute the effects along the line of sight. Using an off-center observer would not give the same result because it would also add an unwanted peculiar velocity for the observer, which has already been removed by considering observational data in the CMB frame.
}

\COM{The validity of this approach is based on the assumption that lensing effects are negligible at low redshifts. Notice that in the LTB metric, an off-centered observer carries a peculiar velocity with respect to the CMB frame. Since in this paper redshifts are already expressed in the CMB frame, we must put the observer at the center of the LTB metric.}

The spherical symmetry and the central position of the observer are not posing any relevant restriction since the radial profile in any given direction is indeed a one dimensional quantity, and different profiles are taken along different directions. There is no fine tuning of the position of the observer, since at low redshift inhomogeneities along the transverse direction are expected to have negligible effects on the observed luminosity distance and there is no assumption about a global spherical symmetry. In other words the local structure can be anisotropic and there is no assumption about the existence of center of spherical symmetry. The local structure is modeled as a set of different radial density profiles along different directions as functions of the radial coordinate with respect to the same point, i.e. the point from which we observe the Universe, which is just at the center of our spherical coordinate system but not a center of symmetry. Analyzing different data in different regions of the sky thus allows to probe the radial density profile in different directions, reconstructing the local structure and its anisotropy.  

However, the presence of the cosmological constant enforces to modify the aforementioned dust model. Since we are interested in the low redshift SNe and Cepheids, the modification can be treated perturbatively. According to \cite{Romano:2016utn}, the cosmological constant at zeroth order only affects the background density, and therefore the growth rate $f$ of the inhomogeneity. This introduces a rescaling of the density contrast $\delta_C \propto f^{-1} = \Omega_{m0} ^{-0.55}$. At first order, one needs to consider two dominant effects. The first one comes from the modification of the deceleration parameter $q_0 = -1+3 \,\Omega_{m0} /2$ that changes the expansion history, resulting in a slight modification on growth rate of the inhomogeneity. The second one comes from the modification of the luminosity distance itself, directly affecting the density profile. Once these 3 effects are taken into account, a pure CDM model suffices to explain the observational data at small redshifts, as further detailed in appendix~\ref{sec:lambda}.

The LTB metric and the associated Einstein field equations (EFEs) can be written as
\begin{align}
\dd s^2 &= \dd t^2 - a^2\left[\left(1+\frac{a,_{r}r}{a}\right)^2 \frac{\dd r^2}{1-k(r)r^2} + r^2 \dd \Omega_2^2\right] \, ,\\
\left(\frac{\dot{a}}{a}\right)^2 &= -\frac{k(r)}{a^2}+\frac{\rho_0(r)}{3a^3}
\quad , \quad
\rho(t,r) = \frac{(\rho_0 r^3)_{, r}}{3 a^2 r^2 (ar)_{, r}} \, , \label{density}
\end{align}
where $a \equiv a(t,r)$ plays the role of a scale factor (and will be called this way in the following), $a r$ is the angular diameter distance, $k(r)$ can be interpreted as the spatial curvature, and the dot refers to the differentiation with respect to $t$. 
We adopt,  without loss of generality, a system of coordinates in which  $\rho _0$ is constant. The solution of the EFEs in eq.~\eqref{density} can be expressed in terms of the conformal time $\eta = \int ^t dt' / a\left(t',r \right)$ as
\begin{align}
a\left(\eta,r \right) &= \frac{\rho _0}{6 k(r)} \left[ 1- \cos \left( \sqrt{k(r)} \eta \right) \right] \,,\\
t\left(\eta,r \right) &= \frac{\rho _0}{6 k(r)} \left[ \eta -\frac{1}{\sqrt{k(r)}} \sin \left(\sqrt{k(r)} \eta \right) \right]+ t_b (r) \,,
\end{align}
where $t_b (r)$ is  called bang function, defining the time of big bang at different locations. In terms of cosmological perturbation theory the bang functions is related to decaying modes and, since these are tightly constrained by early Universe observations such as the cosmic microwave background radiation, we will assume $t_b = 0$.

The luminosity distance $D_L$ for a central observer in a LTB spacetime is  
\begin{align}
D_L (z) &= \left(1+z \right)^2 r(z) \, a \left( \eta \left(z \right) , r \left(z \right) \right) \,, \label{DL}
\end{align}
where $\eta \left(z \right)$ and $r \left(z \right)$ are the solutions of the radial ingoing null geodesic equations, and the redshift z is defined by
\begin{align}
\frac{f_s}{f_o} = \frac{1+ z_s}{1+ z_o} = \exp \left( \partial _{t_s} \int _{s}^{o} \frac{\dd t}{\dd r} \right) \,,
\end{align}
where  $f$ is the frequency, the subscripts $s$ and $o$ stand for the source and the observer positions respectively, and $t_s$ is the time at the source.
To obtain $r \left( z \right)$, $\eta \left( z \right)$ and $k \left( z \right)$, the relation  in eq.~\eqref{DL} needs to be inverted and solved together with the radial null geodesic equations, i.e. we need to solve an inversion problem.

\AER{
In order to fix the initial conditions of the the differential equations of the inversion problem we need to use physically observable quantities and relate them to the functions parametrizing the LTB metric. For this purpose it is convenient to use the low red-shift  Taylor expansion of the luminosity distance for a central observer in a LTB metric \cite{Romano:2009mr} and match it to that of a FRW metric, which in our case we will assume to correspond to the Planck background  cosmological  parameters. The matching of the two Taylor expansions is in fact an approximate solution of the inversion problem at the center, and for this reason provides the correct initial conditions for the numerical solution of the inversion problem. The matching of the first and second order terms of the red-shift expansions  leads to the conditions
\begin{align}
H_0^{FRW} \equiv \left(\frac{d D_L}{d z}\right)^{-1} \Big{|}_{z=0} &= a^{-1} \partial _t a\Big{|}_{z=0} \equiv H_0^{LTB} \, , \\
q_0^{FRW} \equiv 1 -H_0^{FRW} \frac{d^2 D_L}{d z^2}\Big{|}_{z=0} &\approx -a \left( \partial _t a \right)^{-2} \partial _t \partial _t a \Big{|}_{z=0} \equiv q_0^{LTB} \,,
\end{align}
where in the second equation we have assumed that $k'(0) \ll 1$, a condition which is consistent with  observational data and with the reconstructed density profiles, since a large value of $k'(0)$ would lead to an unrealistic central spike of the density contrast. This approximation is indeed quite accurate as shown in figure \ref{fig:linear}. Note that, as discussed in \cite{Romano:2009mr}, $H_0^{\rm LTB}$ and $q_0^{\rm LTB}$ are just convenient mathematical quantities introduced to parametrize the LTB metric and to compute the luminosity distance, but do not have a direct physical meaning.

\COM{
slope of the spatial curvature $\frac{dk}{dz}\big{|}_{z=0} \ll \left( a\big{|}_{z=0} H_{\rm LTB} \right)^2$ and $q_0 $ of order $1$. These two assumptions are valid since $\frac{dk}{dz}\big{|}_{z=0}$ is indeed small for all the model we used in the following analysis, and $q_0 \sim -0.53$ according to \Planck data. With these two conditions and the coordinate choice $a\big{|}_{z=0} = a_0\,$, $r\big{|}_{z=0} = 0\,$,}\COM{Assuming these initial conditions coming from the Taylor expansion of eq. \eqref{DL}:
\begin{gather}
a(z=0) = a_0 \; , \; H_{\rm LTB}(z=0) \equiv [a^{-1} \partial _t a] (z=0) = H_0 \, , \\
q_{\rm LTB} (z=0) \equiv \left[ -a \left( \partial _t a \right)^{-2} \partial _t \partial _t a \right] (z=0) = q_0 \; , \; r(z=0) = 0 \, ,
\end{gather}
}
}
The inversion problem \cite{inversion} can then be reduced to the solution of this system of three 1st-order ODEs:
\begin{gather}
\frac{\dd k}{\dd z} = \frac{ \sqrt{1-S^2} }{3 (1+z) S} \frac{ 2 k \tan(\tau /2) \mathcal{A} }{ 3 -\tau \csc(\tau) ( 2+ \cos(\tau) )} \, , \\
\frac{\dd \eta}{\dd z} = \frac{ 1 }{(1+z) \sqrt{k}} \left( \csc(\tau) \mathcal{B} - \frac{ \sqrt{1-S^2} }{3 S} \mathcal{A} \right) \, , \\
\frac{\dd r}{\dd z} = \frac{ \sqrt{1-S^2} }{(1+z) \sqrt{k}} \left( \frac{ \cos(\tau) +3\tau \csc(\tau) -4 }{ 3 -\tau \csc(\tau) ( 2+ \cos(\tau) )} \frac{ \csc(\tau) \mathcal{A} }{3} +\tan(\tau /2) \right) \, , 
\end{gather}
where we have defined
\begin{gather}
\tau \equiv \sqrt{k} \eta \quad , \quad
S \equiv \sqrt{k} r  \quad , \quad
\mathcal{A} = 1 -\cos(\tau) + \mathcal{B} \, , \\
\mathcal{B} = \frac{2}{S} \frac{ \left( a_0 H_0 \right)^{-3} (1+z) k^{3/2} }{1- \tan(\tau /2) \sqrt{1-S^2}/S} \left( 1 -\frac{1 + 2 q_0}{4 \left( 1+ q_0 \right)^2} \right)^{-1} \, \frac{\dd}{\dd z} \left( \frac{H_0 D_L \left(z\right) }{ \left( 1+z \right)^2 } \right) \, .
\end{gather}
and for simplicity we are denoting $H_0^{\rm LTB},q_0^{\rm LTB}$ as $H_0,q_0$.
A detailed derivation of these equations is given in \cite{inversion}. Notice that $\rho _0$ and the initial conditions $k(z=0)$, $\eta(z=0)$ are fixed from the values of to $a_0$, $H_0$ and $q_0$.
From the solution of this ODEs system, the density in eq.~\eqref{density} can be expressed as
\begin{align}
\rho (z) = \rho \left(t \left( z \right) , r \left( z \right) \right) = \frac{a^{-3} \rho _0}{1 + r  \left. \partial _r \ln a \right| _t} = \frac{a^{-3} \rho _0}{1 + r \left( \left. \partial _r \ln a \right| _\eta - \left. \frac{\partial t}{a \partial r} \partial _\eta \ln a \right| _r \right)} \,,
\end{align}
where $\rho _0 = 3 H_0^2 \left[ 1 -\frac{1 + 2 q_0}{4} \left( 1+ q_0 \right)^{-2} \right]$ is the background density at the current time.

\AER{
In appendix~\ref{sec:lambda} it is shown that the density contrast $\delta_C$ w.r.t. the background homogeneous Universe can be approximated as
\begin{align}
\delta _C = f^{-1} \left( \frac{\rho_{\rm inv} \left(D_L, z \right)}{\rho_{\rm inv} \left(D_L^{\rm Planck}, z \right)}-1 \right) \,.
\label{eq:deltaC}
\end{align}
in which  we denote with $D_L$ the observed luminosity distance\COMT{ and}\HWCR{,} with $D_L^{\rm Planck}$ the theoretical $\Lambda$CDM luminosity distance corresponding to \Planck parameters\HWCR{, and with $f$ the growth rate of the density contrast
\begin{align}
f = \frac{d\ln \delta_C \left( a, \vec{x} \right) }{d\ln a} \equiv \frac{d\ln \left[ D\left( a \right) \delta_C^{\text{now}} \left( \vec{x} \right) \right]}{d\ln a} = \frac{d\ln D\left( a \right)}{d\ln a} \,, \label{eq:f}
\end{align}
where  $\vec{x}$ is the comoving coordinate, $\delta_C \left( a, \vec{x} \right) \equiv D\left( a \right) \delta_C^{\text{now}} \left( \vec{x} \right)$, and $\delta_C^{\text{now}} \left( \vec{x} \right)$ is the density contrast at present time.} 
The density contrast defined here accounts for the cosmological constant through the three corrections proposed at the beginning of this subsection, namely the rescaling of the growth rate $f$, the matching of the deceleration parameter $q_0$ and the normalization w.r.t. the background $\rho_{\rm inv} \left(D_L^{\rm Planck}, z \right)$. More details\HWCR{ about the derivation and the validity of eq.~\eqref{eq:deltaC}} are presented in appendix~\ref{sec:lambda}.

\AER{Notice that while the inversion is mathematically always possible as long as the equations admit a solution, in some cases the density contrast can be smaller than -1. These cases should be interpreted as the indication that no physically viable inhomogeneity could explain the deviation between the observed $D_L$ and $D_L^{\rm Planck}$, and consequently unphysical reconstructed density profiles are excluded from the analysis.}
}

\subsection{\HWC{Confidence bands and invertible bands}}
\label{sec:confidenceband}

\HWC{The confidence band of the distance modulus
, and the corresponding confidence band of the inverted density contrast are obtained according the the scheme described below.}\AER{

\paragraph{\textsc{Step 1}} Sample the parameters space for the distance modulus model using the MCMC algorithm described in section \ref{sec:distModulus}, keeping the samples with $\chi^2$ lower than the threshold defined by the student-t distribution, and obtain the confidence bands. 
\paragraph{\textsc{Step 2}} Reconstruct the density contrast from the distance modulus fitting functions obtained in {\bf\textsc{Step 1}}.
If a fitting function corresponds to an inverted density contrast smaller than -1, then the function is discarded as non-physical, otherwise is accepted and is called invertible. The set of all invertible fitting functions is shown in the plots as the invertible band. The set of inverted density contrasts of all invertible fitting functions is shown in the plots as the confidence band of the inverted density contrast.}
\newline

\COM{Finally, we obtain the confidence bands of the reconstructed density profiles by first fitting the distance observations with MCMC and then applying the inversion to the fitted functions within different confidence levels.
According to the physical density contrast condition in the last paragraph, we define invertible bands of a model as the set of fitted functions belonging to that model which can be inverted into physically acceptable density profiles with $\delta_C$ always greater than -1.}
%

Since some of the samples may be discarded during {\bf\textsc{Step 2}}, \HWC{the}\COM{These} invertible bands are thus narrower than their corresponding \COM{$\chi^2$-based} confidence bands \COM{derived in subsection~\ref{sec:distModulus}}\HWC{obtained in {\bf\textsc{Step 1}}}. In later plots we show both the 68\% and 95\% confidence bands and their associated invertible bands.
\COM{Notice that we always discard models whose best fits are not invertible as the minimal $\chi^2$ of these models would be higher than $\chi^2$ obtained through the fitting procedure once we take the invertibility into account.}

\subsection{Peculiar velocity correction}
\label{sec:NoVeloCorr}
The effect of perturbations on the luminosity distance can be computed using different gauges or methods (c.f. \cite{Misao,Bonvin:2005ps,BenDayan:2012wi,BenDayan:2013gc,Macaulay16}). The dominant effect at low redshift is due to the peculiar velocity of sources (c.f. \cite{BenDayan:2012wi,BenDayan:2013gc,Romano:2016utn}), which can be computed by appropriately correcting the background cosmological redshift. 
The typical whole-sky averaged variance of the distance modulus due to this peculiar velocity is $\gtrsim 0.02$ mag for $z \lesssim 0.2$, $\gtrsim 0.1$ mag for $z \lesssim 0.05$, and even larger when averaging over certain sky windows. At low redshifts, one can remove the effects of the peculiar velocities 
by correcting the observed redshift according to $\zbar = \zobs - \textbf{v}_s \cdot \textbf{n}$, where $\textbf{v}_s$ is the peculiar velocity of the SNe and $\textbf{n}$ is a unit vector in the direction of propagation from the emitter to the observer. The observed redshift $\zobs$ is the \UTPO value (simply called $z$) and $\zbar$ is the adjusted value that corresponds to the background redshift. 
At small redshift the luminosity distance is approximately given by
\be
D_L = \frac{\zobs}{H_0} \left( 1 - \frac{\textbf{v}_s \cdot \textbf{n}}{\zobs} \right) + \Ocal(\zobs^2) = \frac{\zbar}{H_0} + \Ocal(\zbar^2) ~.
\ee
The supernovae redshifts of \UTPO are expressed in the CMB frame, thus our observer peculiar velocity $\textbf{v}_o$ is already taken into account (and as such does not appear in the above equation).

If interested in background cosmological parameters, one could thus correct the luminosity distance 
for the peculiar velocities effects according to the procedure explained above. Relying on the 2M++ catalogue, which is limited in depth, we could define the total peculiar velocity of standard candles as $\textbf{v}_{s}^{\rm pec} = \textbf{v}_{s}^{\rm 2M++}(z<0.067) + \textbf{v}_{\rm s}^{>}(z>0.067)$, 
where the first term corresponds to the velocity field which can be inferred from \TMPP and the second one corresponds to velocities associated to inhomogeneities larger than the depth of \TMPP. Finding the redshift correction of an object, hence deducing its real position, requires an iterative process for which we evaluate the velocity at each intermediary position (see appendix~B of \cite{Carrick15}). The corrected data can then bring a different value of $H_0^{\rm loc}$ with respect to CMB. 

It was shown in \cite{Romano:2016utn} that in the linear perturbative regime the distance to the last scattering surface is not significantly  affected by t local structure because the effect is proportional to the volume average of the density contrast over a sphere of radius equal to the comoving distance, which is asymptotically negligible.  
The same effect can on the contrary be important for objects located inside the inhomogeneity, since it modifies their luminosity distance with respect to a homogeneous Universe, introducing a local modification of the Hubble flow velocity. If the peculiar velocity with respect to the Hubble flow was perfectly known then its non-relativistic effects on the luminosity distance at low redshift could be removed by applying the redshift corrections. On the other hand, if the peculiar velocity field is obtained from density maps extending to scales smaller than the size of the inhomogeneity, like it is the case here with 2M++, then the corresponding contribution to the peculiar velocity cannot be determined and the redshift corrections will not remove the effects of this large scale inhomogeneity.

As a consequence $H_0$ could be miss-estimated, since it would include a contribution from local structure, which on sufficiently larger scales is negligible as explained above, leading to an apparent tension between local and large scale estimations. 
We will thus use uncorrected redshifts and directly compare our density profiles with the ones obtained from the independent observations of \TMPP and \Keenan. We will also apply the \TMPP redshift correction to test if it is enough to remove the effects of inhomogeneities, for example when checking the consistency with other previous analyses such as \Riess. In that study the redshift correction was in fact applied under the assumption that no other structure was present, and that the corrected data were completely free from the effects of local inhomogeneities, and could thus be used to estimate background parameters such as $H_0$. Our analysis shows that \TMPP redshift correction does not completely remove the effects of inhomogeneities, hinting to the existence of inhomogeneities extending on scales larger than its depth.

\section{Setup}
\label{sec:setup}

Before proceeding to our reconstruction of radial density profiles from luminosity distance observations, we need to build the input functions by fitting distance modulus observations according to the methodology explained in section~\ref{sec:method}. A $\Lambda$CDM model with \Planck parameters is chosen as our background and we fit the difference between the observed distance modulus $\mu ^{\rm obs}$ and the background distance modulus $\mu ^{\rm Planck}$ accordingly. Notice that the choice of a different background would directly affect the resulting density contrast. Since the results presented in section \ref{sec:result} and \ref{sec:NOvcYESvd} are always assuming a background with \Planck parameters, we may arrive at different density contrast and different conclusion from \Riess even with a similar dataset. The effect of choosing different backgrounds is discussed in detail in section~\ref{sec:densityrescale}.

The distance modulus dataset we use contains both \Riess Cepheids-hosting galaxies and SNe from \UTPO with $250$ \ks velocity dispersion added to SNe, cut at $\zmax =0.2$ for most cases, as explained in section~\ref{sec:sneia}. The only exception for the redshift cut is the directional analysis along subregion F3, where $\zmax =0.4$ is also considered. Another exception is that in section~\ref{sec:FullSky} we also add varying amounts of velocity dispersion to Cepheids-hosting galaxies.

We perform two separate analyses of the data: one applying peculiar velocity correction of \TMPP, in section~\ref{sec:result} and one without correction in section~\ref{sec:NOvcYESvd}. The motivation for performing both analyses is that the 2M++ density maps may not be correctly normalized with respect to the average density of the Universe, while the inversion method is by construction correctly normalized, so that the comparison of inverted density profiles obtained from uncorrected data can clarify the issue of normalization with respect to \TMPP and \Keenan. 
The corrected distance modulus is denoted as $\mu ^{\rm cor}$, different from the uncorrected $\mu ^{\rm obs}$.

We recall that a fitting model is given by the set of parameters $( N_0,N_{-1},N_{\rm NL} )$, as explained in Sec. \ref{sec:distModulus}, \HWC{which we use to identify our models.} 
Notice that a $(1,0,0)$ model, i.e. $f (z)=\mu ^{\rm obs}(z) - \mu ^{\rm Planck}(z) = w_0$ , corresponds to a locally homogeneous model with an apparent value of the Hubble parameter given by
\be
H_0^{\rm loc} \equiv H_0^{\rm Planck} {10}^{- f (z=0) /5} = H_0^{\rm Planck} {10}^{- w_0 /5} = {10}^{- w_0 /5} (\, 66.93 ~ \text{\ksm} ) \,. \label{eq:Hloc}
\ee
In the rest of the paper we will call these models locally homogeneous, but it is important to note that these are local inhomogeneities with respect to the background density of the Universe corresponding to the larges scale \Planck cosmological parameters. They are homogeneous in the sense that their radial density profile is constant within the inhomogeneity, which in the case of spherical symmetry are also called Hubble Bubble.

However for models more complicated than a $(1,0,0)$ model, the local Hubble parameter defined in eq.~\eqref{eq:Hloc} represents the observed Hubble \HWC{parameter at $z=0$}\COM{flow near the local cluster}, and is therefore different from the one in \Riess where SNe with $0.0233<z<0.15$ are considered.

In section~\ref{sec:NOvcYESvd} a selection of outliers is necessary in order to find sufficiently smooth distance modulus fits to be inverted into density profile. We do this selection considering the list of potential outliers \HWC{$\mathcal{L} = \{$NGC 4536, NGC 4424, NGC 3447, NGC 3370, NGC 3021, 1999cl, 2007bz, 2006x, 2005ag, 2004gu, 2001v, 2008bf\! $\}$}
and select the best model according to a F-test {\it Threshold} around $95\%$. Additionally, since we are trying to obtain the confidence band of the inverted density contrast, the invertibility of the best fit is also considered as a physical requirement.



\section{Fitting of the distance modulus with  \TMPP peculiar velocity corrections} 
\label{sec:result}

In this section we demonstrate that even after considering the \TMPP peculiar velocity corrections (PVC), some evidence of  additional anisotropy and inhomogeneity of the local Universe still persist, especially along the subregion F3. This is a hint that \TMPP density maps cannot fully explain the observed luminosity distance, contrary to the \COM{usual} assumption of \Riess.
Notice that the evidence of this inhomogeneity not captured by \TMPP 
relies on the Cepheids data of \Riess, and their superior precision compared to SNe.

\subsection{Full sky analysis}

\label{sec:FullSky}

As a preliminary consistency test of our method, we fit the data of the 20 Cepheid-hosting galaxies from \Riess together with \UTPO dataset with $\zmax =0.2$ after the calibration given in eq.~\eqref{eq:mucorr}, and the \TMPP peculiar velocity correction.
 
\subsubsection{Locally Homogeneous fits}

Assuming local homogeneity, i.e. as previously explained a $(1,0,0)$ model for the distance modulus fit, and adding the effect of the $250$ \ks velocity dispersion to both SNe and Cepheid-hosting galaxies, we obtain \HWC{$H_0^{\rm loc} = 75.17 \pm 2.12 \, \text{(stat.)}$} \ksm, as shown in figure~\ref{fig:FullSky_Mu_Riess}, in good agreement with the $73.24 \pm 1.61 \, \text{(stat.)} \pm 0.66 \, \text{(sys.)}$ \ksm value of \Riess.
This can be considered a consistency check of our data analysis method but it is not an evidence of homogeneity, since other inhomogeneous models fit better the data as we will show later, depending on the value of the velocity dispersion of the Cepheids-hosting galaxies $v_{c}$.

Considering that every Cepheids-hosting galaxy of \Riess contains hundreds of Cepheids scattered over the whole galaxy plane, the same 250 \ks velocity dispersion should not be applied to both Cepheids-hosting galaxies and SNe. In fact, as explained in section~\ref{sec:2m++}, a $v_c=$ 40 \ks dispersion is estimated from the intra-filament motion (local sheet in \cite{Tully:2007ue} and Leo Spur in \cite{2015ApJ...805..144K}). We therefore consider it in our analysis, in addition with the case of $v_c=$ 0 \ks as a reference.

As shown on the left column of figure~\ref{fig:FullSky}, when gradually reducing  $v_c$ the lower redshift data points become more and more dominant, leading to an increasing $\chi ^2 _R$ from \HWC{1.48 ($v_c=$ 250 \ks), to 4.12 ($v_c=$ 40 \ks) and to 12.1} ($v_c=$ 0 \ks) for the $(1,0,0)$ model.

\subsubsection{Inhomogeneous fits}

\begin{figure}

\subfigure[Dataset with $v_c = 250$ \ks]{
\includegraphics[width=.48\textwidth]{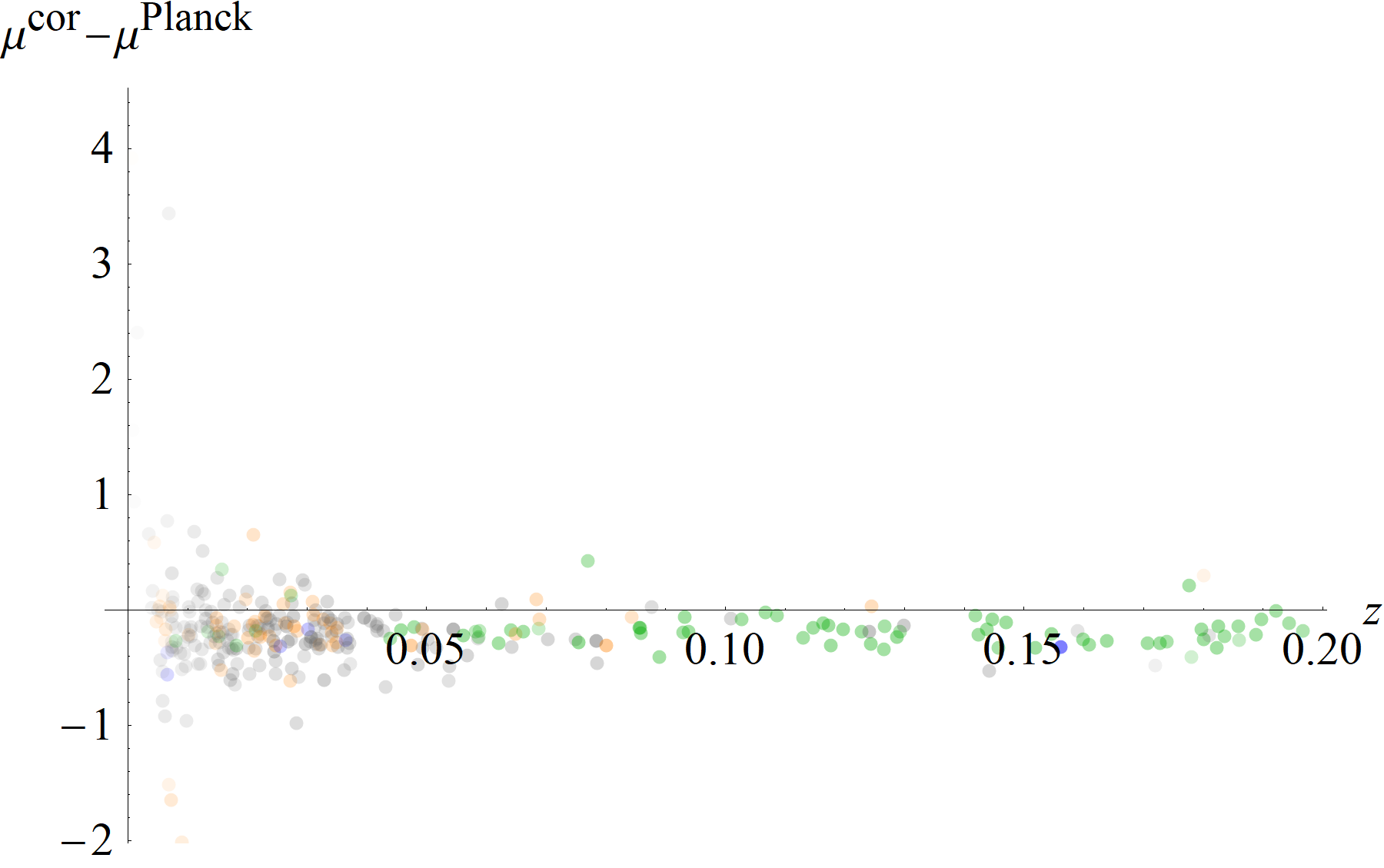}\llap{\raisebox{.19\textwidth}{\includegraphics[width=.24\textwidth]{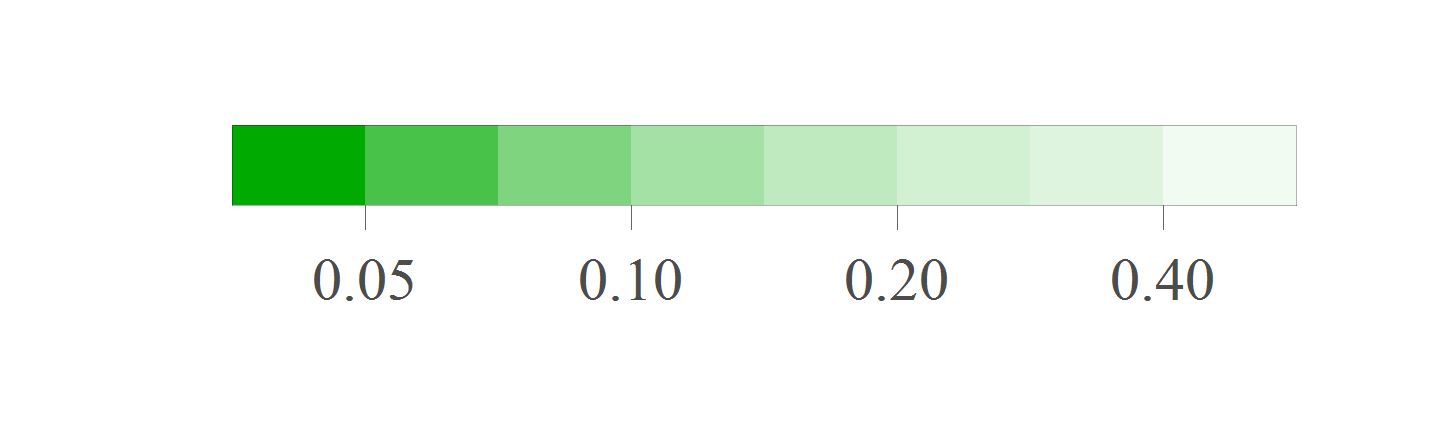}}}
\label{fig:FullSky_data_Riess}
}
\subfigure[Best fit of \ref{fig:FullSky_data_Riess} with $(1,0,0)$ model ]{
\includegraphics[width=.48\textwidth]{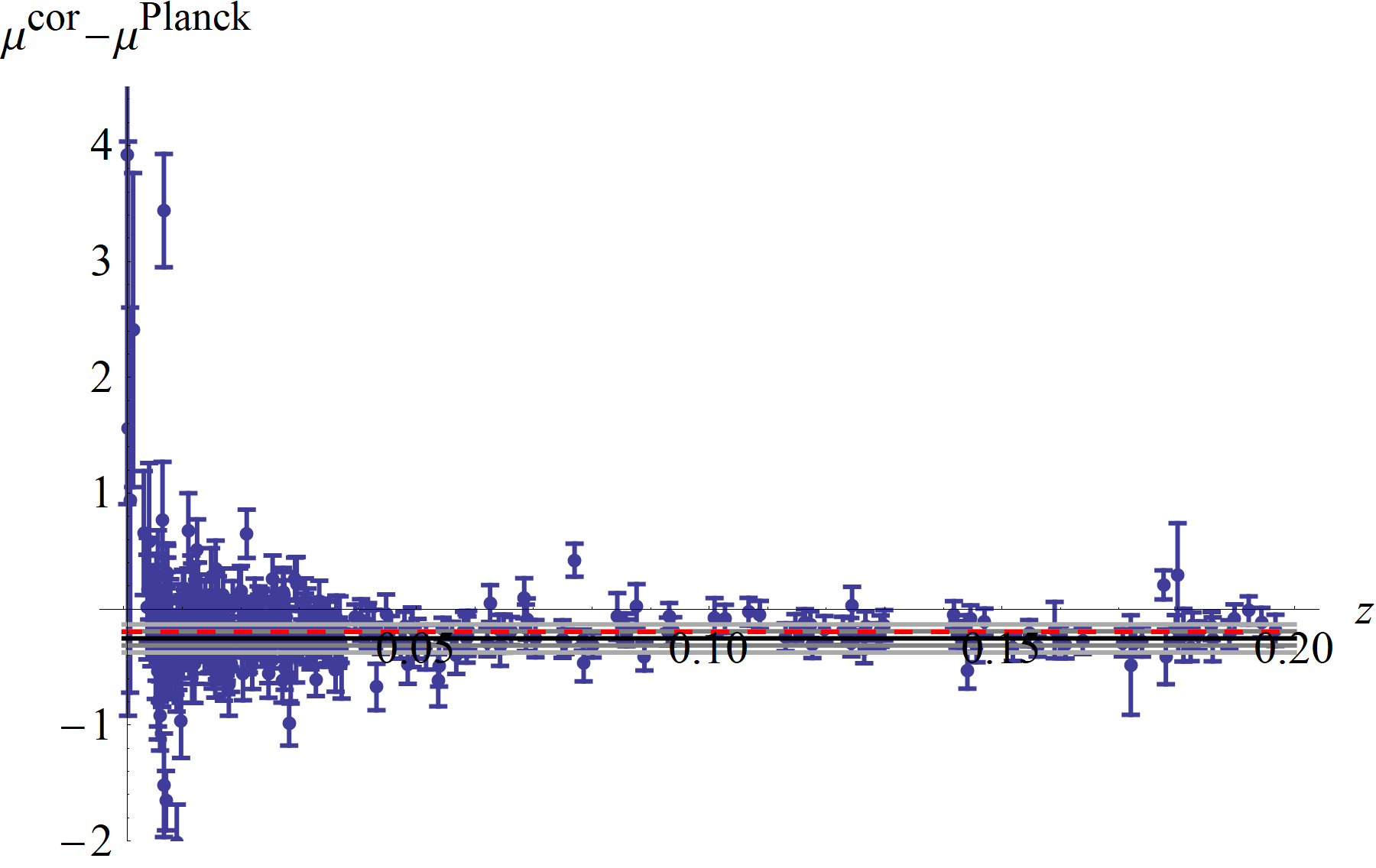}
\label{fig:FullSky_Mu_Riess}
}
\subfigure[Dataset with $v_c = 40$ \ks]{
\includegraphics[width=.48\textwidth]{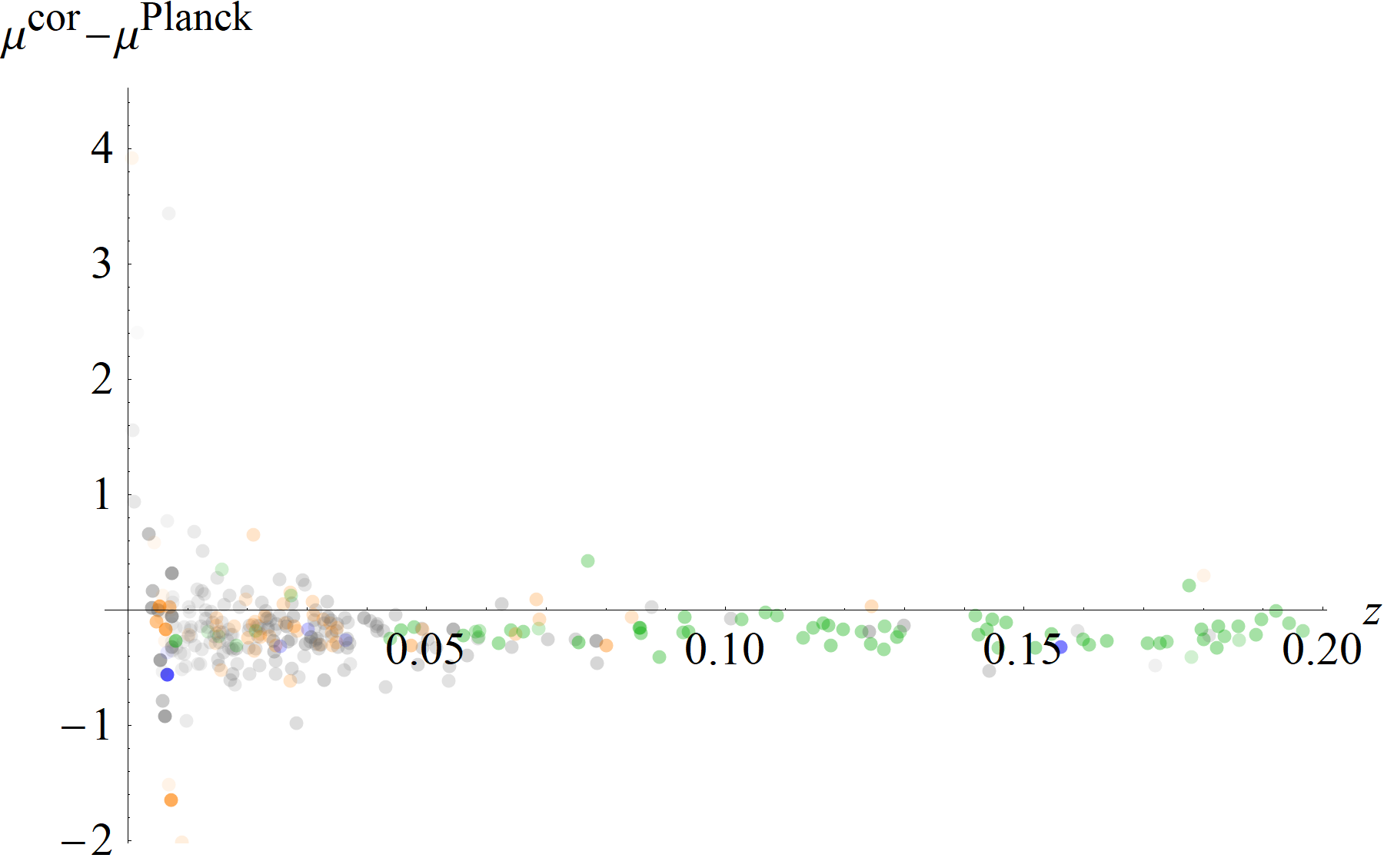}\llap{\raisebox{.19\textwidth}{\includegraphics[width=.24\textwidth]{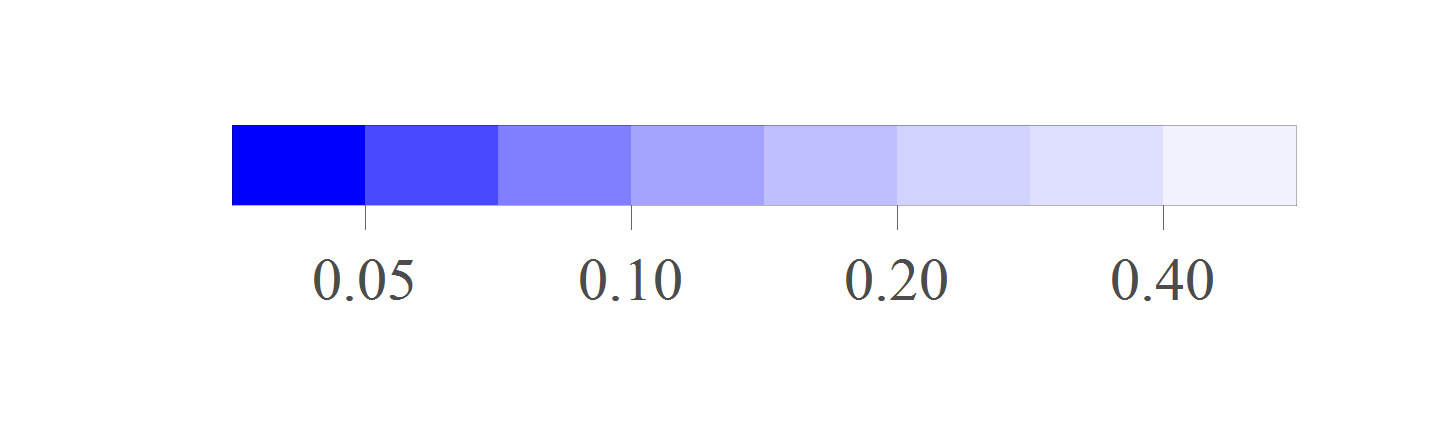}}}
\label{fig:FullSky_data_Cluster}
}
\subfigure[Best fit of \ref{fig:FullSky_data_Cluster} with $(0,1,3)$ model]{
\includegraphics[width=.48\textwidth]{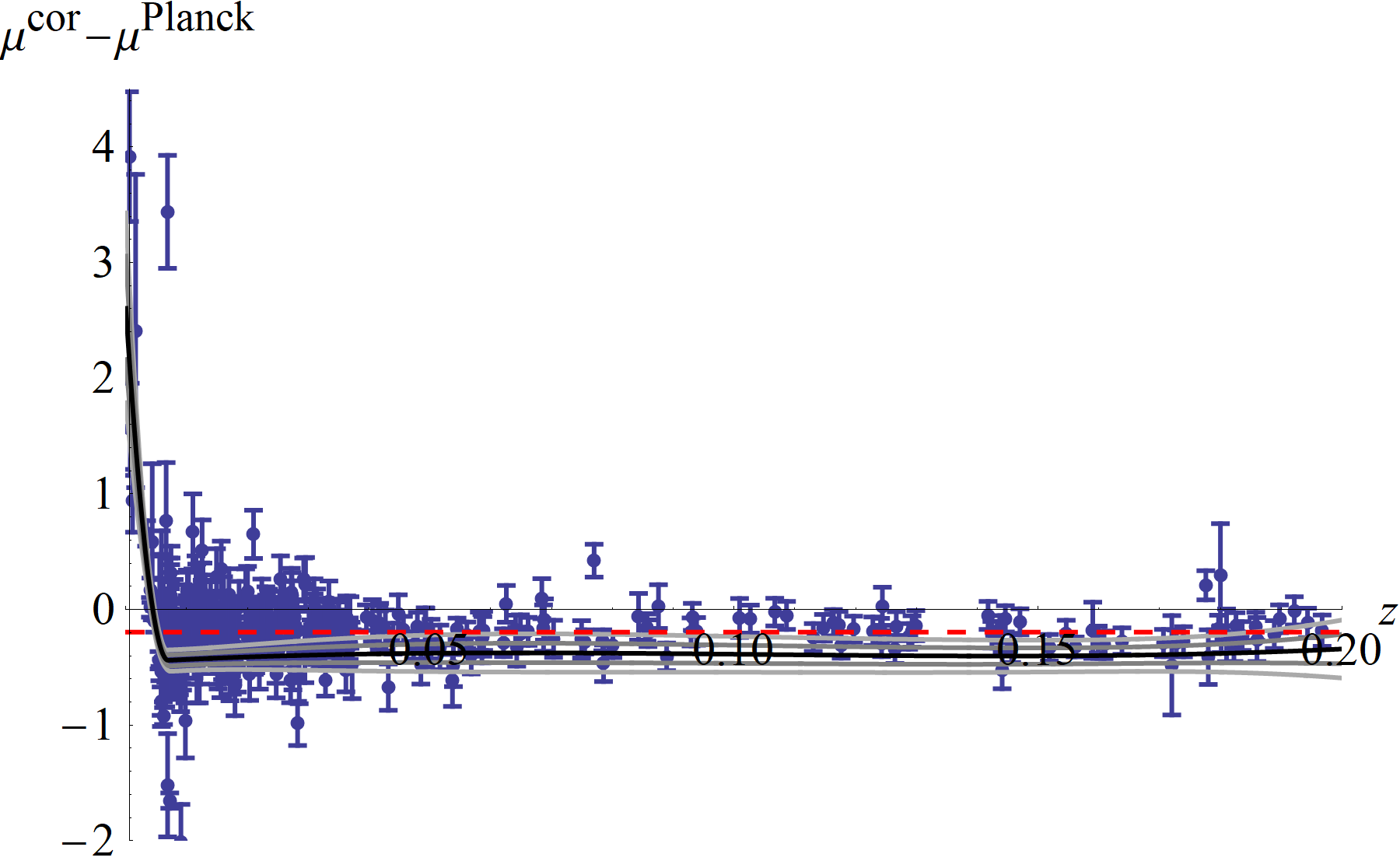}
\label{fig:FullSky_Mu_Cluster}
}
\subfigure[Dataset with $v_c = 0$ \ks]{
\includegraphics[width=.48\textwidth]{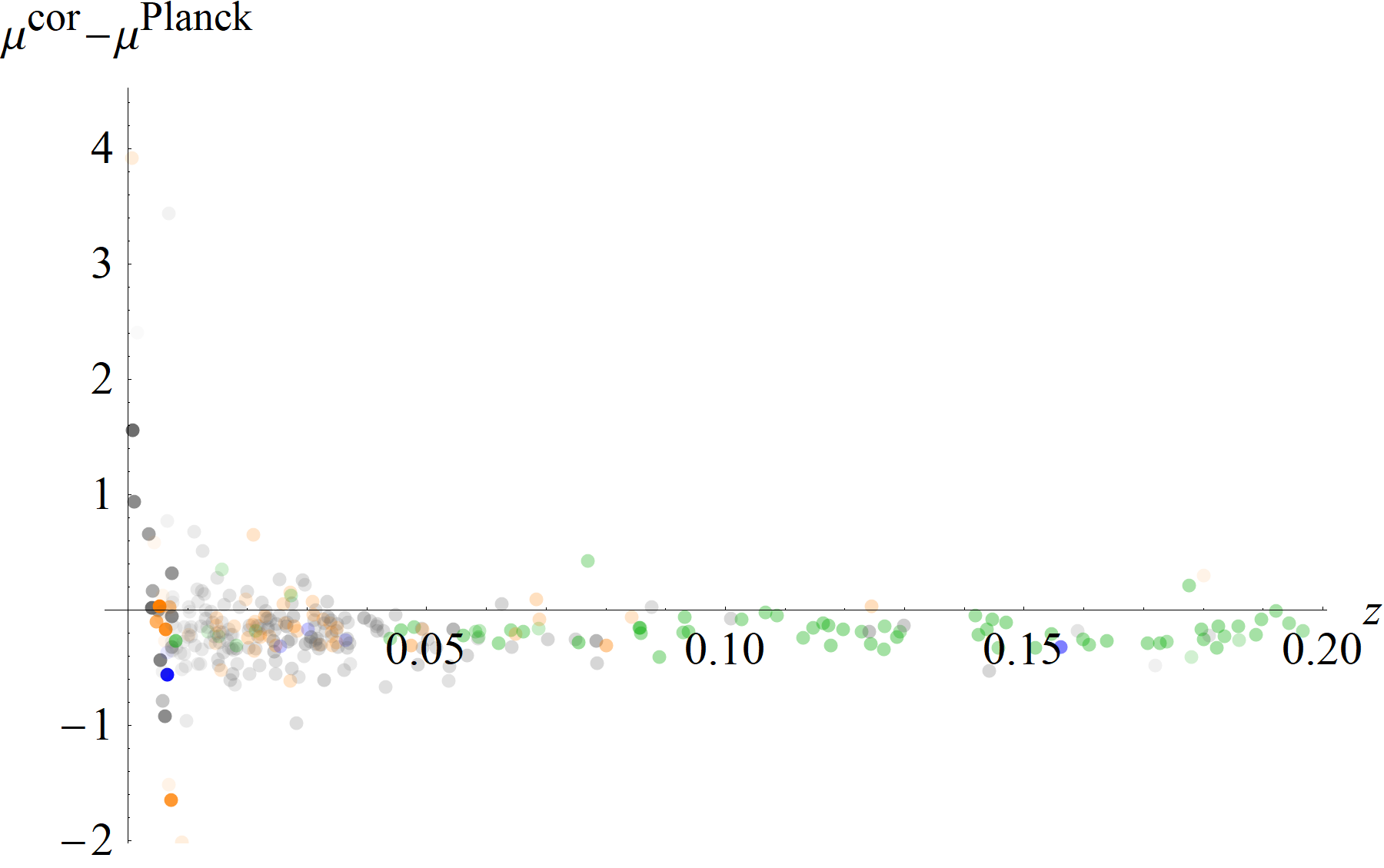}\llap{\raisebox{.19\textwidth}{\includegraphics[width=.24\textwidth]{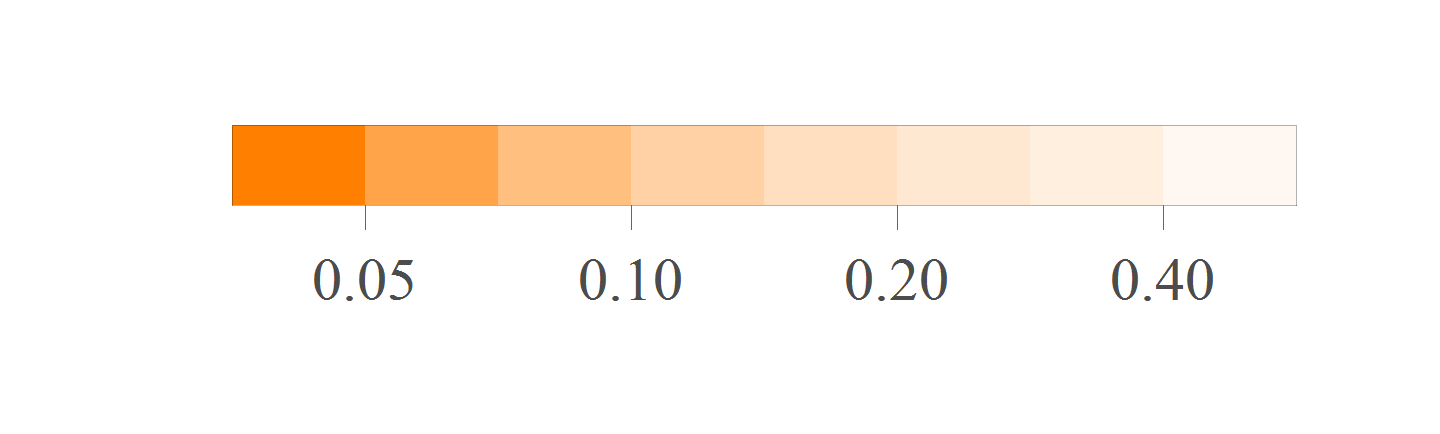}}}\llap{\raisebox{.12\textwidth}{\includegraphics[width=.24\textwidth]{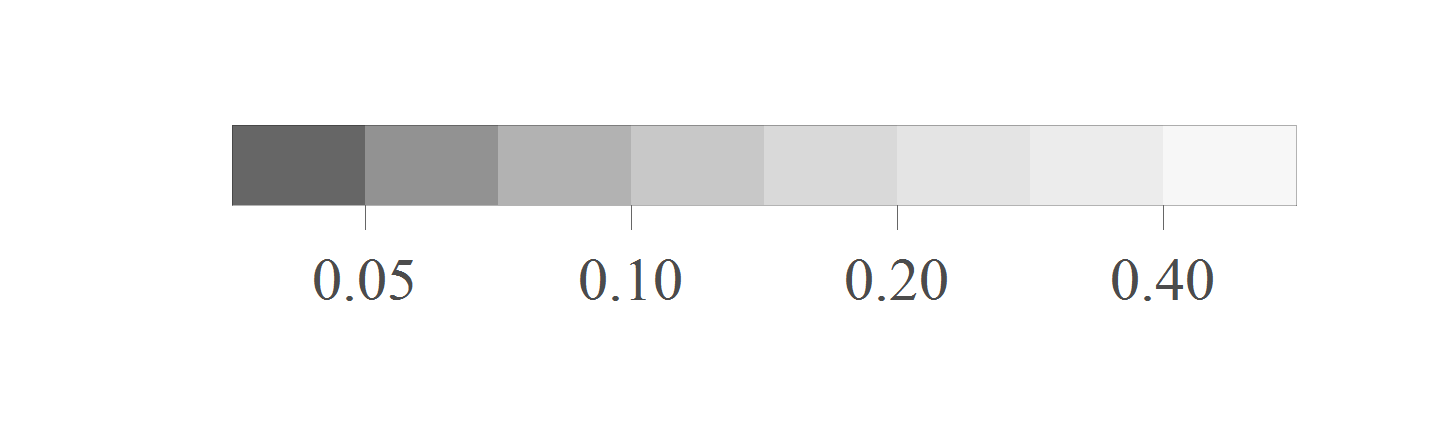}}}
\label{fig:FullSky_data_Normal}
}
\subfigure[Best fit of \ref{fig:FullSky_data_Normal} with $(1,1,3)$ model]{
\includegraphics[width=.48\textwidth]{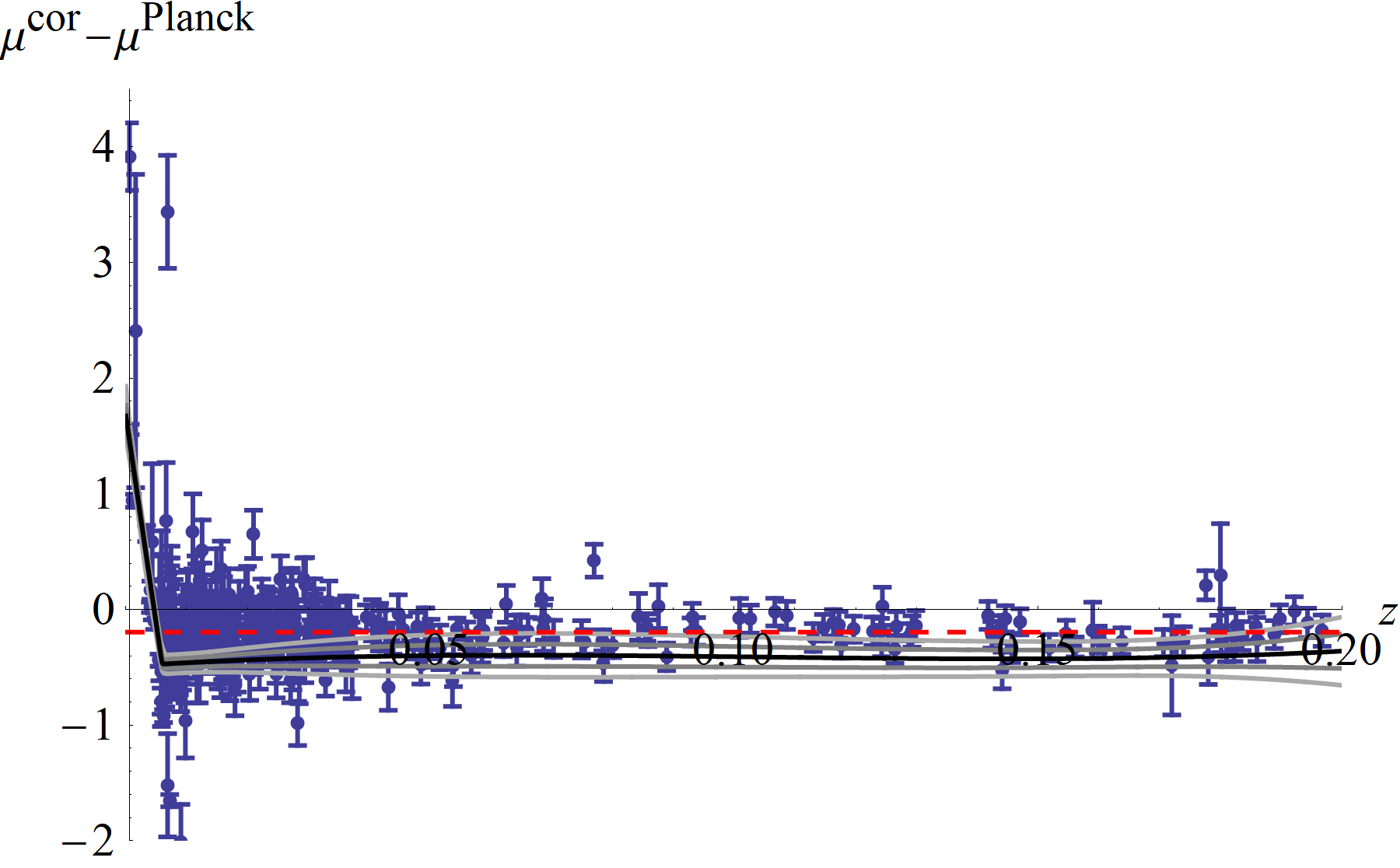}
\label{fig:FullSky_Mu_Normal}
}
\caption{\label{fig:FullSky}
The effects of different Cepheids-hosting galaxies velocity dispersions $v_c$ are shown in different cases.
\emph{Left}: Full sky datasets up to $\zmax = 0.2$, with peculiar velocity corrections, a $250$ \ks velocity dispersion added to SNe, and a varying amount of $v_c$. More opaque/transparent points correspond to standard candles with lower/higher \HWC{$\Delta \mu$}\COM{error} (according to the legends shown on the top right corner of the figures), while different colors represent the sky position: subregions F1 (green), F2 (blue), F3(orange), and none of the above (gray).
\emph{Right}: Best models with respect to the corresponding datasets on the left, according to the F-test. The best fit (black), 68{\%} (gray) and 95{\%} (light gray) confidence bands are shown. The dashed red line is plotted as a reference and corresponds to $\mu ^{\rm Riess} - \mu ^{\rm Planck} = -5\log _{10} \left(73.24 / 66.93 \right)$.
}
\end{figure}

When no a-priori assumption about the model is made and inhomogeneous models are included in the analysis, the F-test always gives preference to inhomogeneous models for $v_c<$ 250 \ks.
Examples are given in figures \ref{fig:FullSky_Mu_Cluster} and \ref{fig:FullSky_Mu_Normal}. It is difficult to identify a best fit model because adding new parameters keeps increasing their F-test likelihood. 
This kind of behavior hints to two important conclusions:
\begin{itemize}
\item the homogeneous model is clearly not the best model even after applying redshift correction, which implies that some other structure which 2M++ cannot detect is affecting luminosity distance observations,
\item the monopole component of local structure is not sufficient to model the observed data, and for this reason the best fit of the spherically symmetric model is difficult to identify.
\end{itemize}
These are good motivations to proceed further with directional analysis.

\subsection{Directional analysis}


To investigate the anisotropy of the local structure we consider 3 particular subsets of SNe + Cepheids-hosting galaxies defined in section~\ref{sec:sneia}: F1 up to $\zmax = 0.2$, F3 up to $\zmax = 0.2$ and $\zmax = 0.4$. The choice of these regions is made in order to compare with the previous luminosity density analysis of \Keenan along the same directions, pointing to the existence of inhomogeneities with sizes larger than the scales probed by \TMPP. As a consequence these structures could affect the luminosity distance even after \TMPP redshift corrections. Making use of Cepheids-hosting galaxies precision and the evidence from last section for $v_c < 250$ \ks, we chose $v_c = 0$ \ks rather than $40$ \ks in order to appreciate their full contribution. We in fact find evidence of these residual effects of the local structure, since the inhomogeneous models fit better the data.

\begin{figure}
\centering
\subfigure[F1 $( 1, 0, 0 )$]{
\includegraphics[width=.48\textwidth]{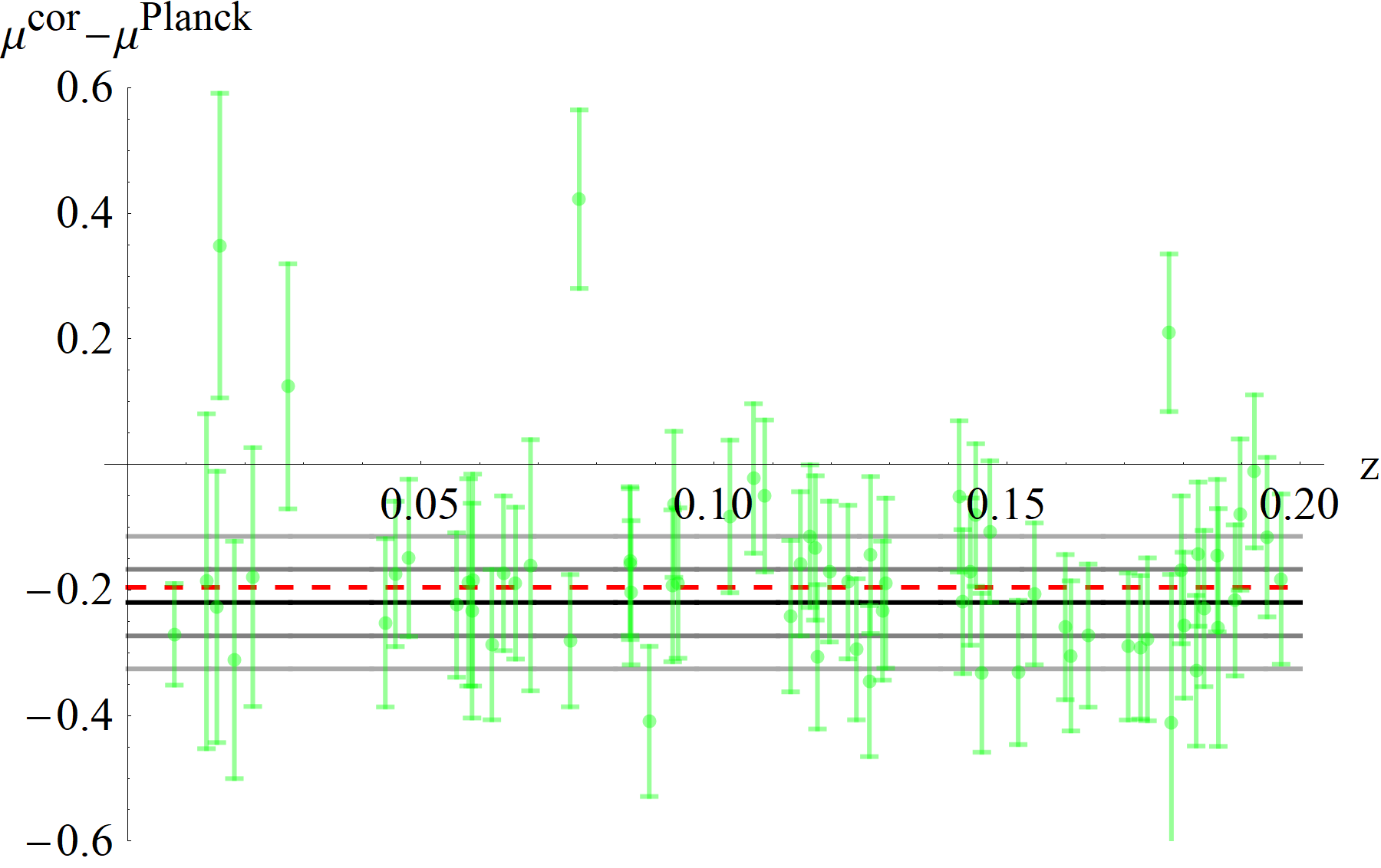}
\label{fig:yvc_F1_Mu}
}
\subfigure[F3 with $\zmax=0.2$ and NGC4536 removed $( 0, 1, 3 )$]{
\includegraphics[width=.48\textwidth]{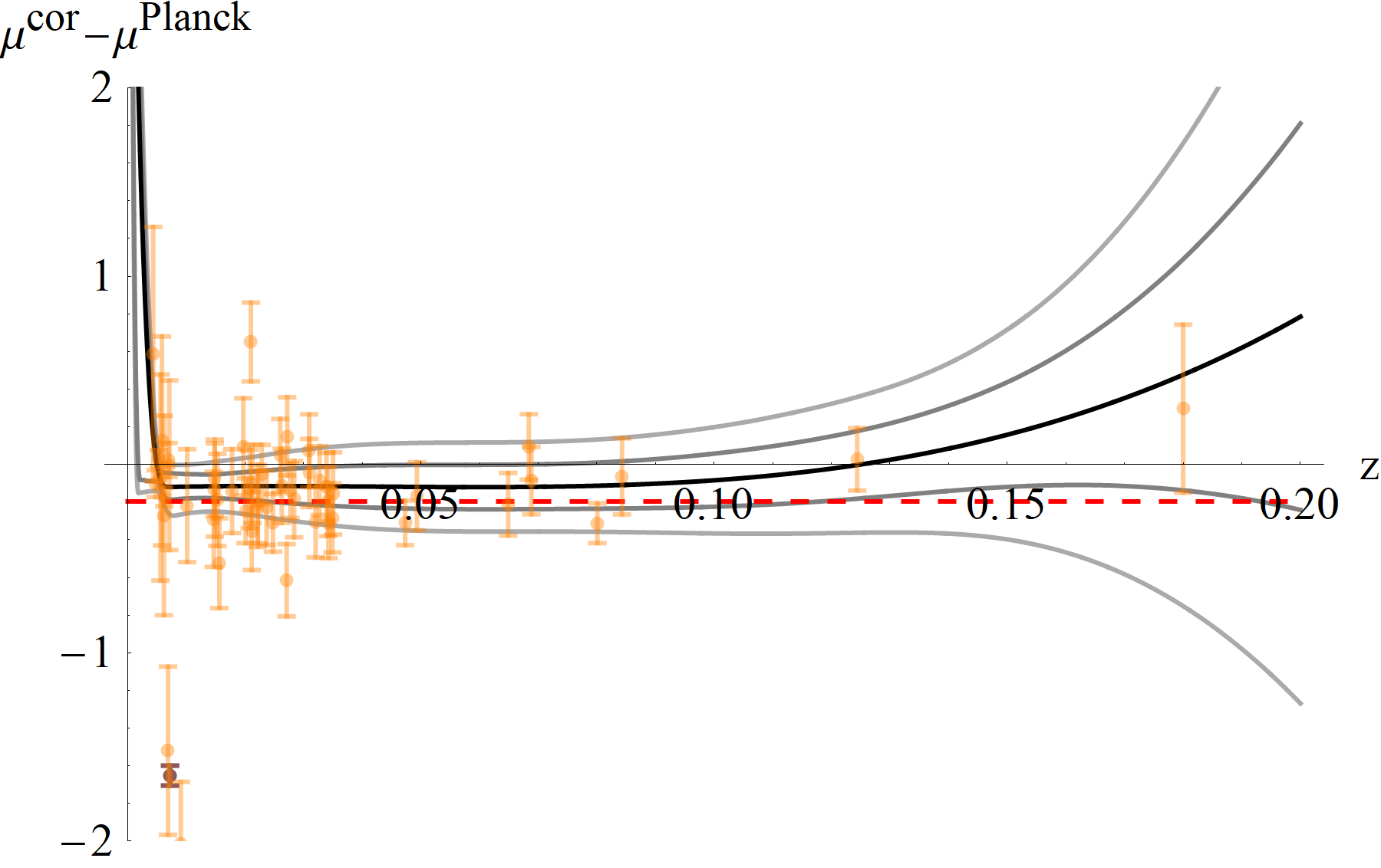}
\label{fig:yvc_Mu_1}
}
\caption{\HWC{
F1/F3 standard candles distance modulus data (up to $\zmax = 0.2$, with peculiar velocity corrections, and with 250 \ks velocity dispersion for SNe) are plotted with their best fit (black), 68{\%} (gray) and 95{\%} (light gray) confidence bands according to the method of section~\ref{sec:method}.
The dashed red line is plotted as a reference and corresponds to $\mu ^{\rm Riess} - \mu ^{\rm Planck} = -5\log _{10} \left(73.24 / 66.93 \right)$.
The field, removed data points (if any), and the model parameters are shown in the sub-captions.
\emph{Right}: The removed outlier is shown as a darker data point. NGC 4424 is outside of the top left corner of the plot.}
}
\end{figure}

For F1, the best fit we get is from a simple $(1,0,0)$ model with \HWC{$H_0^{\rm loc} = 74.06 ~\pm~ 1.81$ \ksm and $\chi_R^2$ = 1.00}, presented in figure~\ref{fig:yvc_F1_Mu}. The next best one \HWC{(F-test {\it Threshold} $< 40\%$) is given by a $(1,1,10)$ model with $\chi_R^2 = 0.72$}.

For F3, as shown in table~\ref{tab:yvcyvdMuRhoChi2}, the best model is a $(0,1,6)$ model regardless of taking $\zmax = 0.2$ or $\zmax = 0.4$, with an associated $\chi_R^2$ = \HWC{2.88} in $\zmax = 0.2$ case and a $\chi_R^2$ = \HWC{6.02} in $\zmax = 0.4$ case. \HWC{For $\zmax = 0.4$ case a $(1,0,0)$ model is also suggested when the F-test {\it Threshold} is above 85\%, with $\chi_R^2=17.6$ and $H_0^{\rm loc}=78.73\pm3.56$.} The staggeringly \HWC{high $\chi_R^2$ of the homogeneous model and the suggestion of} complicated model\HWC{s are}\COM{ is} mainly due to NGC 4536.
After removing it the resulting best case is a \HWC{$( 0, 1, 3 )$} model for both $\zmax = 0.2$ and $\zmax = 0.4$ cases, as shown in figure~\ref{fig:yvc_Mu_1}.
A close inspection reveals that NGC 4424 is the main cause of the preference for this inhomogeneous model. Since after peculiar velocity correction its redshift is extremely close to zero, we naturally consider removing it as well. Unsurprisingly, after removing these 2 outliers the best model is finally a $(1,0,0)$ model with \HWC{$H_0^{\rm loc} = 69.14 \pm 1.16$ \ksm and an associated $\chi_R^2$ = 1.92 for $\zmax = 0.2$ case, and $H_0^{\rm loc} = 69.15 \pm 1.14$ \ksm with $\chi_R^2$ = 1.87 for  $\zmax = 0.4$}.

The fact that two fields lead to two $H_0^{\rm loc}$ \COM{differed by}\HWC{with} $\sim 3 \sigma$ \HWC{difference} suggests that the luminosity distance is affected by some additional structure which has size larger than the scale probed by \TMPP , which is in fact evident even after peculiar velocity correction. \HWC{To test the claim we first compute $\chi_R^2$ of the combined dataset of F1 and F3 using a $(1,0,0)$ model, and compare it to the combined $\chi_R^2$ of a $(1,0,0)$ model for F1 and a $(0,1,6)$ model for F3. The resulting $\chi_R^2$ is $8.88$ for the simple $(1,0,0)$ model and $1.71$ for the combined model, suggesting that indeed there is a directional inhomogeneity.}
\HWC{In total, there are 4 evidences suggesting that \TMPP velocity correction is not enough to remove completely the effects of local structure:
\begin{itemize}
\item The directional fit is much better than a full sky fit.
\item There is a $\sim 3\sigma$ difference between F1 and F3 in terms of $H_0^{\rm loc}$ of the $(1,0,0)$ model.
\item For F3 the $(0,0,0)$ model leads to extremely high $\chi^2_R$ and inhomogeneous models are clearly preferred  by the F-test {\it Threshold}.
\item Though not evident in this section, by comparing same models in tables \ref{tab:yvcyvdMuRhoChi2} (with peculiar velocity correction) and \ref{tab:MuRhoChi2} (without), the peculiar velocity correction does not decrease $\chi^2_R$.
\end{itemize}
According to the four arguments above}
\COM{Since we have obtained evidence that \TMPP velocity correction is not enough to remove completely the effects of local structure}, 
in next section we will reconstruct the density profile directly from the distance modulus, without applying any peculiar velocity correction.
\begin{table}[ht!]
\centering
\footnotesize

\begin{tabular}{|c|cccc|c|}
\hline
$\zmax$				&$\chi_R^2$& \it Threshold	& PRESS	& \modelparam	& Removal	\\
\hline
\multirow{6}{*}{0.2}&19.1& Not Preferred		& 1456	& $78.72\pm3.72$&			\\
					&2.88& $20  \sim 100 \%$	& 1518	& $(0, 1, 6)$	&			\\
\cline{2-6}
					&5.31& $94.2\sim 100 \%$	& 292.3	& $(0, 0, 0)$	&\multirow{2}{*}{NGC 4536}	\\
					&2.04& $23  \sim 94.1\%$	& 308.7	& $(0, 1, 3)$	&			\\
\cline{2-6}
					&2.02& $94.3\sim 100 \%$	& 107.1	& $(0, 0, 0)$	&\multirow{2}{*}{+ NGC 4424}	\\
					&1.92& $66  \sim 94.2\%$	& 113.5	& $69.14\pm1.16$&			\\
\hline
\multirow{6}{*}{0.4}&17.6& $86  \sim 100 \%$	& 1462	& $78.73\pm3.56$&			\\
					&6.02& $16  \sim 85  \%$	& 3979	& $(0, 1, 6)$	&			\\
\cline{2-6}
					&4.98& $95.0\sim 100 \%$	& 293.7	& $(0, 0, 0)$	&\multirow{2}{*}{NGC 4536}	\\
					&2.00& $22  \sim 94.9\%$	& 298.2	& $(0, 1, 3)$	&			\\
\cline{2-6}
					&1.96& $94.8\sim 100 \%$	& 113.8	& $(0, 0, 0)$	&\multirow{2}{*}{+ NGC 4424}	\\
					&1.87& $63  \sim 94.9\%$	& 120.1	& $69.15\pm1.14$&			\\
\hline


\end{tabular}
\caption{
\label{tab:yvcyvdMuRhoChi2}
Distance modulus best fit model parameters with progressive removal of the outliers for F3, with peculiar velocity corrections from \TMPP. The {\it Threshold} column shows the F-test threshold of the model. In the \modelparam column, if the model is homogeneous, i.e. a $( 1, 0, 0 )$ model, we give the value of $H_0^{\rm loc}$ as defined in eq.~\eqref{eq:Hloc} and its standard deviation.
}
\end{table}


\section{Reconstruction of density profiles using standard candles distance moduli}
\label{sec:NOvcYESvd}

As discussed in the previous sections the large improvement of $\chi^2_R$ from homogeneous to inhomogeneous models shows that applying peculiar velocity corrections from \TMPP cannot remove completely the effects of inhomogeneities on the distance modulus. We have also noted in section~\ref{sec:NoVeloCorr} that peculiar velocity corrections based on \TMPP cannot account for the effects of structures as the one detected in \Keenan. 

We now employ the fitting method on data uncorrected from peculiar velocities, and compare our inverted radial density profile with 2M++ and \Keenan. In order to compare with \TMPP we perform an angular average of the \TMPP density fields within the regions of interest (F1 or F3). 
However as shown in figure~\ref{fig:2mpp}, \COM{since }there are substructures within each of the 2 fields \HWC{which would affect the luminosity distance,} and the angular-averaged density does not take that into account\HWC{. Thus one should not expect the inverted density to reproduce every feature of \Keenan and \TMPP density profiles.}\COM{, one should not quantitatively compare \Keenan and \TMPP with our inverted profiles}. The angular averages of the density maps  from  galaxy catalogs are rather serving as a useful quality and consistency test of the SNe/Cepheids data and inversion algorithm.
We include in the analysis the effects of a velocity dispersion of $250$ \ks to account for SNe host rotation, and no dispersion for Cepheids-hosting galaxies, i.e. $v_c=0$. It can also be interesting to estimate how the results change \COM{when this additional dispersion is removed}\HWC{if $v_c=40$}, and we report this in appendix~\ref{sec:NOvcNOvd}. 

\subsection{Subregion F1}
\begin{figure}[ht!]
\centering
\includegraphics[width=.48\textwidth]{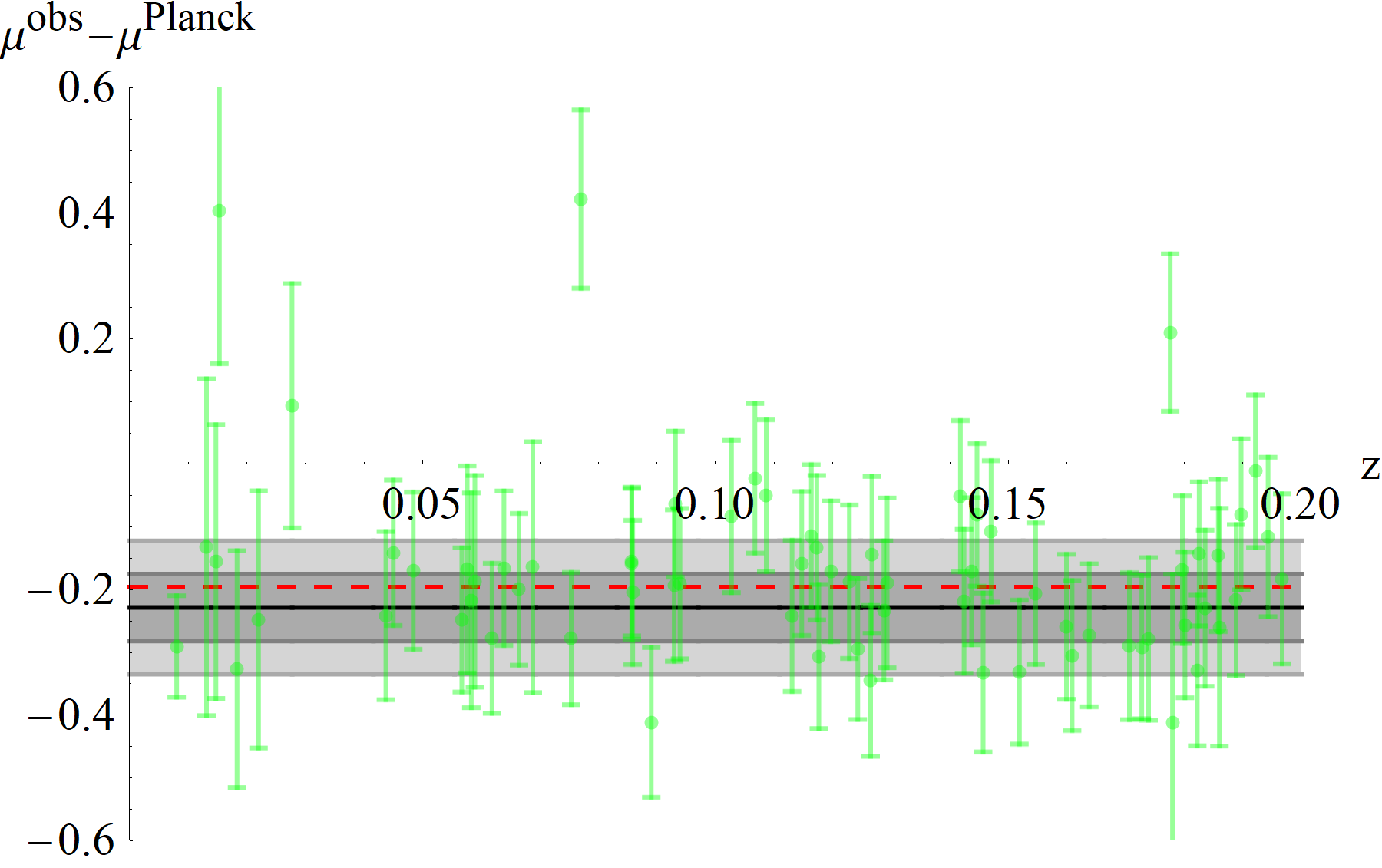}~
\includegraphics[width=.48\textwidth]{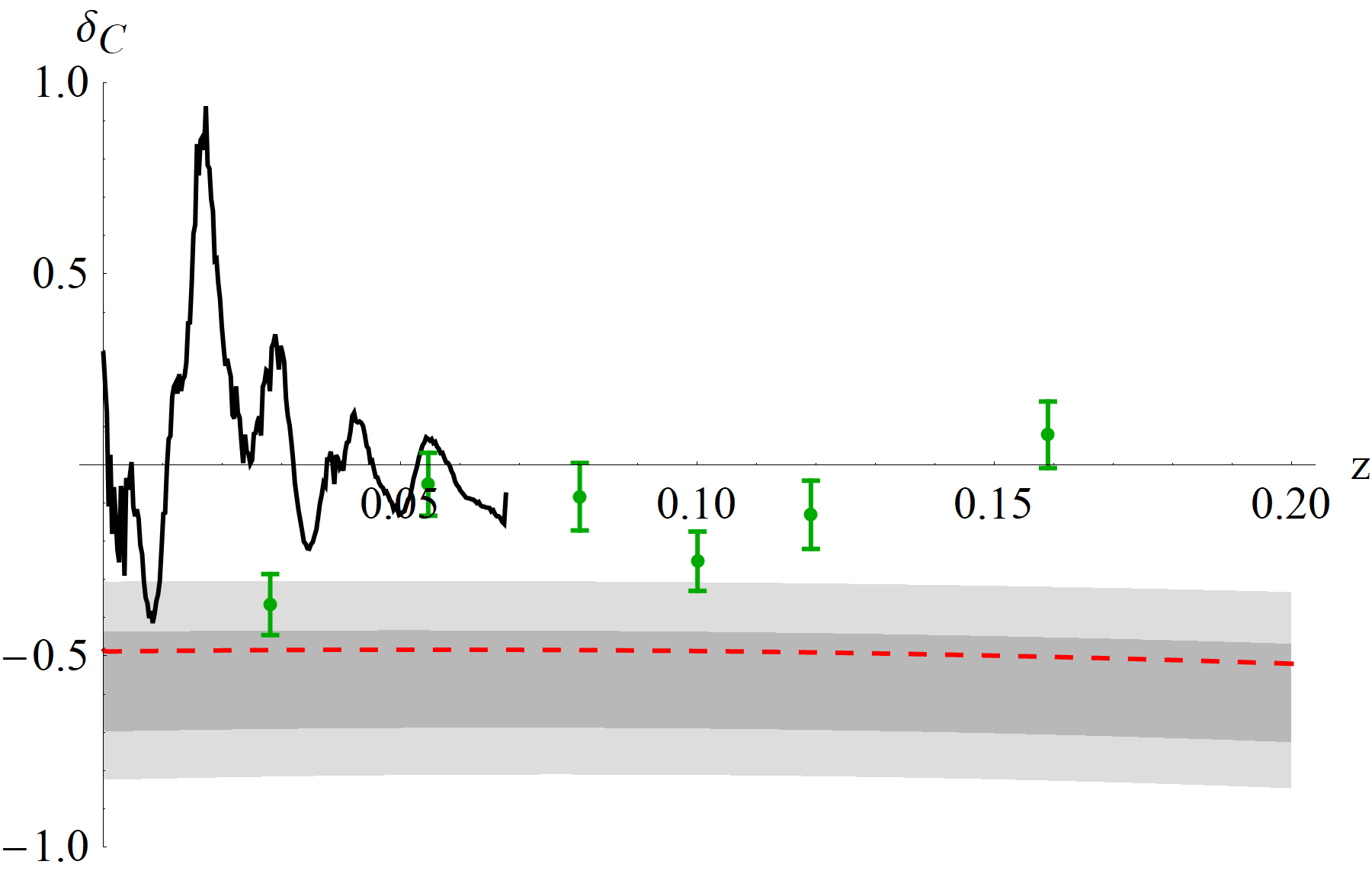}
\caption{ \label{fig:F1MuRhoZmax0pt2}
The (1,0,0) best fit model is plotted for F1 and $\zmax = 0.2$, with no peculiar velocity corrections, and a $250$ \ks velocity dispersion for SNe.
\emph{Left}: Standard candles distance modulus data are plotted with their best fit (black), 68{\%} (gray) and 95{\%} (light gray) confidence bands according to the method of section~\ref{sec:method}. The invertible bands are shown as the shaded region (68{\%}-gray shade, 95{\%}-light gray shade).
The dashed red line is plotted as a reference and corresponds to $\mu ^{\rm Riess} - \mu ^{\rm Planck}$. \emph{Right}: The confidence bands of the inverted density contrast corresponding to the invertible bands of the distance modulus are shown (68{\%}-gray, 95{\%}-light gray). The data points of \Keenan are plotted in green, the \TMPP density contrast averaged over F1 as a solid black curve, and the dashed red line is for density contrast that would lead to a local Hubble parameter $H_0^{\rm loc}=H_0^{\rm Riess}$ assuming a large scale $H_0^{LS}=H_0^{\rm Planck}$.
}
\end{figure}

For F1, the best fitting model we get without removing any outlier and with $\zmax = 0.2$, is a \HWC{locally} homogeneous $(1,0,0)$ model with \HWC{$H_0^{\rm loc} = 74.35 \pm 1.83$ \ksm with F-test {\it Threshold}  $>37\%$}, shown in figure~\ref{fig:F1MuRhoZmax0pt2}, while the second best model is an inhomogeneous \HWC{$(1,1,6)$}  model. 
The $\chi_R^2 \sim 0.75$ of the inhomogeneous model is not so much lower than that of the homogeneous model, which has $\chi_R^2 \sim 1.02$.
As we can observe, the constant density profile that we obtain is below the \TMPP average along the F1 direction. This can be partially understood from the top plot of figure~\ref{fig:2mpp} in which we see that SNe and Cepheids are located far from the highest density regions and that the angular average of the \TMPP density field has a large variance.

It should be noted that we are correctly normalizing our reconstructed density profile with respect to the background since we are assuming cosmological background parameters obtained from large scale observations such as the Planck mission, which are insensitive to local structure as shown for the luminosity distance in \cite{Romano:2016utn}. On the contrary \TMPP is not normalized with respect to the average density of the Universe but with respect to the average within its depth, and consequently its normalization can be wrong if \TMPP is embedded in a larger structure. 

The same normalization problem can arise for \Keenan analysis as well since the background is again assumed to be the averaged luminosity density over the data set, not over the all Universe.
This shows that our method can be very useful to establish the correct normalization with respect to the average density of the Universe, which could otherwise be incorrectly fixed.

We \COM{try to determine} \FN{discuss} the normalization factor which should be applied to \TMPP and \Keenan in order to match our reconstructed density profile \COM{and show the results} in section~\ref{sec:densityrescale}, with some good qualitative agreement of the profiles. 

\subsection{Subregion F3}
\begin{figure}[ht!]
\centering
\includegraphics[width=.48\textwidth]{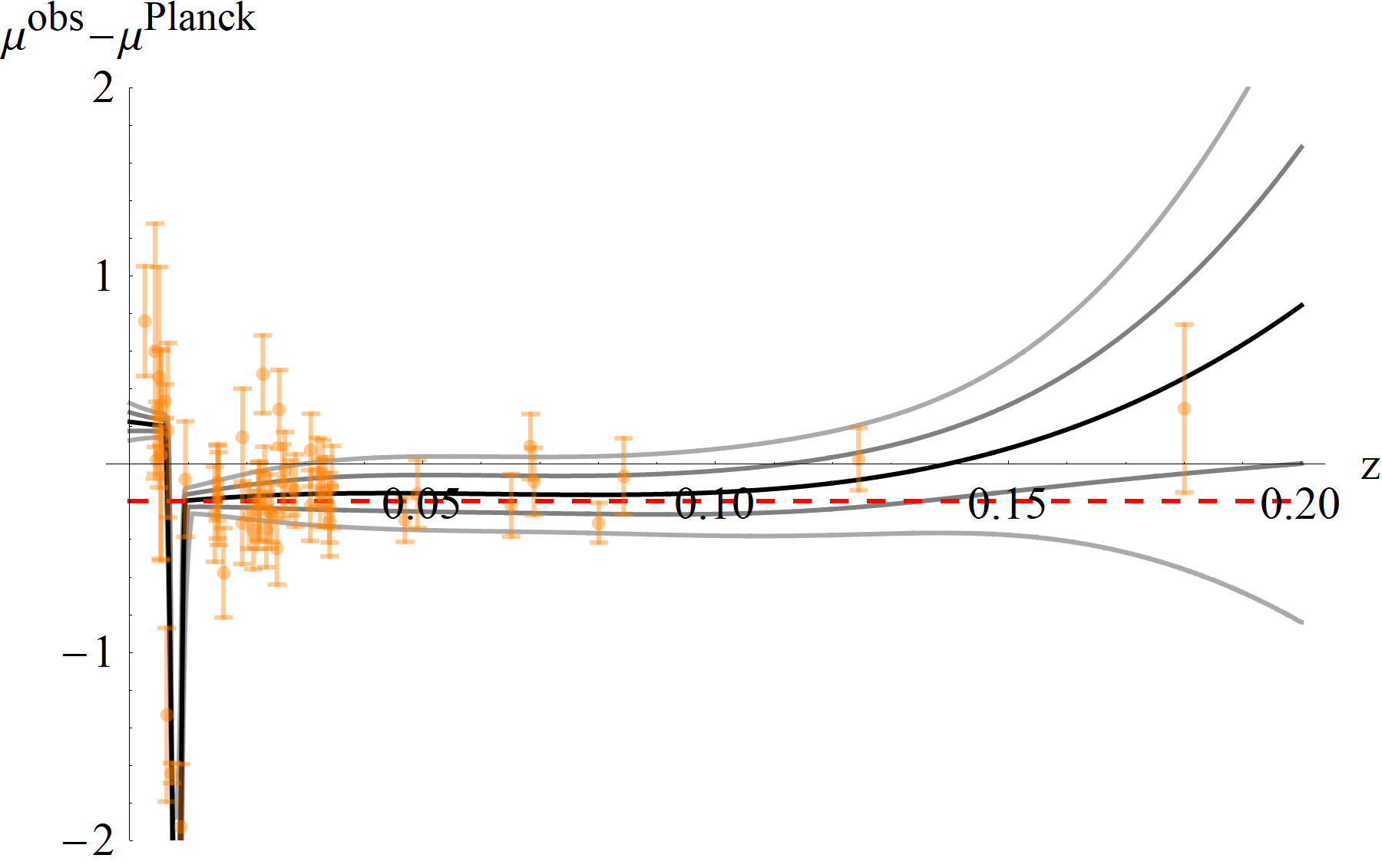}~
\includegraphics[width=.48\textwidth]{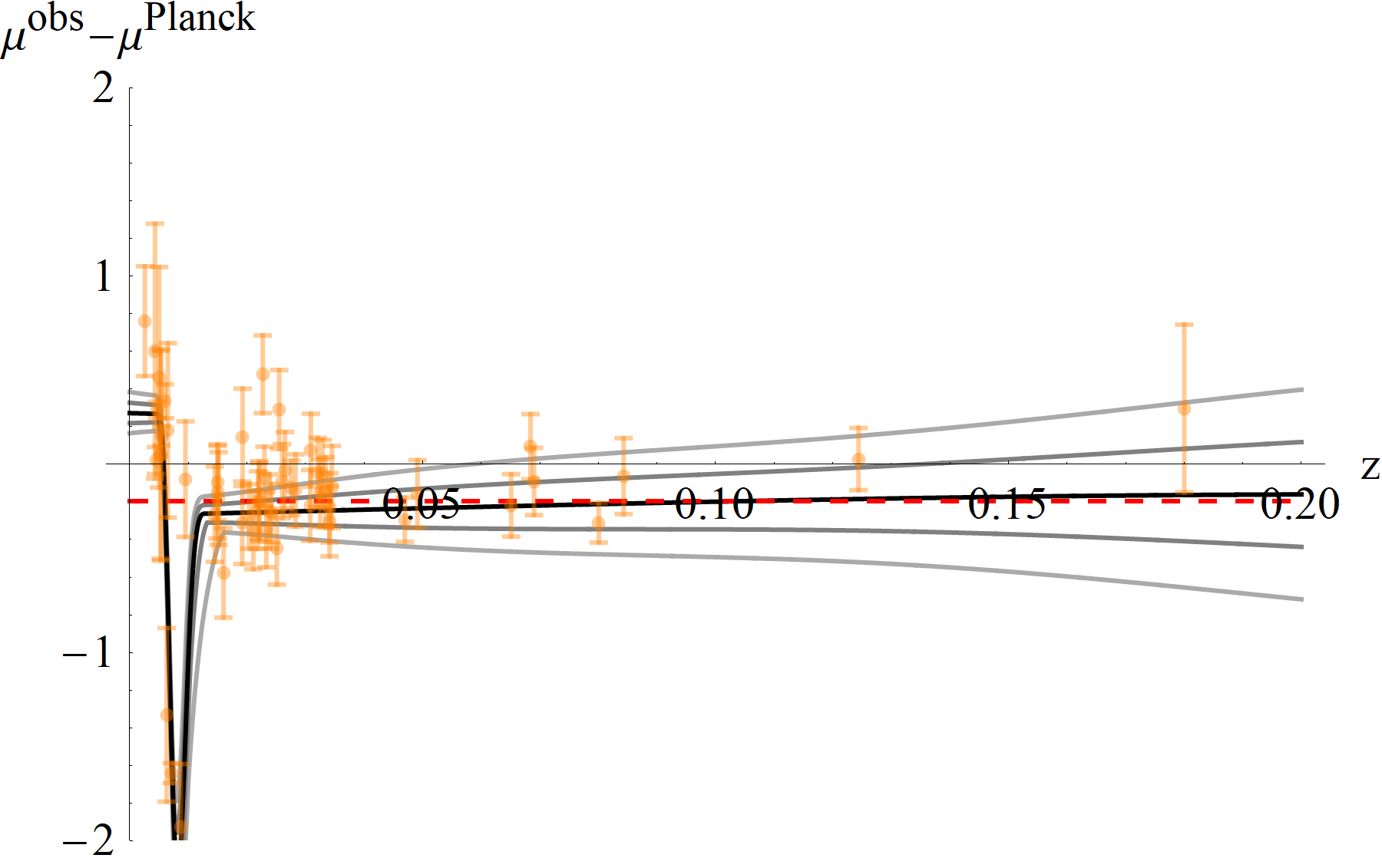}
\caption{\label{fig:Mu_0}
F3 standard candles distance modulus data (without peculiar velocity correction, with 250 \ks velocity dispersion for SNe, and without outlier removal) are plotted with their best fit (black), 68{\%} (gray) and 95{\%} (light gray) confidence bands according to the method of section~\ref{sec:method}, with $\zmax = 0.2$ and a $( 0, 0, 5 )$ model on the left, $\zmax = 0.4$ and a $( 0, 1, 5 )$ model on the right. \HWC{The dashed red line is plotted as a reference and corresponds to $\mu ^{\rm Riess} - \mu ^{\rm Planck}$.} In both cases the fits are not invertible.}
\end{figure}

For subregion F3, we present the results for two different redshift intervals, $\zmax = 0.2$ and $\zmax = 0.4$. In both cases the range of the plots is $z \in [0,0.2]$ to allow an easier comparison between the two. The \COM{fits}\HWC{best models} we obtain\COM{ correspond respectively to $(0,0,5)$ and $(0,1,5)$}\HWC{, regardless of $\zmax$, are $(0,0,5)$} models, shown in figure~\ref{fig:Mu_0}.
\HWC{The homogeneous model is not preferred with a F-test {\it Threshold}$>0.9999$, suggesting that it is necessary to have some inhomogeneity. Unfortunately some v}\COM{V}ery low $z$ data points cannot be explained as the effects of inhomogeneities, because they would lead to unphysical negative energy densities, and are probably due to large intrinsic peculiar velocities not related to the local structure.

\begin{figure}[ht!]
\centering
\subfigure[NGC 4536 removed $( 1, 0, 0 )$]{
	\includegraphics[width=.48\textwidth]{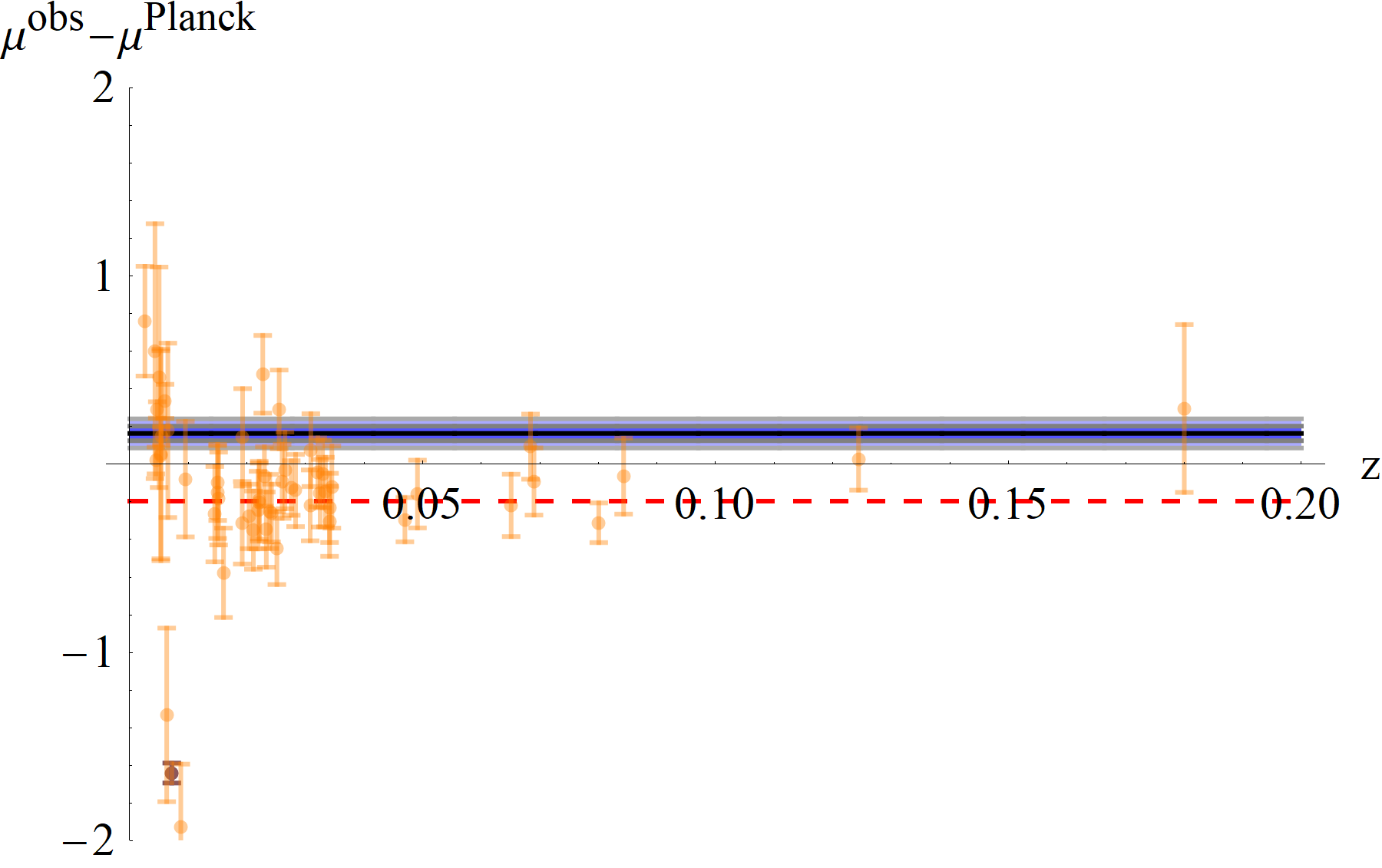}
	\label{fig:Z02Mu1a}
 }
\subfigure[ NGC 4536 removed $( 0, 0, 1 )$]{
	\includegraphics[width=.48\textwidth]{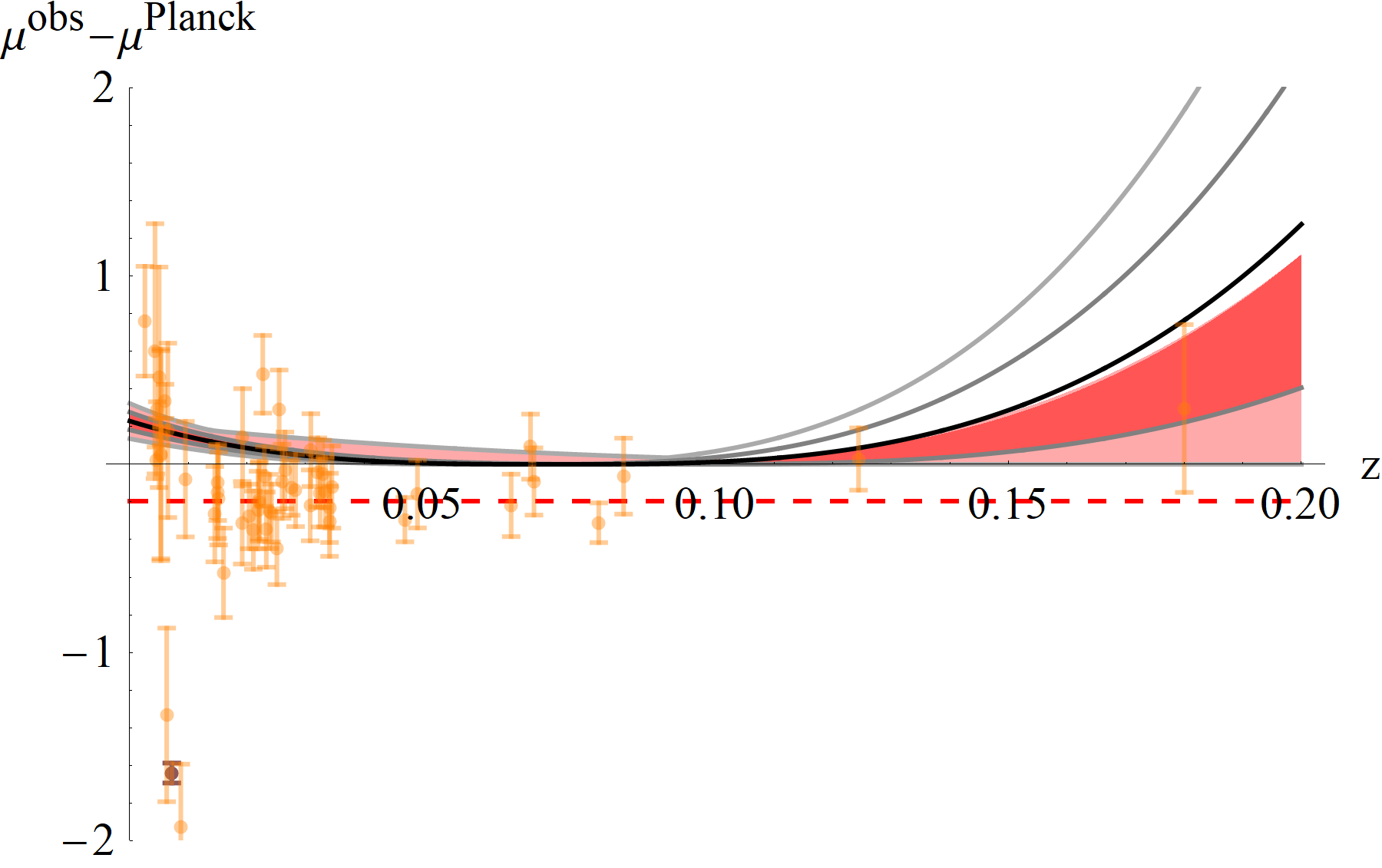}
	\label{fig:Z02Mu1c}
 }
\subfigure[\bf NGC 4536 removed $\mathbf{( 1, 0, 1 )}$]{
	\includegraphics[width=.48\textwidth]{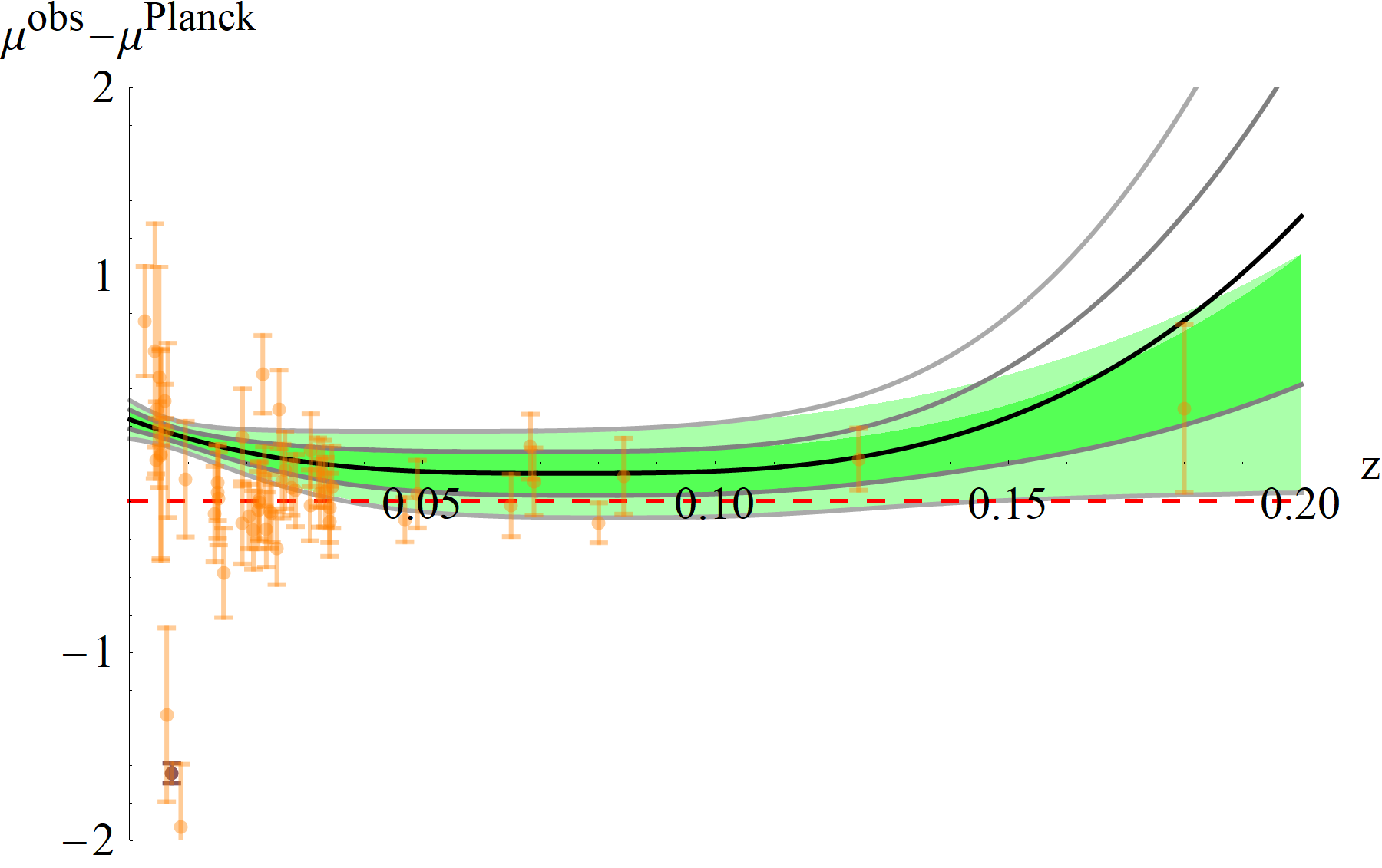}
	\label{fig:Z02Mu1b}
 }
\subfigure[Inverted density]{
	\includegraphics[width=.48\textwidth]{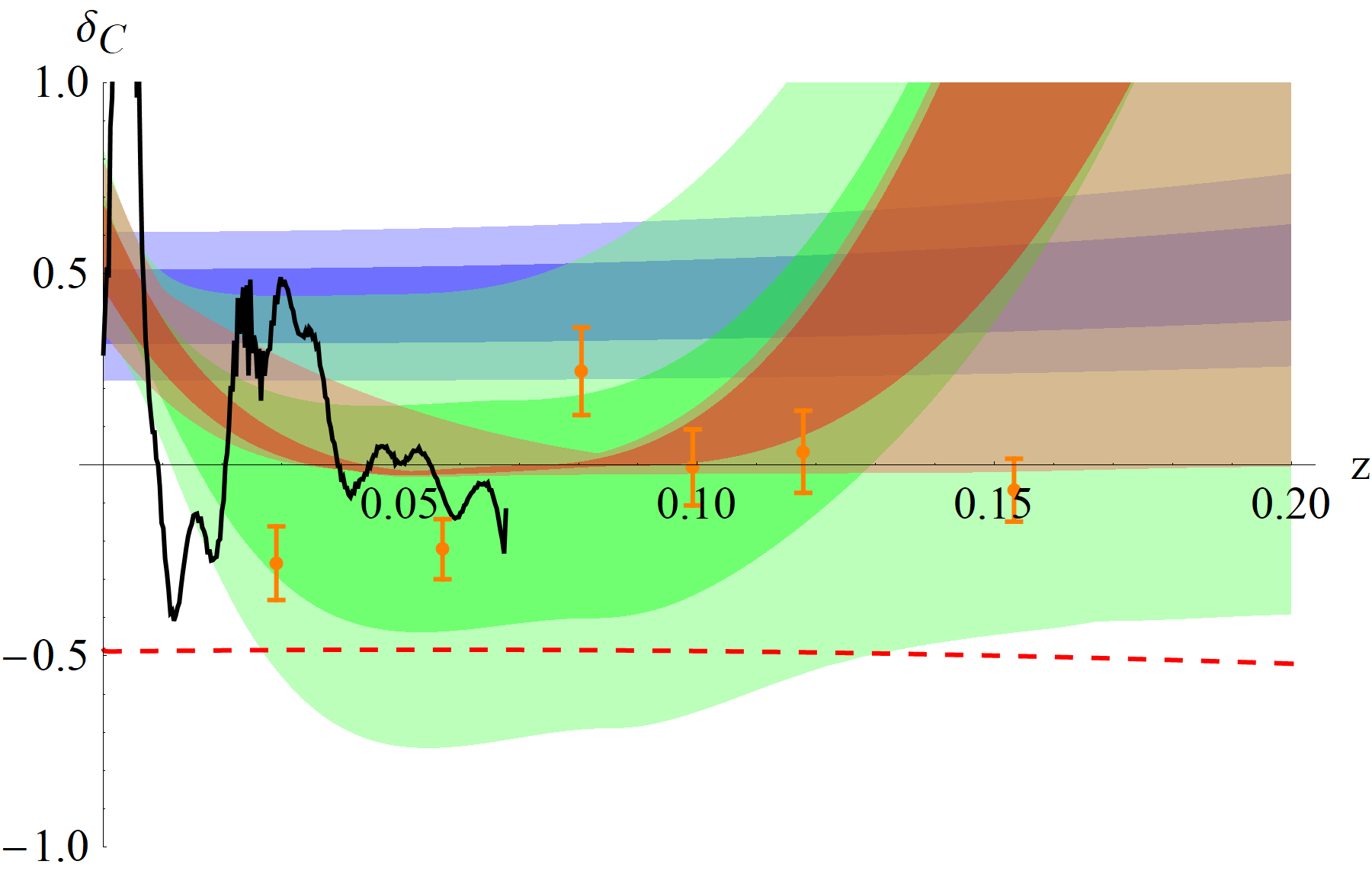}
	\label{fig:Z02Rho}
 }
\caption{\label{fig:MuRhoZmax0pt2}
Distance modulus best fit models are plotted for F3 with $\zmax = 0.2$, without peculiar velocity corrections and with a 250 \ks velocity dispersion for SNe. The model parameters and the removed data points are shown in the sub-captions. \emph{(a, b, c)}: Standard candles distance modulus data are plotted with their best fit (black), 68{\%} (gray) and 95{\%} (light gray) confidence bands according to the method of section~\ref{sec:method}. The invertible bands are shown as the shaded region (68{\%}-darker color, 95{\%}-lighter color), while the removed outliers are shown as darker data points. \HWC{The dashed red line is plotted as a reference and corresponds to $\mu ^{\rm Riess} - \mu ^{\rm Planck}$.} Case (c) is preferred. \emph{(d)}: The confidence bands of the inverted density contrast corresponding to the invertible bands of the distance modulus are shown (68{\%}-darker color, 95{\%}-lighter color). The data points of \Keenan are plotted in orange, the \TMPP density contrast averaged over F3 as a solid black curve, and the dashed red line is for density contrast that would lead to a local Hubble parameter $H_0^{\rm loc}=H_0^{\rm Riess}$ assuming a large scale $H_0^{LS}=H_0^{\rm Planck}$. The bands are color coded case by case with blue (a), red (b), green (c).
}
\end{figure}

\begin{figure}[ht!]
\vspace{-0.3cm}
\centering
\subfigure[NGC 4536 removed $( 1, 0, 0 )$]{
	\includegraphics[width=.48\textwidth]{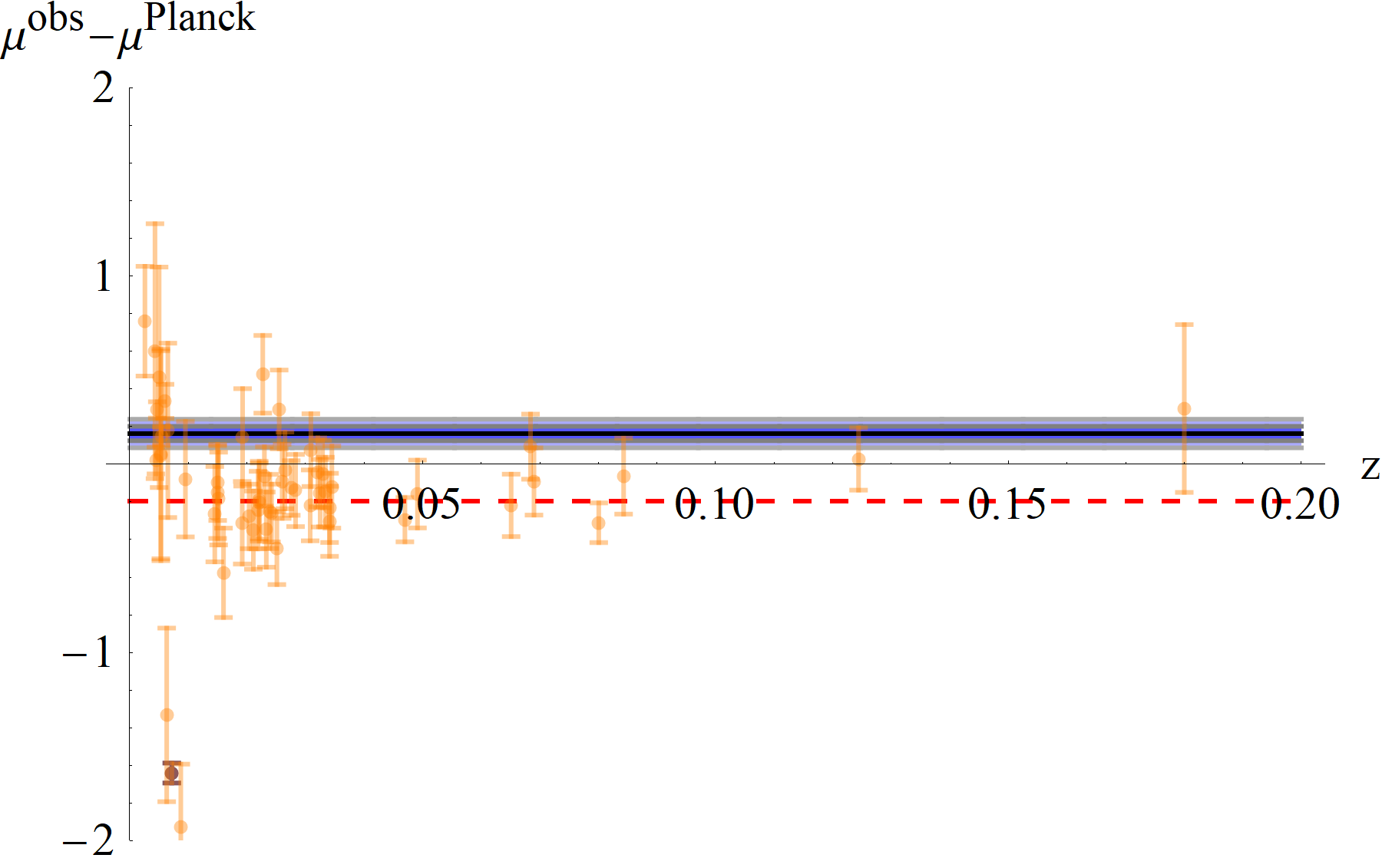}
	\label{fig:Z04Mu1a}
 }
\subfigure[NGC 4536 removed $( 0, 0, 1 )$]{
	\includegraphics[width=.48\textwidth]{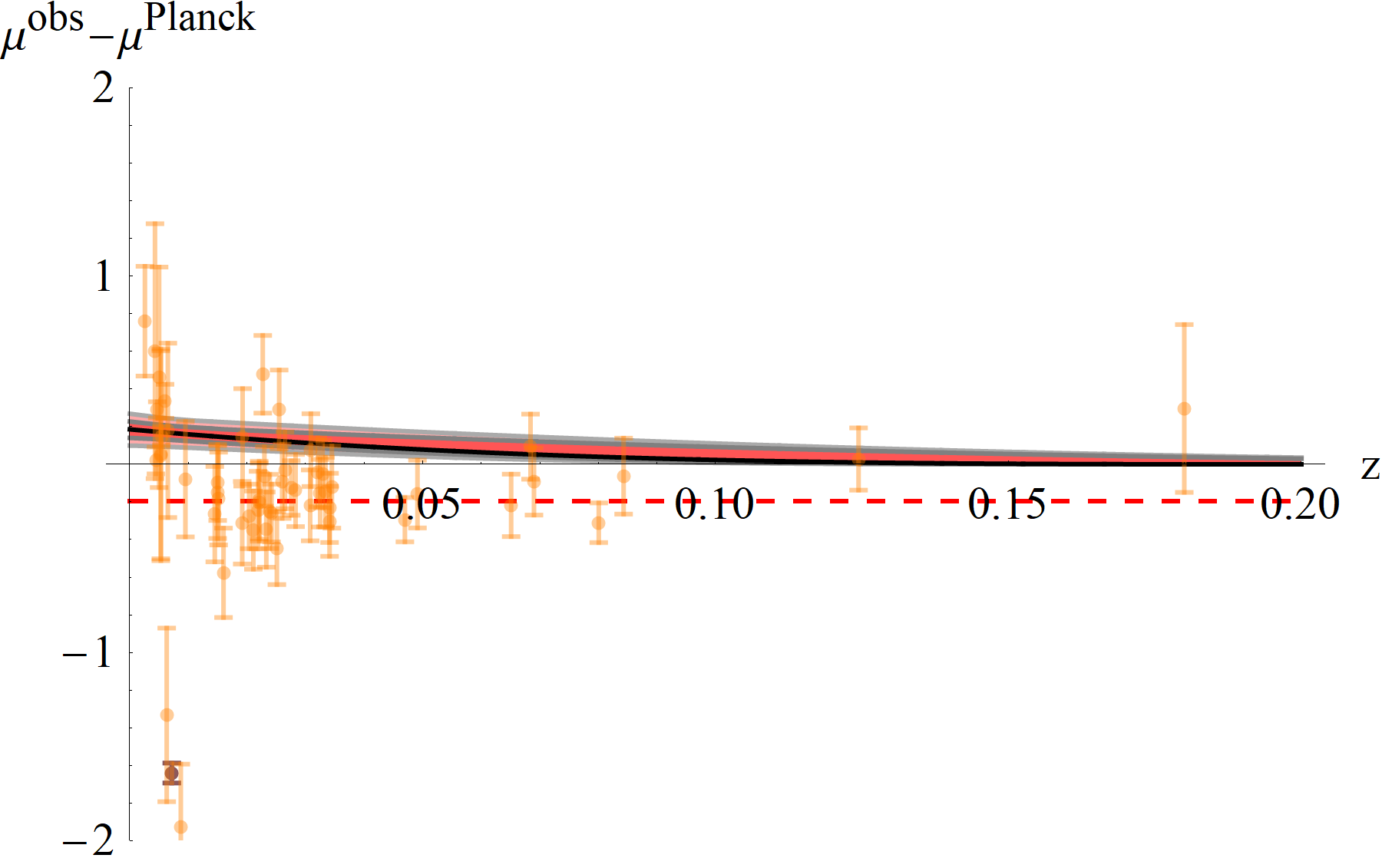}
	\label{fig:Z04Mu1c}
 }
\subfigure[{\bf NGC 4536 removed} $\mathbf{( 1, 0, 1 )}$]{
	\includegraphics[width=.48\textwidth]{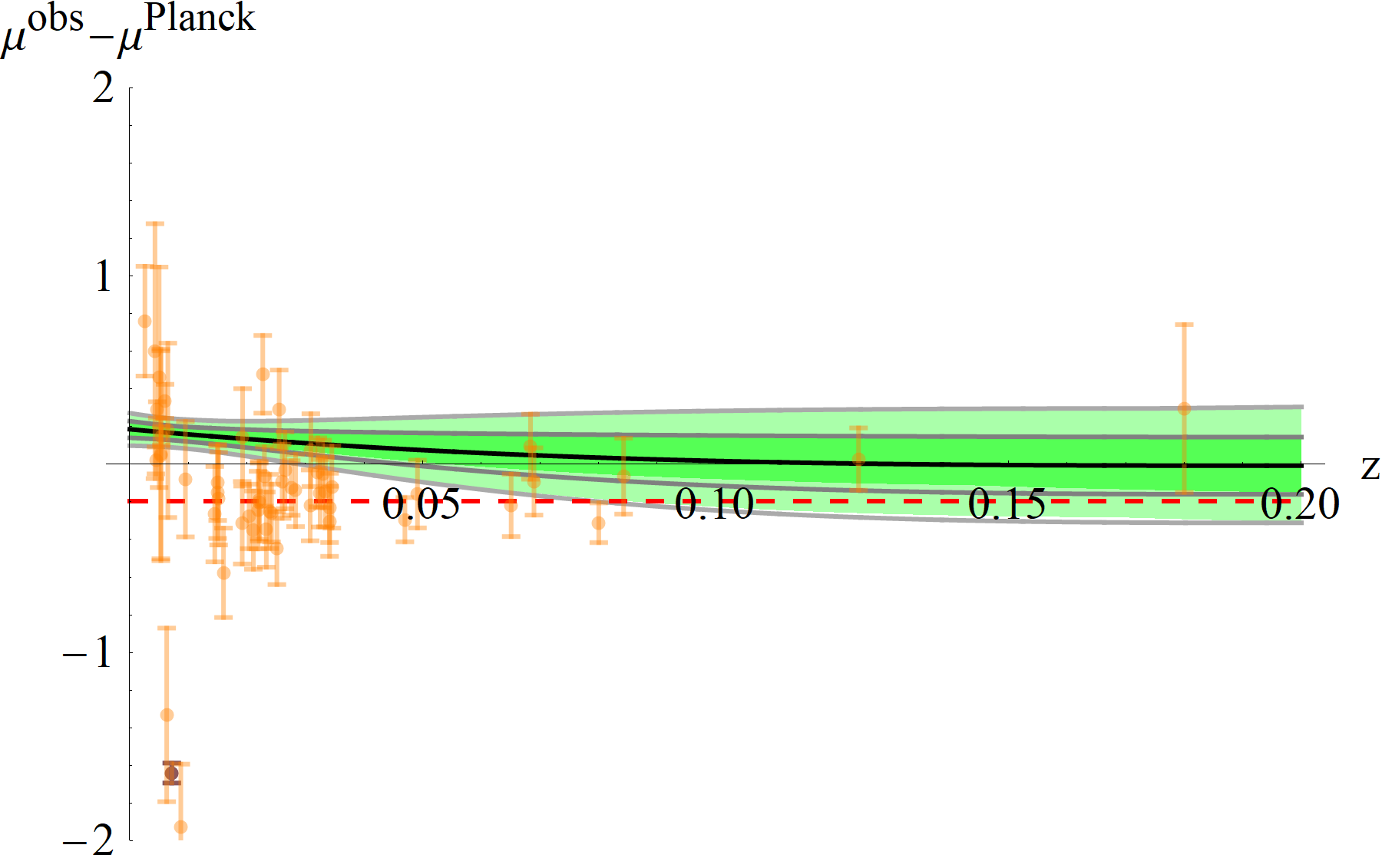}
	\label{fig:Z04Mu1b}
 }
\subfigure[Inverted density]{
	\includegraphics[width=.48\textwidth]{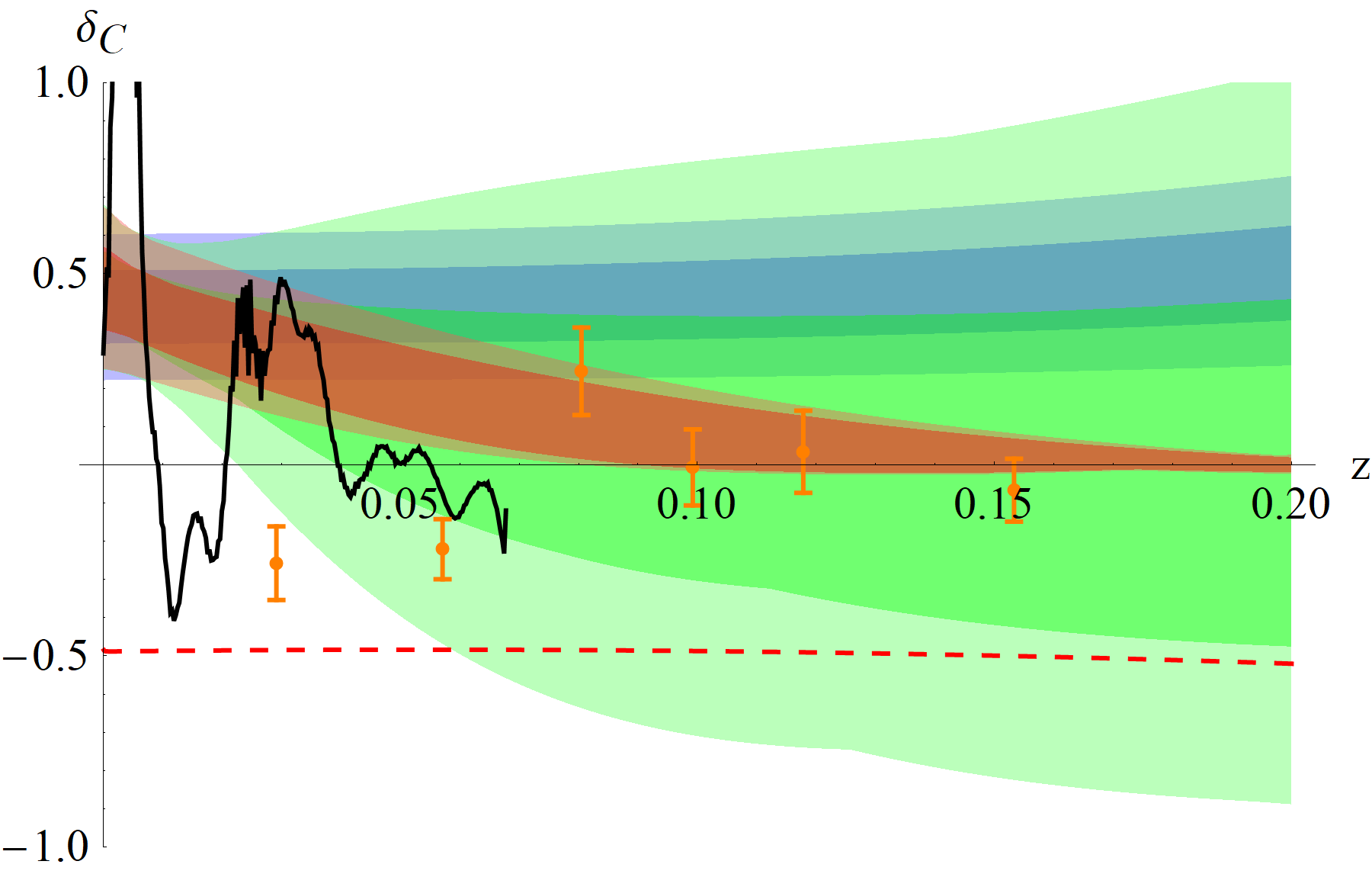}
	\label{fig:Z04Rho}
 }
\caption{\label{fig:MuRhoZmax0pt4}
Distance modulus best fit models are plotted for F3 with $\zmax = 0.4$, without peculiar velocity corrections and with a 250 \ks velocity dispersion for SNe. The model parameters and the removed data points are shown in the sub-captions. \emph{(a, b, c)}: Standard candles distance modulus data are plotted with their best fit (black), 68{\%} (gray) and 95{\%} (light gray) confidence bands according to the method of section~\ref{sec:method}. The invertible bands are shown as the shaded region (68{\%}-darker color, 95{\%}-lighter color), while the removed outliers are shown as darker data points. \HWC{The dashed red line is plotted as a reference and corresponds to $\mu ^{\rm Riess} - \mu ^{\rm Planck}$.} Case (c) is preferred. \emph{(d)}: The confidence bands of the inverted density contrast corresponding to the invertible bands of the distance modulus are shown (68{\%}-darker color, 95{\%}-lighter color). The data points of \Keenan are plotted in orange, the \TMPP density contrast averaged over F3 as a solid black curve, and the dashed red line is for density contrast that would lead to a local Hubble parameter $H_0^{\rm loc}=H_0^{\rm Riess}$ assuming a large scale $H_0^{LS}=H_0^{\rm Planck}$. The bands are color coded case by case with blue (a), green (b), red (c).
}
\end{figure}

Hence, we  progressively remove different outlier candidates as shown in table~\ref{tab:MuRhoChi2}. The \COM{two} most relevant outlier\COM{s} we find \COM{are}\HWC{is} NGC 4536\COM{ and SN 1999cl}, and the different fits 
are displayed in figures \ref{fig:MuRhoZmax0pt2} and \ref{fig:MuRhoZmax0pt4}.
We can see that once NGC 4536 \COM{and SN 1999cl are}\HWC{is} removed, \COM{more than half of the distance-modulus curves defining the $1\sigma$ and $2\sigma$ bands can be inverted}\HWC{the suggested models become invertible} in both $\zmax = 0.2$ and $\zmax = 0.4$ cases.

\HWC{The model\COM{s} suggested for $\zmax = 0.2$ \COM{are either a homogeneous $(1,0,0)$ model with an unusually low $H_0^{\rm loc} = 62.05 \pm 1.10$ \ksm and a F-test {\it Threshold} $>69\%$, or}\HWC{is} a $(0,0,1)$ model with F-test {\it Threshold} between 69 and 96.6\%. $(1,0,0)$ and $(1,0,1)$ models have similar performance in terms of PRESS \HWC{and $\chi_R^2$}, and thus \COM{both}\HWC{all} of them are shown in table \ref{tab:MuRhoChi2} and figure \ref{fig:MuRhoZmax0pt2}. In the following discussion we mainly focus on the $(1,0,1)$ model \HWC{as it matches with the \Keenan result quite well}. The next outlier selected according to F-test, SN 1999cl, does not have much effect on either model selection or the bands themselves, and thus should not be removed. 

Changing $\zmax$ from 0.2 to 0.4 however makes the $(1,0,0)$ model more preferable. We believe it is due to the fact that at larger scales the universe should appear homogeneous. Figure \ref{fig:MuRhoZmax0pt4} supports the argument as all the locally inhomogeneous models approach homogeneity at higher redshift.} 

\COM{Since we require the invertibility of the best fit, the (1,0,1) model is not selected in both cases. Consequently, the best models are respectively a $(0,0,2)$ model with $\chi^2_R \sim 1.47$ for $\zmax = 0.2$, and a $(0,0,4)$ model with $\chi^2_R \sim 1.37$ for $\zmax = 0.4$.}
We show the inverted density profiles of the different fits in figures \ref{fig:Z02Rho} and \ref{fig:Z04Rho}. All the inverted profiles point to the presence of \HWC{an over-density around $z=0.02$ in F3, which seems to be connected to the peak in \TMPP, i.e. the Virgo cluster. In addition, there seems to be} a large scale under-density \HWC{for the $(1,0,1)$ model with $\zmax = 0.2$}\COM{in F3}, as discussed in more details in section~\ref{sec:densityrescale}.

\begin{table}[ht!]
\centering
\footnotesize
\begin{tabular}{|c|cccccc|}
\hline
$\zmax$		&Fig.			&$\chi_R^2$&{\it Threshold} (\%)& PRESS	& \modelparam	& Removal	\\
\hline
\multirow{7}{*}{0.2}& N/A		& 19.4	& Not Preferred		& 1621	& $72.94\pm3.48$&			\\
			& \ref{fig:Mu_0}	& 1.41	& $27  \sim 100 $	& 248.6	& $(0, 0, 5)$	&			\\
\cline{2-7}
			& \ref{fig:Z02Mu1a}	& 2.19	& $96.6\sim 100 $	& 131.0	& $62.05\pm1.10$& \multirow{3}{*}{NGC 4536}	\\
			& \ref{fig:Z02Mu1c}	& 2.04	& $69  \sim 96.6$	& 126.5	& $(0, 0, 1)$	&			\\
			& \ref{fig:Z02Mu1b}	& 2.26	& Not Preferred		& 131.5	& $(1, 0, 1)$	&			\\
\cline{2-7}
			& N/A				& 1.67	& $99.6\sim 100 $	& 101.1	& $62.00\pm0.96$& \multirow{2}{*}{+ 1999cl}	\\
			& N/A				& 1.45	& $69  \sim 99.5$	& 92.8	& $(0, 0, 1)$	&			\\
\hline
\multirow{7}{*}{0.4}& N/A		& 17.9	& Not Preferred		& 1627	& $72.95\pm3.33$&			\\
			& \ref{fig:Mu_0}	& 2.53	& $13  \sim 100 $	& 934.2	& $(0, 0, 5)$	&			\\
\cline{2-7}
			& \ref{fig:Z04Mu1a}	& 2.12	& $78  \sim 100 $	& 138.1	& $62.07\pm1.09$& \multirow{3}{*}{NGC 4536}	\\
			& \ref{fig:Z04Mu1c}	& 2.04	& $49  \sim 77  $	& 141.9	& $(0, 0, 1)$	&			\\
			& \ref{fig:Z04Mu1b}	& 2.14	& Not Preferred		& 145.5	& $(1, 0, 1)$	&			\\
\cline{2-7}
			& N/A				& 1.65	& $93  \sim 100 $	& 108.4	& $62.02\pm0.96$& \multirow{2}{*}{+ 1999cl}	\\
			& N/A				& 1.59	& $62  \sim 92  $	& 110.3	& $(0, 0, 1)$	&			\\
\hline


\end{tabular}
\caption{\label{tab:MuRhoChi2}
Distance modulus best fit model parameters with progressive removal of the outliers for F3, without peculiar velocity correction and with 250 \ks velocity dispersion for SNe. The {\it Threshold} column shows the F-test threshold of the model. 
}
\end{table}

\section{Discussions}
\label{sec:discuss}

We here discuss the implications and the limitations of our study and describe some possible ways to \COM{improve} \HWC{increase} its precision \AER{to further  investigate the existence of inhomogeneities.} 

\subsection{Peculiar velocity and galaxy surveys}
\label{sec:discPecVdensity}

The 2M++ catalogue \cite{Carrick15} gives the density field and the predicted peculiar velocities within 200 $\Mpc/h$ ($z<0.67$). The under-density claimed by \Keenan along F2 and F3 has a size of $\sim 300 \, \Mpc/h_{70}$ ($z \sim 0.07$) and is thus roughly outside the 2M++ catalogue. Also the boundary of the under-density, i.e., an over-dense filament-like structure at $z\sim 0.08$ (Sloan Great Wall), is slightly outside the 2M++ window. If this ``super void + filament" structure exists, the depth limitation of 2M++ does not allow a full reconstruction of the velocity field that takes the structure into consideration.

Another independent galaxy survey, Cosmicflows-2 of \cite{Tully2014}, suggests that we are part of a super-structure (basin of attraction) called ``Laniakea''. As shown in figure 1 of \cite{Tully2014}, along the center of F3 (corresponding to the +Y direction in supergalactic coordinates) there seems to exist a large void which would confirm \Keenan's finding, also in \FN{possible} agreement with\COMT{ case (c) of our} figure~\ref{fig:Z02Mu1b}.
In principle, such an inhomogeneity should be detectable in 2M++, Cosmicflows-2, and \Keenan. In practice though, it is hard to compare these surveys since for example \Keenan only uses GAMA DR1 \cite{2011MNRAS.413..971D} as an anchor and does a comparison with its own analysis of the old 2M++ dataset \cite{Lavaux}, while \cite{Carrick15} uses its own average density as the background. The different normalization introduces additional scaling to the density contrast that may jeopardize the whole comparison of their density contrasts. The under-density that we find in F3 is not aligned with the CMB dipole, and thus does not seem explainable from the bulk flow of the Local Group.



\subsection{Choice of background and the associated background density} 
\label{sec:densityrescale}

\begin{figure}[ht!]
\centering
\subfigure[Subregion F1 $( 1, 0, 0 )$]{
	\includegraphics[width=.48\textwidth]{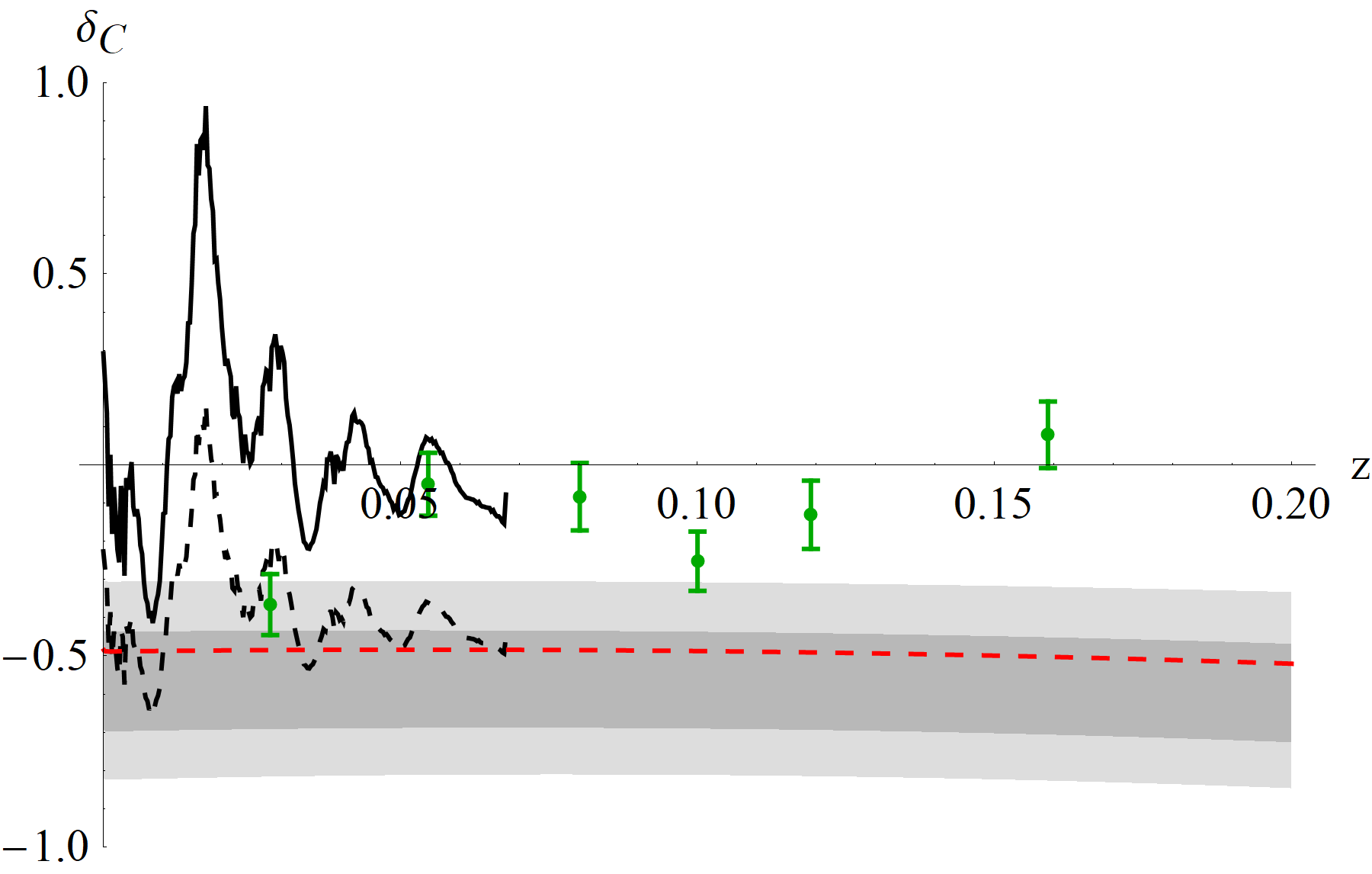}
	\label{fig:F1Rho_rescaled}
 }\\
\subfigure[Subregion F3 $\zmax = 0.2$ with NGC 4536 \COM{and 1999cl} removed $( 1, 0, 1 )$]{
	\includegraphics[width=.48\textwidth]{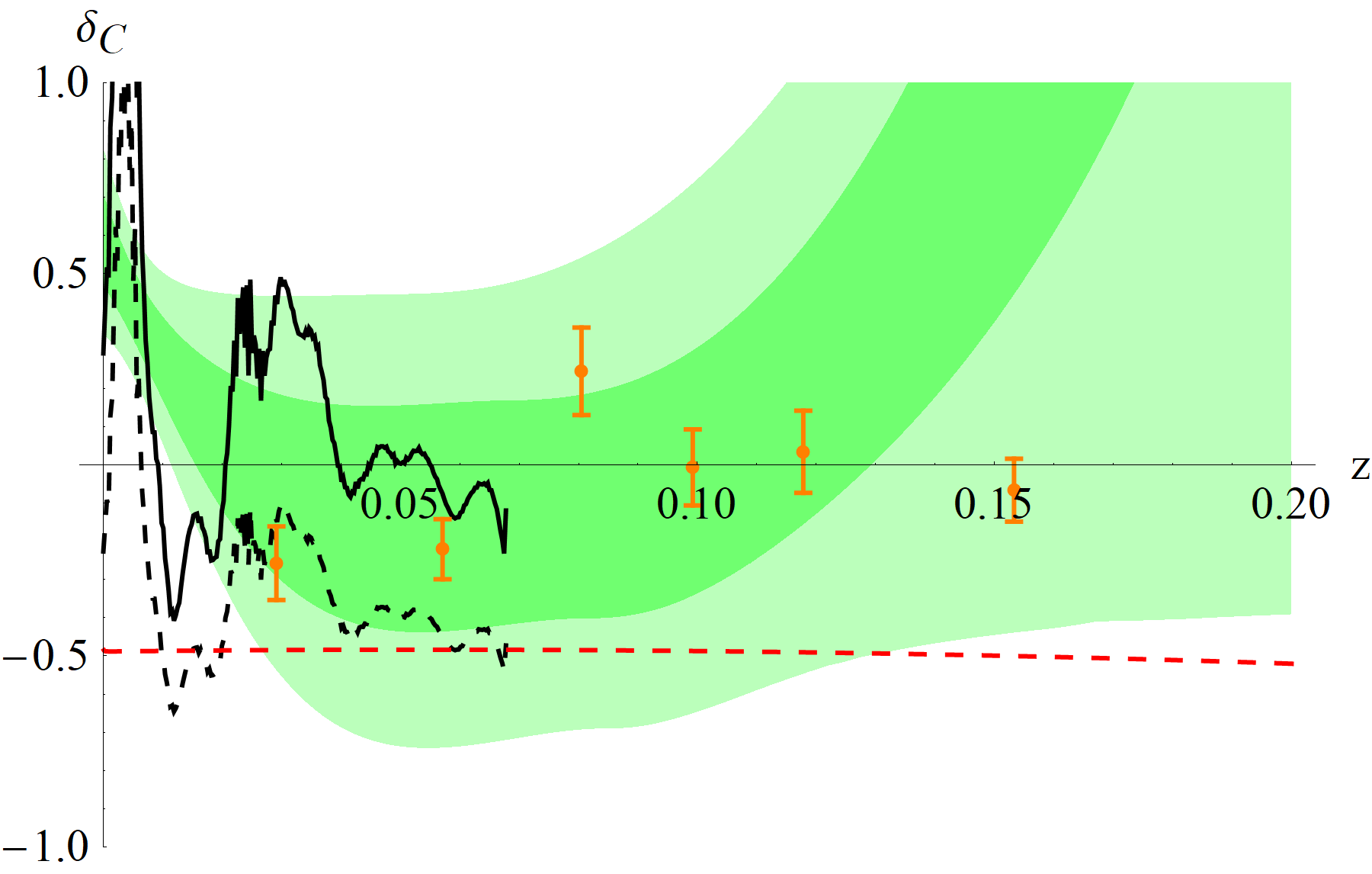}
	\label{fig:Z02Rho2_rescaled}
}
\subfigure[Subregion F3 $\zmax = 0.4$ with NGC 4536 \COM{and 1999cl} removed $( 1, 0, 1 )$]{
	\includegraphics[width=.48\textwidth]{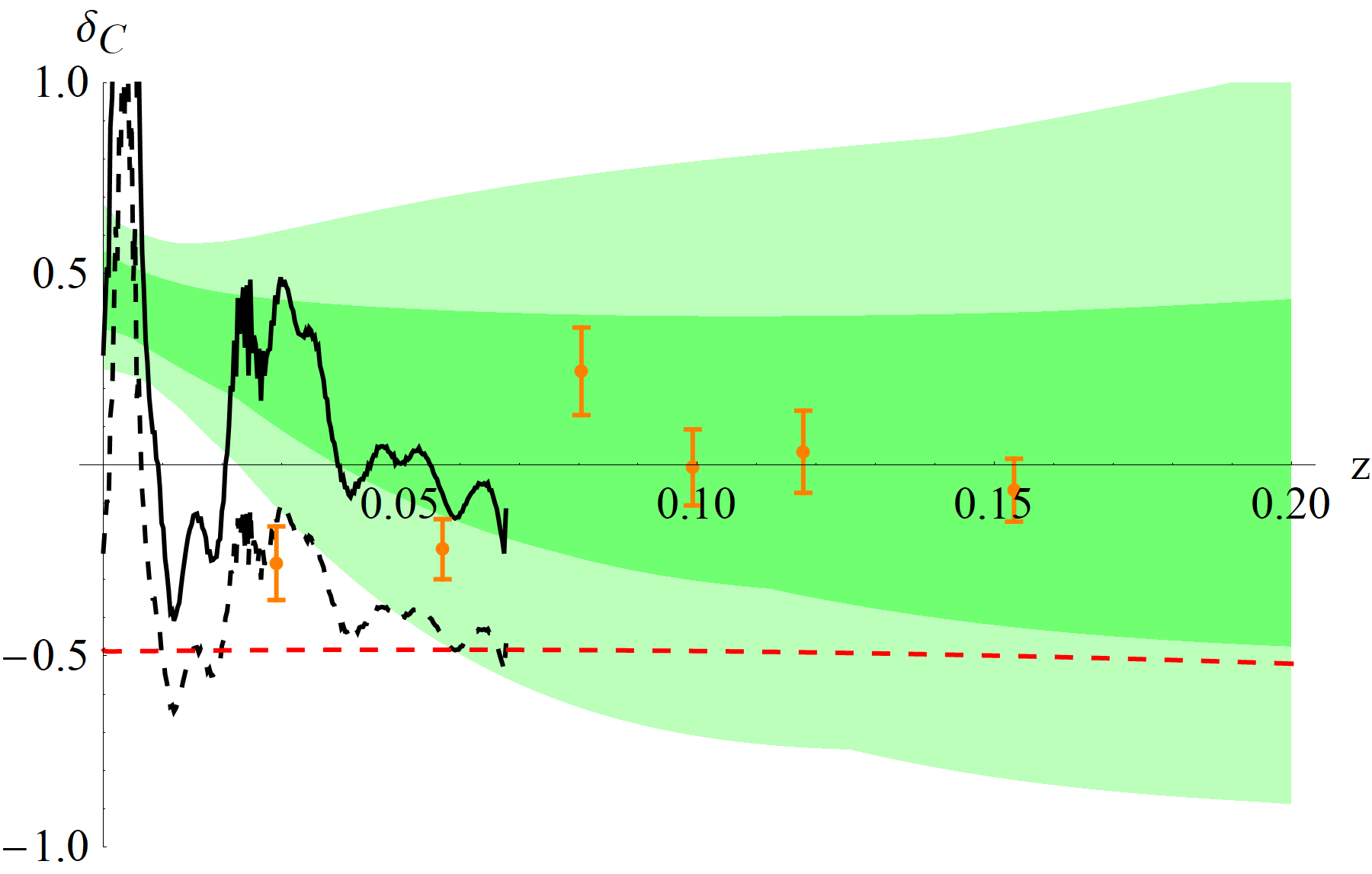}
	\label{fig:Z04Rho2_rescaled}
 }
\caption{\label{fig:Rho_rescaled}
The confidence bands of the inverted density contrast for the standard candles distance modulus data without peculiar velocity corrections and with a 250 \ks velocity dispersion for SNe are shown (68{\%}-darker color, 95{\%}-lighter color). The model parameters and the cuts of the dataset are shown in the sub-captions. In addition the data points of \Keenan are plotted in orange for F3 and green for F1, the \TMPP density contrast averaged over the subregion specified in the sub-captions as a solid black curve, its rescaled version by a factor of $0.6$ according to \Keenan as a dashed black curve, and the dashed red line is for density contrast that would lead to a local Hubble parameter $H_0^{\rm loc}=H_0^{\rm Riess}$ assuming a large scale $H_0^{LS}=H_0^{\rm Planck}$.
}
\end{figure}

As discussed before, one plausible explanation for the discrepancy between the observed $H_0$ values of \Riess and \Planck is that the assumed background density in \TMPP, i.e. the averaged density within its observation depth, is not the real background density. If we take \Keenan background density we would have to rescale the density contrast of \TMPP as
\begin{align}
\delta _C^{\rm cor} = \frac{\tilde{\rho}_\text{\TMPP}}{\tilde{\rho}_\text{\Keenan}} (1+ \delta _C)-1,
\end{align}
where $\delta _C^{\rm cor}$ is the rescaled density contrast, while $\tilde{\rho}_\text{\TMPP}$ and $\tilde{\rho}_\text{\Keenan}$  are the assumed background density of \TMPP and \Keenan. According to \Keenan the needed rescaling is a factor of $\sim 0.6$. As shown in figure~\ref{fig:Rho_rescaled} after the rescaling \HWC{of the density contrast} in both subregion F1 and F3 the \TMPP averaged density matches quite well with the inverted density that we obtain in section~\ref{sec:NOvcYESvd}\HWC{. This, along with the finding that the inverted density in F3 is compatible with the density in \Keenan,} indicat\HWC{es}\COM{ing again} that the existence of the $\sim 300$ Mpc inhomogeneity  of \Keenan could explain the tension between \Riess and \Planck estimation of $H_0$.

Alternatively we may also assume a different background $H_0^\text{\Riess}=73.24$ \ksm. In that case all the inverted densities shown in section~\ref{sec:NOvcYESvd} need to be rescaled as (up to the 
zeroth order in redshift)
\begin{align}
\delta _C^{\rm cor} \sim \frac{H_0^\text{\Riess}}{H_0^{\rm Planck}} (\frac{3}{f} + \delta _C) - \frac{3}{f},
\end{align}
where $f \sim \Omega _{m0}^{0.55}$ is the growth rate, while $H_0^\text{\Riess}$ and $H_0^{\rm Planck}$ are the Hubble parameter of \Riess and \Planck. The resulting shift on the inverted density contrast is $\sim 0.5$, again consistent with the observed density contrast of \TMPP.

Both explanations could explain simultaneously the luminosity distance data of standard candles and the luminous density data. However, a higher $H_0$ value would create inconsistencies with the CMB observations, and since there are solid theoretical reasons to expect that high redshift observations are insensitive to local structure \cite{Romano:2016utn}, it seems more justified to interpret these results as the need for a proper renormalization of \TMPP rather than invoking an hypothetical early Universe physics modification which may affect CMB observations independently from local structure, as proposed for example in \cite{Bernal:2016gxb}. In any case, the difference in the reconstructed density profile for the two different directions is an evidence of anisotropy which cannot be explained by considering different values of $H_0$, and would remain even assuming $H_0^\text{\Riess}$.

\subsection{Impact of Supernovae data quality on the reconstruction of density}
\label{sec:discSNe}

Our analysis uses 70 (145) SNe or Cepheids-hosting galaxies in F1 and 55 (60) in F3 for $\zmax = 0.2$ ($\zmax = 0.4$). Despite the seemingly ``large'' numbers for such a small redshift range, the redshift distribution of the data has some  limitations. In F2 there are no SNe, while \Keenan also reports an under-density profile in that field of view. Along F1 SNe are distributed evenly in redshift, but along F3 most of SNe and Cepheids-hosting galaxies are at $z < 0.04$ (only 9 SNe between $0.04<z<0.2$ and only 5 SNe between $0.2<z<0.4$). So we may  be missing the peak of \Keenan if it exists. 
As our method consists in a 1-dimensional fit,  SNe  with the same redshift but different angles are fitted together, implying that some angular regions can have more weight than others if more SNe are located therein. This can explain the deviation of our reconstructed profiles with respect to the angular average of the 2M++ density profile. In addition, SNe are affected by intrinsic dispersion and, as shown in the past, their intrinsic color can play an important role in assessing the existence of a Hubble bubble (c.f. \cite{Jha2006,Conley2007,Wang2007}).

Nevertheless we can still draw some conclusions from our results. First, according to figures \ref{fig:Z02Rho2_rescaled} and \ref{fig:Z04Rho2_rescaled}, the inverted density contrasts could not reveal the over-dense regions of \TMPP very well, except at very small redshifts ($z \lesssim 0.02$). According to figure~\ref{fig:2mpp}, all SNe actually lie in regions with density very close to the background, which shows that apparently the SNe sample does not probe well extreme density variations. We can argue that this phenomenon is due to the fact that\HWC{, firstly the selected models are too simple to accommodate fine structures, secondly} most star-forming galaxies are lower-mass galaxies outside the densest regions. According to \cite{Odderskov16} the cutoff scale for SNe-producing galaxy mass is $10^{11} M_{\astrosun}$. Therefore, qualitatively SNe underestimate the small scale density fluctuations and that makes SNe more suitable for probing density fluctuations on large\COM{r} scales \COM{than}\FN{compared to} galaxies.


\subsection{Statistical significance of the fits}
\label{sec:BMC}

There are recent publications (c.f. \cite{Cardona2016,Wei2016,Wang2016}) using model-independent techniques like Gaussian process or the Bayesian hyper-parameters method to fit data. Since the multiple models we have are selected using step-wise regression and ranked using $\chi ^2$, it would be interesting to consider those methods as well. 

In order to test the statistical significance of our results, we ran \COM{7100}\HWC{10000} Monte Carlo simulations assuming a homogeneous universe  with \Planck cosmological parameters, with each simulation having the same number of data points as in F3, randomly generated with the same errors of the observational data in F3.

Using a stepwise regression method with a 95\% F-test {\it Threshold} 
we found that is 99.5\% of the best fit models are homogeneous (either a $(0,0,0)$ or a $(1,0,0)$ model). Given the low 0.5\% misidentification rate, the selection of an inhomogeneous model as the best fit model is unlikely to be a statistical fluke. 
In addition, we may also consider the full 3-$D$ fit of the luminosity distance by extending our fitting method to a multi-dimensional version. In that case we may apply a density map reconstruction technique to recover the full-sky density map, and compare it directly to galaxy surveys like \TMPP and Cosmicflows-2, without depending on window averaging.
Notice that "outliers" removal may be an artifact due to window averaging. For example, NGC 4536, 2006x and 1999cl are right behind the Virgo cluster and NGC 4424 is in front of it, making them prone to the extreme density fluctuations within the Virgo cluster. These four objects in fact happen to be \COM{the} most frequently \COM{removed}\HWC{selected as} outliers in our analysis. 


\section{Conclusions}
\label{sec:conclusion}

We have  applied a new method to extract information about the large scale structure from the observed distance modulus of supernovae (SNe) and Cepheids. We used a combination of the \UTPO Type Ia SNe of \cite{Union2.1} (with an added 250 \ks velocity dispersion) and the Cepheid calibrators of \cite{Riess2016} (\Riess). The inversion method we utilize requires the input luminosity distance as a smooth function of the redshift, so we fit the observational data with a set of radial basis functions (RBFs), without any prior on the local structure (except for meta-fitting parameters such as the explicit form of RBF). Using this method, any deviation of the observed luminosity distance from its homogeneous Universe ($\Lambda$CDM) prediction can be used to reconstruct the local structure. Rather than fitting the complete dataset with SNe from the whole sky, we have analyzed the density profile along different directions. Under the assumption that lensing effects at low redshift are negligible the radial profiles in different angular directions have been modeled as independent LTB radial profiles.

Note that there is no fine tuning of the position of the observer since we are reconstructing the radial density profile in different directions separately. In other words, we are not assuming isotropy and the center of the coordinate system where the observer is located is not a center of spherical symmetry. The density profile in a given direction is a function of the radial coordinate and as such can be mapped into the geometry of a solution depending on a single function of the radial coordinate such as the LTB, since this is a one dimensional problem. Assuming the lensing effects are negligible at low redshift  should be a good approximation. In other words the use of a LTB solution is just a computational tool and we are not assuming a spherically symmetric model of the local Universe, since we reconstruct radial profiles in different directions independently.

We focused on further investigating the existence of inhomogeneities with a size of several hundred $\Mpc$, which was previously studied by \cite{Keenan} (\Keenan) using observed luminosity density in different fields of observations. \Keenan studied three different regions (which we call F1, F2, F3), while we reconstruct the density profile only in F1 and F3, where the number of SNe is high enough to allow a statistical analysis. 

Our results are in good agreement with the rescaling of \TMPP proposed by \Keenan for both F1 and F3.
\COM{At low redshift t}\HWC{T}he agreement of our reconstructed density profiles for \COM{F1 and} F3 is also good with respect to  \Keenan, while \HWC{in F1} at higher redshift there is some difference \COM{which could be due to the smaller number of higher-redshift SNe in our dataset}. 
The density profile along F1 and F3 directions are different and this clearly shows the existence of an \emph{anisotropy} not detectable by \TMPP. 
 
The inhomogeneity detection depends crucially on the velocity dispersion of Cepheids-hosting galaxies. This is naturally expected since large values of the velocity dispersion can introduce noise in the data which dominate over the effects of inhomogeneities. 
As a confirmation of the importance of the velocity dispersion, we find that the very low-redshift peak in \TMPP, corresponding to the Virgo cluster, is well reconstructed from the luminosity distance data when we consider a small velocity dispersion for Cepheids-hosting galaxies, but disappears for larger dispersion.
This suggests that the Cepheids-hosting galaxies data should be analyzed assuming a value for the velocity dispersion smaller than the one used in \Riess, and is supported by the observations of nearby clusters \cite{Tully:2007ue,2015ApJ...805..144K}.  Large values of the velocity dispersion could introduce an artificially strong noise in the analysis that contaminates the real signal of large scale structures on the velocity field.

According to our analysis, in some directions the size of the inhomogeneity is larger than the depth of the \TMPP survey. Consequently, the normalization of the latter with respect to the average density of the Universe may require a rescaling, which we find to be very close to what \Keenan obtained. This could in fact play a very important role in explaining the apparent discrepancy between the local and large scale estimation of $H_0$ \cite{Romano:2016utn}, due to the fact that about 40\% of low redshift SNe used to estimate $H_0^{loc}$ are affected by the inhomogeneities we found along F1 and F3.
In the future this method could be used to correctly  normalize density maps with respect to the average density of the Universe, a procedure which can be especially important when galaxy catalogues have a depth smaller than the size of the large scale structure inside which they are embedded, as it seems the case for \TMPP. 

We also checked that our method does not depend significantly on meta-parameters. The accuracy of our distance-modulus fitting is obviously limited by the SNe data precision, and a higher number of events would reduce the size of the confidence bands. The inversion and the comparison with the luminosity density profiles are also limited by the non-uniform angular and redshift distribution of SNe. For instance, F3 does not have a lot of data points at $z \sim 0.08$ (with less than ten SNe for $0.04 < z < 0.2$), so the size of the inhomogeneity cannot be precisely confirmed. Future data could overcome these statistical limitations.

Our analysis could be extended to a larger dataset, for example the carefully calibrated SNe data presented in \Riess (see also \cite{Courtois:2012kg,Sorce:2012pk,Sorce:2013wt} for SNe) and the density field of Cosmicflows \cite{Tully2013}. We could also use the Nearby Supernova Factory data of \cite{Aldering2002} (see \cite{Feindt2013} for example) or other future surveys such as WFIRST\footnote{\url{https://jet.uchicago.edu/blogs/WFIRST/}}, in order to find out whether or not we can reconstruct a more accurate large scale structure map and compare it to what is obtained using galaxy catalogues. Once new data will be available supernovae could become a very important source of information about large scale structure, especially useful at high redshift where other astrophysical objects are difficult to observe, allowing to overcome the limits of depth and angular limitation of galaxy catalogues.
In the future it could be also interesting to develop a fully non linear inversion method which could  be applied when the assumptions we made are not satisfied, such as at high red-shift, or when lensing is important.


\section*{Acknowledgments}

The authors of this paper particularly want to thank Prof. Adam G. Riess for his insightful comments. They also want to thank Prof. Lucas M. Macri and Dr. Ryan Keenan for giving us details on their publications, and Dr. Guilhem Lavaux for pointing out the presence of an external dipole in the \TMPP velocity map. FN is grateful to Prof. Pierre Astier, Dr. S\'{e}bastien Bongard, Dr. Marc Betoule and Dr. Jenny Sorce for interesting discussions about supernovae data. FN and HWC also want to thank Dr. Peter Scicluna and Dr. Konstantina Kontoudi for their advice on data fitting. FN and HWC researches are supported by the Leung Center for Cosmology and Particle Astrophysics (LeCosPA) of the National Taiwan University (NTU). This research has made use of the NASA/IPAC Extragalactic Database (NED) which is operated by the Jet Propulsion Laboratory, California Institute of Technology, under contract with the National Aeronautics and Space Administration. 
This work was supported by the UDEA Dedicacion exclusiva and
Sostenibilidad programs and the CODI projects 2015-4044 and 2016-10945.


\appendix
\begin{appendices}

\section{Effect of cosmological constant on density contrast}
\label{sec:lambda}

As briefly explained in the section~\ref{sec:theory}, the effect of the cosmological constant on the density contrast is non-negligible  at the very small redshift because the growth rate can play an important role despite the effects of dark energy  on the luminosity distance are not important. Here we derive the  effects of $\Lambda$ on $\delta_C$ comparing with  perturbation theory. We then test the inversion method applying it to luminosity distance  computed numerically using  a $\Lambda$LTB solution  showing that the reconstructed density contrast defined in eq.~\eqref{eq:deltaC} is  in good agreement with the numerical computation of the corresponding density profile.

According to \cite{Romano:2016utn}, assuming a spherically symmetric universe, the density contrast $\delta _C$ in the pertubative regime is given by

\begin{align}
\delta _C &= \frac{\chi}{3} \frac{d\bar{\delta}}{d\chi} + \bar{\delta} \quad , \quad \bar{\delta} = \frac{3}{f} \left( \frac{D_L^{\rm obs}}{\overline{D}_L} -1 \right) \left( 1-z \right)^{-1} \,, \label{lineardc}
\end{align}
where $\chi$ is the unperturbed comoving distance, f the growth rate defined in eq.~\eqref{eq:f}, $z$ the observed redshift at $\chi$, $D_L^{\rm obs}$ the observed and $\overline{D}_L$ the background luminosity distances at $z$, and $\bar{\delta}$ is the comoving-volume-averaged density contrast $\bar{\delta} = \left( 4\pi \chi ^3 /3 \right)^{-1} \int ^{\chi} 4\pi {\chi'}^2 \delta_C \left( \chi' \right)  d\chi'$.
\COMT{For a perturbed ($\Lambda$)CDM model with}\HWCR{For a ($\Lambda$)CDM model with a spherically symmetric density contrast perturbatively solved up to the first order, i.e. a linearized ($\Lambda$)LTB model, and} cosmological background parameters estimated from large scale observations\COMT{, i.e. a linearized ($\Lambda$)LTB model}, the density contrast at first order in z can be expressed as
\begin{align}
\delta _C^{\Lambda \rm LTB} (\Delta) &= \frac{3}{f} \Delta + \frac{z}{f} \left( \frac{d \Delta}{d z} + 4 \Delta \right) \,, \label{eq:linearrho} \\
\delta _C^{        \rm LTB} (\Delta) &=       3     \Delta +       z     \left( \frac{d \Delta}{d z} + 4 \Delta + 3\Omega _{\Lambda 0} \left( 1 + \Delta \right) \right) \,,
\end{align}
where $\Delta = \frac{D_L^{\rm obs}}{D_L^{\text{Hom}}} -1 \equiv 10^{\left(\mu ^{obs} -\mu ^{\text{Hom}} \right)/5}-1$ is the relative difference between observed luminosity distance and the luminosity distance $D_L^{\text{Hom}}$ of a homogeneous $\Lambda$CDM model with cosmological background parameters estimated from large scale observations, such as \Planck. 

There are two differences between LTB and $\Lambda$LTB models. First, there are additional terms proportional to $\Omega _{\Lambda 0}$ in $\delta _C^{\rm LTB}$ coming from $D_L^{\text{Hom}}$. 
Second, the differences in the evolution of the Hubble flow and of the background matter density imply a different growth rate values. 
The density contrast we propose in eq.~\eqref{eq:deltaC}
\begin{align}
\delta _C &= f^{-1} \left( \frac{\rho_{\rm inv} \left(D_L^{obs}, z \right)}{\rho_{\rm inv} \left(D_L^{\rm Planck}, z \right)} -1 \right)
\approx f^{-1} \left( \frac{1 + \delta _C^{\text{LTB}} \left( \Delta \right) }{1 + \delta _C^{\text{LTB}} \left( 0 \right) } -1 \right) \nonumber \\
&\approx f^{-1} \left( 3 \Delta + z \left( \frac{d \Delta}{d z} + 4 \Delta - 6 \Omega _{\Lambda 0} \Delta \right) \right)
\approx \left( 1-2 \Omega _{\Lambda 0} z \right) \delta _C^{\Lambda\text{LTB}} (\Delta)\,,
\end{align}
takes care of all the factors mentioned above, and is approximately the density contrast in a $\Lambda$LTB model up to first order in redshift. The term $-2 \Omega _{\Lambda 0} z \delta _C^{\Lambda\text{LTB}} (\Delta)$ is actually at the second order and we do not consider it for simplicity.

\begin{figure}[ht!]
\centering
\includegraphics[width=.488\textwidth]{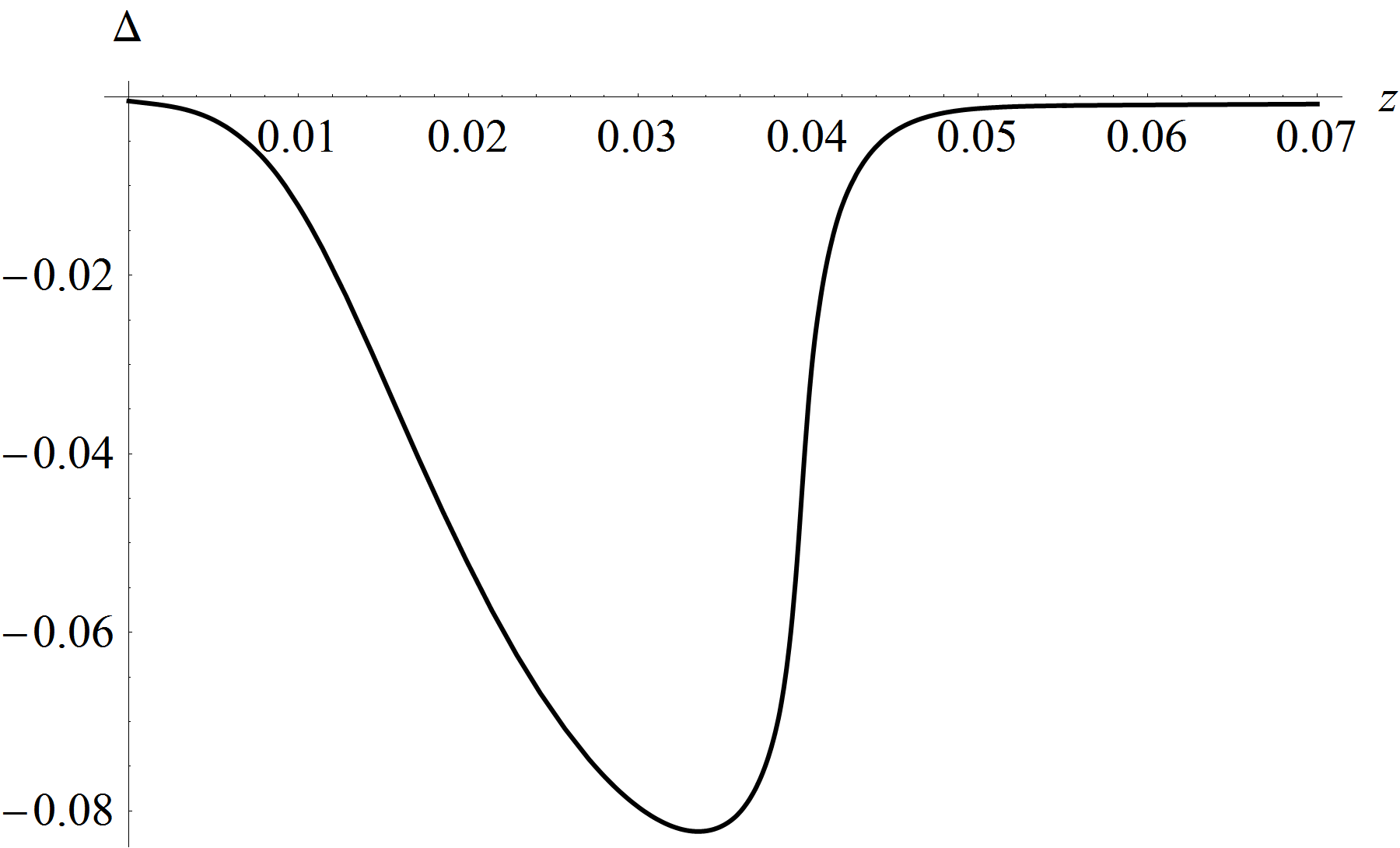}~\includegraphics[width=.488\textwidth]{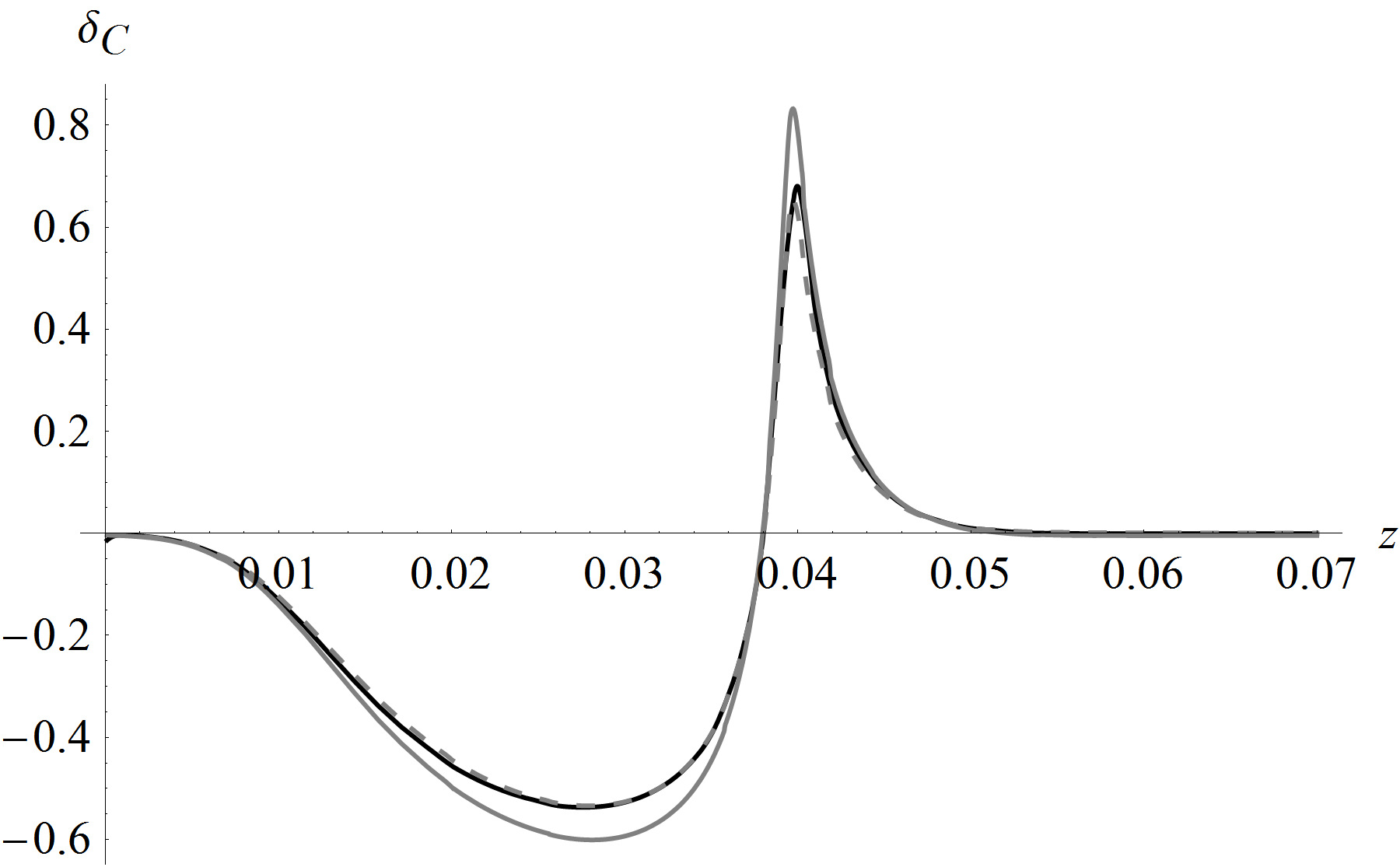}
\caption{\label{fig:linear}
\emph{Left}: The plot of $\Delta = \frac{D_L}{D_L^{\rm Planck}} -1$ for the model from \cite{Romano:2014iea}.
\emph{Right}: The plot of density contrast for the model from \cite{Romano:2014iea}. The black curve is the original density contrast of the $\Lambda$LTB model. The  grey curve is the density profile reconstructed from the luminosity distance of the   $\Lambda$LTB model   using the linear method defined in eq.~\eqref{lineardc}. The  dashed  curve is the  density profile reconstructed using the non linear method defined in eq.~\eqref{eq:deltaC}. The profile reconstructed  using the non linear method matches well  the original profile, while the pertubative formula is not as accurate in some regions due to non linear effects.
}
\end{figure}

To test the validity of eq.~\eqref{eq:deltaC}, we consider a specific setup (compensated void, i.e. $\lim _{z\to \inf} \bar{\delta} \to 0$) presented in \cite{Romano:2014iea}. 
As shown in fig. ~\ref{fig:linear}, the reconstructed density contrast obtained using eq.~\eqref{eq:deltaC} is in very good agreement with the numerical calculation using the  $\Lambda$LTB solution, indicating that eq.~\eqref{eq:deltaC} is indeed a good approximation.

Nevertheless there are some limitations related to the inversion derived above, since it is based on the low red-shift expansion of eq.~\eqref{lineardc}. 
The first limitation is related to the fact that at high red-shift eq.~\eqref{eq:deltaC} may require higher order corrections in the red-shift expansion.
The second has to do with the fact that eq.~\eqref{eq:deltaC} is assuming that the Doppler effect is the dominant contribution, but in the fully non linear regime it is possible that other  contributions such as  lensing or  ISW could become important.
Consequently from a purely mathematical point of view it cannot be claimed that the inversion method in eq.~\eqref{eq:deltaC} is fully non linear,  but  it can at least be stated that is better than the perturbative method as shown in fig.~\ref{fig:linear}, and quite accurate as long as the Doppler term is dominating, which is a reasonable assumption at low red-shift. 

As explained in appendix \ref{sec:DirectionalLTB} lensing observations support the assumption that the Laplacian of the gravitational potential in the directions orthogonal to the line of sight is negligible compared to the Laplacian in the radial direction, so  when applied to low red-shift data, the method should be accurate, but from a purely mathematical point of view it will be less accurate when applied to high red-shift data, or for  systems with large shear or  large non Doppler effects.  

\section{LTB inversion and anisotropic matter distributions}
\label{sec:DirectionalLTB}

To justify the application of LTB inversion method we will demonstrate that up to first order in perturbation theory and assuming, as supported by observations\cite{Hikage:2018qbn,Troxel:2017xyo,Kohlinger:2017sxk,Hildebrandt:2016wyi,Kilbinger:2014cea,VanWaerbeke:2013eya}, a negligible lensing effect, it is possible to express the effects of inhomogeneities on the luminosity distance only in terms of $\delta_C$ along the line of sight. 
 According to eq.~(B1) of \cite{Bonvin:2005ps}, the luminosity distance can be expressed as
\begin{align}
d_L \left( z_S, \vec{n} \right) &= \left( 1+z_S \right) \Delta\eta \Bigg{\{} 1 + \frac{\Psi_O  - \Psi_S - \left(\vec{v}_O - \vec{v}_S \right) \cdot \vec{n}}{\mathcal{H}_S \Delta\eta} - \Psi_S - \vec{v}_S \cdot \vec{n} \nonumber\\
& + \int_{\eta_S}^{\eta_O}d\eta \left[ \frac{2}{\Delta\eta} - \frac{2}{\Delta\eta} \left( \frac{1}{\mathcal{H}_S} - \int_{\eta_S}^{\eta}d\eta' \right) \nabla_{\vec{n}} - \frac{\left( \eta - \eta_S \right) \left( \eta_O - \eta \right)}{\Delta\eta} \nabla_\bot^2 \right] \Psi \Bigg{\}} \,, \label{eq:dLperturbed}
\end{align}
where subscripts ${}_O$ and ${}_S$ stand for observer and source, $\nabla_\bot^2 \equiv \triangle - \nabla_{\vec{n}} \nabla_{\vec{n}} + 2 \left( \eta_O - \eta \right)^{-1} \nabla_{\vec{n}}$ is the screen-space Laplacian\cite{Bolejko:2012uj}, $A \cdot B = A^i B_i$ is the inner product with $i$ the spatial index running from $1$ to $3$, $\nabla_{\vec{n}} = \vec{n} \cdot \vec{\nabla}$ is the covariant derivative along $\vec{n}$
, $\triangle = \vec{\nabla} \cdot \vec{\nabla}$ is the 3-Laplacian, $\vec{v}$ is the peculiar velocity, $\vec{n}$ is the unit 3-vector along the photon direction, $\eta$ is the conformal time along the photon trajectory, $\Delta \eta \equiv \eta_O - \eta_S$, $\Psi$ is the gravitational potential, and $\mathcal{H}$ is the conformal Hubble parameter.
Only the second derivative term explicitly depends on quantities off the line of sight, and corresponds to the \COMT{lensing effect}\HWC{convergence in the context of the weak lensing}\cite{Bolejko:2012uj,Kilbinger:2014cea}, which at low red-shift should be negligible, as supported both by theoretical calculations \cite{Bolejko:2012uj} and observation \cite{Hikage:2018qbn,Troxel:2017xyo,Kohlinger:2017sxk,Hildebrandt:2016wyi,Kilbinger:2014cea,VanWaerbeke:2013eya}. 

Assuming  $\nabla_\bot^2 \Psi \approx 0$, zero pressure for the background and perturbations, and that the peculiar velocity field is irrotational, the perturbed Einstein's equations give \cite{baumann2015cosmology} 

\begin{align}
&\frac{2}{3} \Omega_M \mathcal{H}^2 \delta_C = \triangle \Psi 
 = \nabla_\bot^2 \Psi + n^i \nabla_i \left( n^j \nabla_j \Psi \right) - 2 \left( \eta_O - \eta \right)^{-1} n^i \nabla_i \Psi 
 \approx r^{-2} \partial_r \left( r^2 \partial_r \Psi \right) \,,\label{B2}\\
&\frac{2}{3} \Omega_M \mathcal{H}^2 n^i v_i = - n^i \partial_i \left[ \left( \partial_\eta +\mathcal{H} \right) \Psi \right] \equiv \left( \partial_\eta +\mathcal{H} \right) \partial_r \Psi = f \mathcal{H} \partial_r \Psi \,,\label{B3} 
\end{align}
where $\Omega_M$ is the matter density parameter, $r \equiv \eta_O - \eta$ the comoving distance from the observer, $\partial_r = - n^i \partial_i$ the partial derivative of $r$, $\partial_\eta$ the conformal time derivative,\HWCR{ $a$ is the scale factor,} and\HWCR{ $f$ is the growth rate defined in eq.~\eqref{eq:f}}. 
We will show later that lensing observations indeed support the assumption $\nabla_\bot^2 \Psi \approx 0$.
In a fully non linear regime this approximation may not be valid anymore, and a non spherically symmetric approach should be adopted. Nevertheless based on our analysis we expect the non linear effects to be negligible at low red-shift, as supported also by other observations \cite{Hikage:2018qbn,Troxel:2017xyo,Kilbinger:2014cea}.

\HWCR{
It is important to check under which conditions the orthogonal part of  the Laplacian $\nabla_\bot^2\Psi$ can be neglected,
and this can be done by using its relation to the 
the convergence
\begin{equation}
\kappa \left[ \eta_S \right] \equiv - \int_{\eta_S}^{\eta_O}d\eta \frac{\left( \eta_O - \eta \right) \left( \eta - \eta_S \right)}{\Delta\eta} \nabla_\bot^2 \Psi \,. \label{eq:kappa}
\end{equation}
Solving eq.~\eqref{B2} without assuming $\nabla_\bot^2\Psi \approx 0$, we obtain
\begin{align}
\Psi = \Psi_O + \int r^{-2} dr \int r'^2 dr' (\triangle \Psi - \nabla_\bot^2\Psi) \approx \Psi_{\text{iso}} + \kappa \,,
\end{align}
where $r$ is the comoving distance from the observer, $\Psi_\text{iso}$ is the potential obtained assuming isotropy, i.e. the solution of $r^{-2} \partial_r \left( r^2 \partial_r \Psi_{ISO} \right)=\frac{2}{3} \Omega_M \mathcal{H}^2 \delta_C$. 

From  the identity $\frac{d}{d\eta}\Psi \left( \eta, \vec{x}\left( \eta \right) \right) = \frac{\partial}{\partial \eta}\Psi + \nabla_{\vec{n}} \Psi \equiv s^{-1} \Psi + \nabla_{\vec{n}} \Psi$, we have
\begin{align}
\int_{\eta_S}^{\eta_O}d\eta \frac{\Psi}{s} = \Psi \bigg{|}_{\eta_S}^{\eta_O} -&  \int_{\eta_S}^{\eta_O}d\eta \nabla_{\vec{n}} \Psi \sim \left( \Psi  - s \nabla_{\vec{n}} \Psi \right) \bigg{|}_{\eta_S}^{\eta_O} - s \int_{\eta_S}^{\eta_O}d\eta \nabla_\bot^2 \Psi \,,\\
\int_{\eta_S}^{\eta_O}d\eta \frac{\nabla_{\vec{n}} \Psi}{s} \sim& \nabla_{\vec{n}} \Psi \bigg{|}_{\eta_S}^{\eta_O} + \int_{\eta_S}^{\eta_O}d\eta \nabla_\bot^2 \Psi \,,\\
\int_{\eta_S}^{\eta_O}d\eta \int_{\eta_S}^{\eta}d\eta' \frac{\nabla_{\vec{n}} \Psi}{s} =& -\Delta\eta {\nabla_{\vec{n}} \Psi}_S + \int_{\eta_S}^{\eta_O}d\eta \left( 1 - \int_{\eta_S}^{\eta}d\eta' \nabla_{\vec{n}} \right) \nabla_{\vec{n}} \Psi \nonumber \\
\sim s \nabla_{\vec{n}} \Psi \bigg{|}_{\eta_S}^{\eta_O} &- \Delta\eta {\nabla_{\vec{n}} \Psi}_S + \int_{\eta_S}^{\eta_O}d\eta \left( s + \int_{\eta_S}^{\eta}d\eta' \right) \nabla_\bot^2 \Psi \,,
\end{align}
where $\sim$ denotes that terms depending only on the density contrast $\delta_C$ along the line of sight, are dropped and we have defined  $s^{-1} = \left( f-1 \right) \mathcal{H}$, and approximated it as a constant, which should be reasonable within the regime of interest ($z<0.2$).

According to  eq.~\eqref{eq:dLperturbed} the relative  fractional difference between the perturbed and background luminosity distance can be split into the sum of a radial $\Delta_r$ and transverse $\Delta_{\bot}$ part as
\begin{align}
\frac{d_L \left( z_S, \vec{n} \right)}{\left( 1+z_S \right) \Delta\eta} -1 &= \Delta_r+ \Delta_{\bot}=\Delta_r - \frac{1}{\Delta\eta} \int_{\eta_S}^{\eta_O}d\eta \left[ \frac{2s}{\mathcal{H}_S} + \int_{\eta_S}^{\eta}d\eta' \left( \eta' - \eta_S -2s \right) \right] \nabla_\bot^2 \Psi \,.
\end{align}
The first part $\Delta_r$ contains terms depending only on observables along the line of sight.
The term $\Delta_{\bot}$ is associated to the transverse part of the Laplacian, it is the one which is being neglected when assuming isotropy, and as such can be interpreted as the error due to such an approximation. Using eq.~\eqref{eq:kappa} we can conveniently write $\Delta_{\bot}$ in terms of the convergence as
%
\begin{align}
\Delta_{\bot} &= \frac{2s}{\mathcal{H}_S} \left( \frac{\kappa'}{\Delta\eta} - 2\mathbf{E} \left[ \frac{\kappa'}{\eta_O - \eta} \right] \right) +2s \left( \kappa' - \mathbf{E} \left[ \kappa' \right]\right) +  \kappa \,,
\end{align}
where $\kappa' \left[ \eta \right] = \frac{\partial}{\partial \eta} \kappa \left[ \eta \right]$ and $\mathbf{E}\left[ F \right] = \frac{1}{\Delta\eta} \int_{\eta_S}^{\eta_O}d\eta F$.

Given that the shear power spectrum $P_{\gamma}$ is the same as the convergence power spectrum $P_{\kappa}$ (c.f. e.g. eq.~(30) of \cite{Kilbinger:2014cea}), we may estimate the error due to assuming isotropic by using  weak lensing observations. According to \cite{Hikage:2018qbn,Troxel:2017xyo,Kohlinger:2017sxk,Hildebrandt:2016wyi,Kilbinger:2014cea,VanWaerbeke:2013eya} lensing observations give a value of the shear power spectrum of $\sim 10^{-4} z^2$, implying that the error in neglecting $\Delta_{\bot}$ is indeed negligible for any practical application of the inversion method.} Furthermore the convergence power spectrum is theoretically predicted \cite{Bernardeau:1996un} to grow as $z_S^{1.5} \sim z_S^{2.5}$, which implies that the dominating part of $\Delta_{\bot}$ decays approximately as  $1/\Delta\eta$,  confirming the validity of the isotropic approximation in the linear regime.

\section{Results with an additional dispersion added to Cepheids-hosting galaxies}
\label{sec:NOvcNOvd}

We present here the results of our analysis in F1 and F3, without peculiar velocity corrections, and as opposed to section~\ref{sec:NOvcYESvd}\HWC{, with $v_c = 40$}\COM{ without adding a velocity dispersion of $250$ \ks to the SNe data}. The results are very similar to the case \COM{including}\HWC{excluding} this dispersion, but are still differing slightly due to the induced small \COM{increase}\HWC{decrease} of \COM{SNe}\HWC{Cepheids} importance (\HWC{higher}\COM{lower} uncertainty).

For F1, where most high redshift SNe are located, without removing any outlier and with redshift cut $\zmax = 0.2$, the fit we get is almost identical to figure~\ref{fig:F1MuRhoZmax0pt2}, with either a $(1,0,0)$ model with $H_0^{\rm loc} = 74.18 \pm 1.89$ \ksm for a F-test {\it Threshold} $> 38\%$, or a inhomogeneous $(1,1,9)$ model when the F-test {\it Threshold} is lower. The reduced chi square $\chi_R^2 \sim 0.74$ of the inhomogeneous model is not so much lower than that of the $(1,0,0)$ model ($\chi_R^2 \sim 1.01$).

For F3, without outliers removal, a $(0,0,5)$ model with both $\zmax = 0.2$ and $\zmax = 0.4$ cases is preferred. \COM{Despite the absence of $w_{-1}$ for $\zmax = 0.4$ with respect to figure~\ref{fig:Mu_0},}\HWC{Again} the fits are very similar \HWC{to what we obtain in section \ref{sec:NOvcYESvd}.} \COM{(due to the low $z$ data points dominating the fit). Again these fits are not invertible into radial density profiles and we have to use progressive removal of outliers. The}\HWC{Other} fits are presented in table~\ref{tab:nvd_Chi2} and are very similar to those presented in table~\ref{tab:MuRhoChi2}. \COM{In the case of $\zmax = 0.2$ with NGC 4536 and SN 1999cl considered as outliers, a $(0,0,2)$ model is preferred again. The corresponding $\chi^2_R$ is $\sim 1.77$, as compared to $\sim 1.47$ in section~\ref{sec:NOvcYESvd}, which indicates a lower fit quality and shows the relevance of adding peculiar velocity dispersion to the analysis. For the case of $\zmax = 0.4$ with removal of the same two outliers a $(0,0,4)$  model is again preferred, with $\chi^2_R \sim 1.58$, higher than $\sim 1.37$ obtained in section~\ref{sec:NOvcYESvd}.}

\begin{figure}[ht!]
\centering
\subfigure[$\zmax=0.2$ NGC4536 removed $( 1, 0, 0 )$]{
	\includegraphics[width=.48\textwidth]{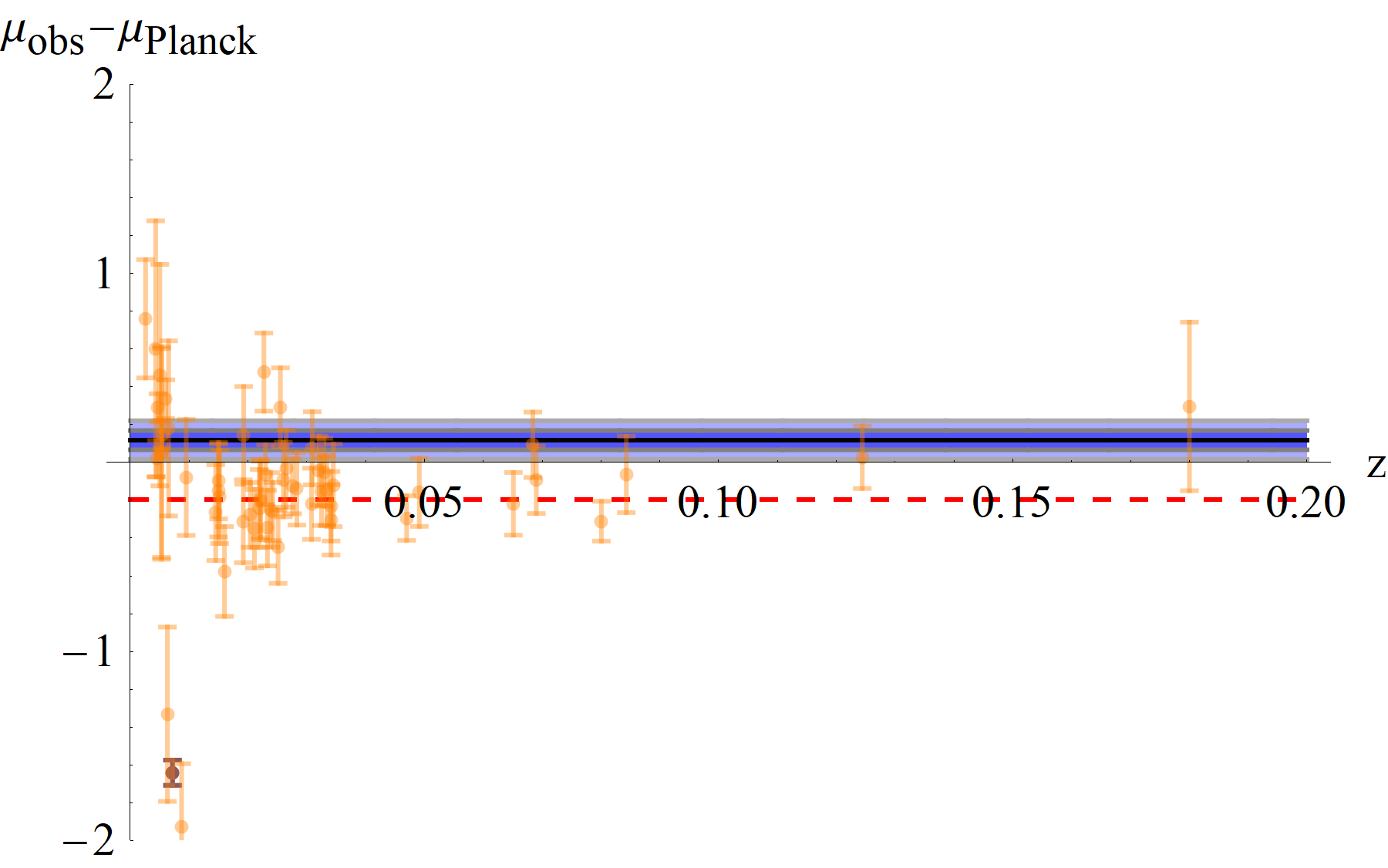}
	\label{fig:svd_0pt2_Mu_1}
 }
\subfigure[$\zmax=0.2$ NGC4536 and 1999cl removed $( 1, 0, 1 )$]{
	\includegraphics[width=.48\textwidth]{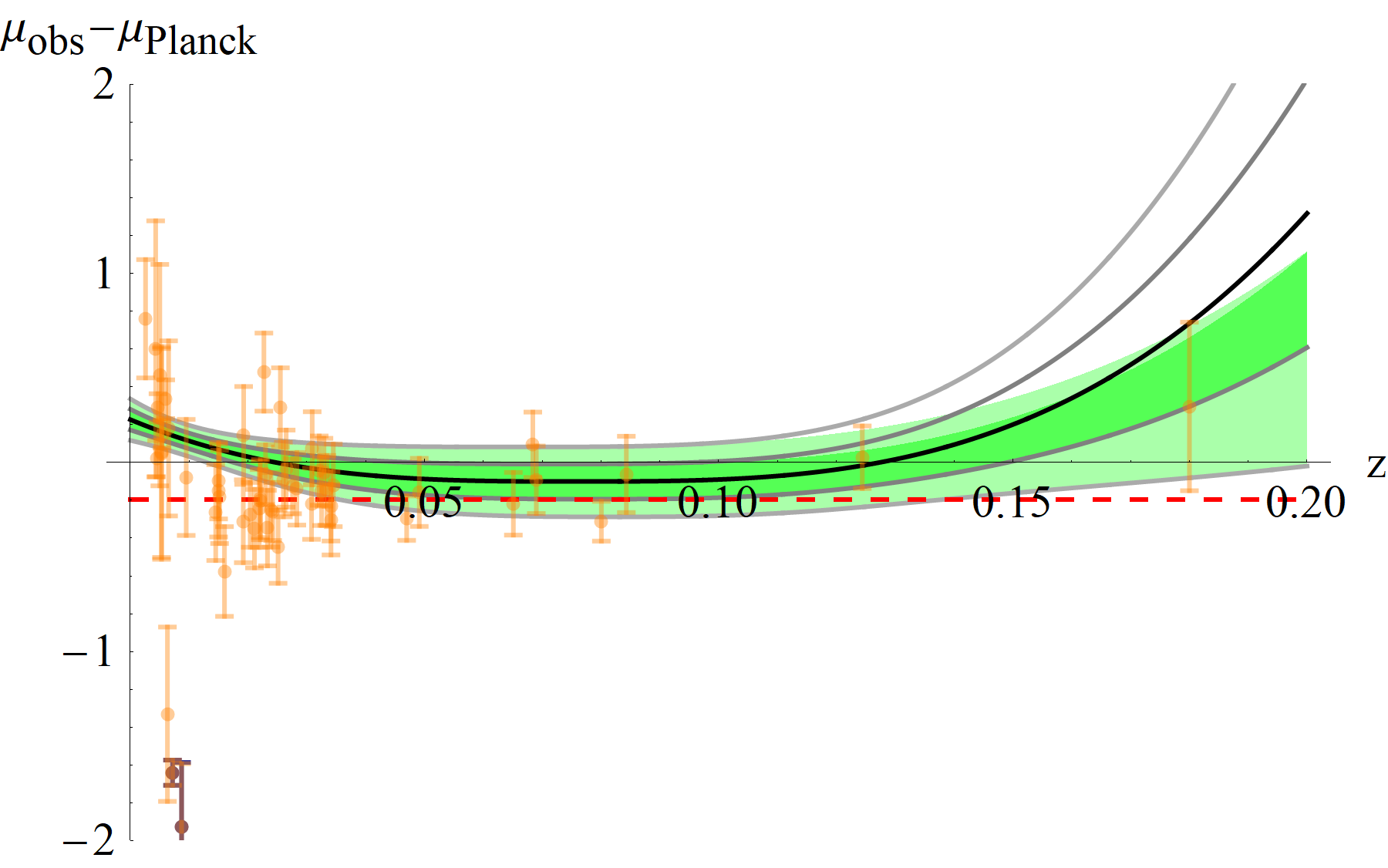}
	\label{fig:svd_0pt2_Mu_2}
 }
\subfigure[$\zmax=0.4$ NGC4536 removed $( 1, 0, 1 )$]{
	\includegraphics[width=.48\textwidth]{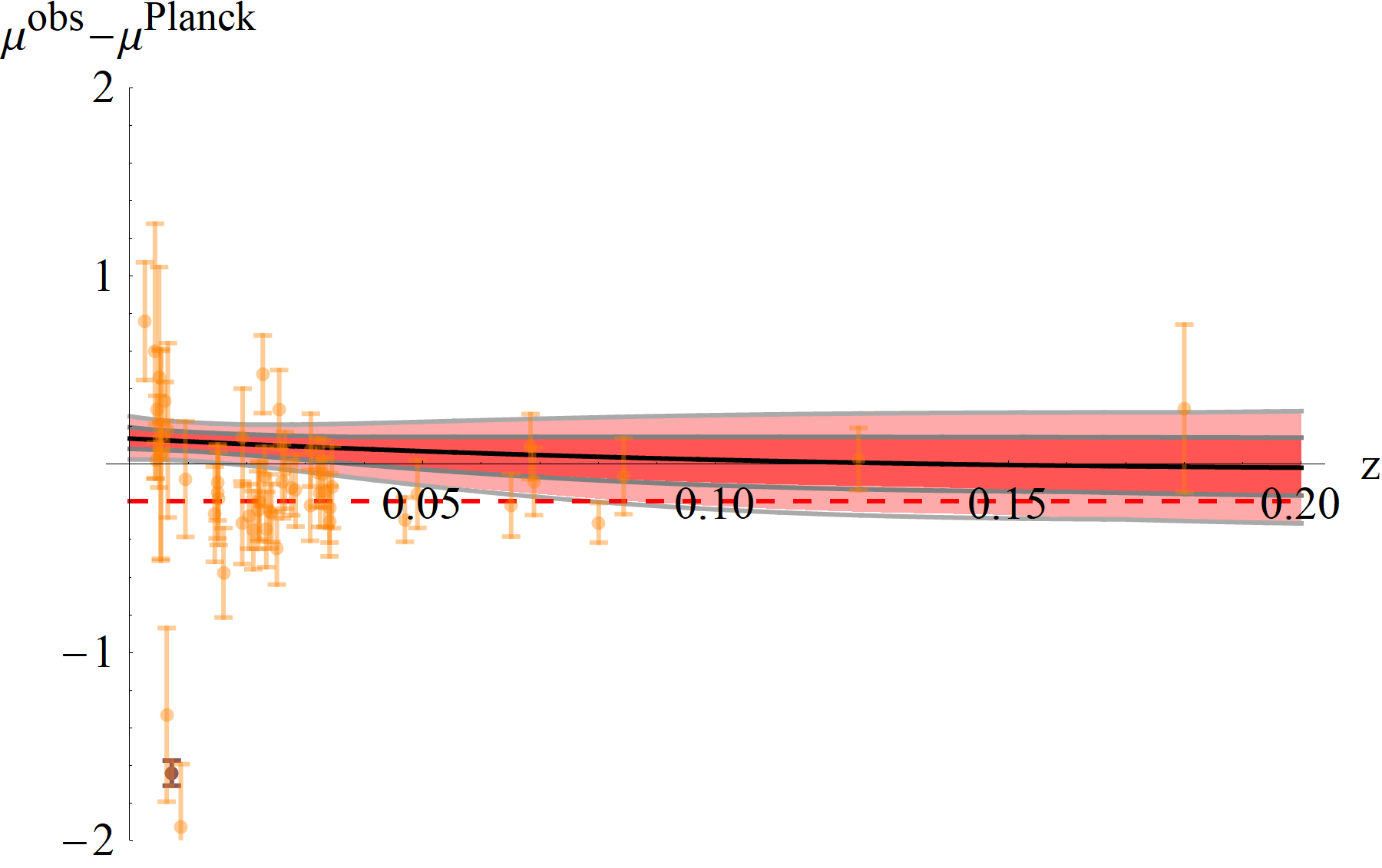}
	\label{fig:svd_0pt4_Mu_1}
 }
\subfigure[Inverted density]{
	\includegraphics[width=.48\textwidth]{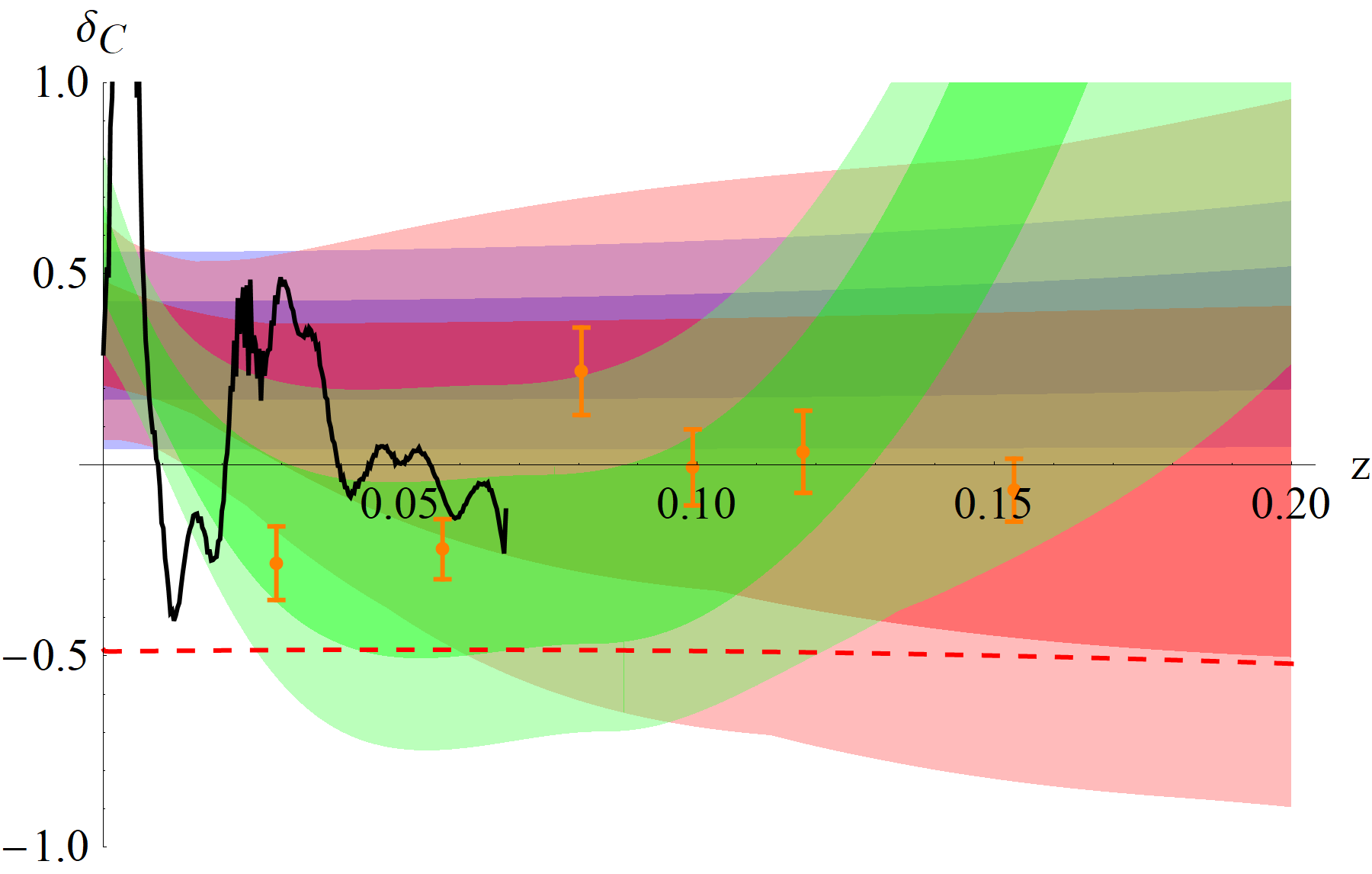}
	\label{fig:svd_Rho_stack}
 }
\caption{\label{fig:nvd_0pt2}
Distance modulus best fit models are plotted for F3 without peculiar velocity corrections and with $v_c = 40$ \ks. The redshift cut $\zmax$, model parameters and the removed data points are shown in the sub-captions. \emph{(a, b, c)}: Standard candles distance modulus data are plotted with their best fit (black), 68{\%} (gray) and 95{\%} (light gray) confidence bands according to the method of section~\ref{sec:method}. The invertible bands are shown as the shaded region (68{\%}-darker color, 95{\%}-lighter color), while the removed outliers are shown as darker data points. \HWC{The dashed red line is plotted as a reference and corresponds to $\mu ^{\rm Riess} - \mu ^{\rm Planck}$.} \emph{(d)}: The confidence bands of the inverted density contrast corresponding to the invertible bands of the distance modulus are shown (68{\%}-darker color, 95{\%}-lighter color). The data points of \Keenan are plotted in orange, the \TMPP density contrast averaged over F3 as a solid black curve, and the dashed red line is for density contrast that would lead to a local Hubble parameter $H_0^{\rm loc}=H_0^{\rm Riess}$ assuming a large scale $H_0^{LS}=H_0^{\rm Planck}$. The bands are color coded case by case with blue (a), green (b), red (c).
}
\end{figure}

Finally one can observe that the different fits for $\zmax = 0.2$ and $\zmax = 0.4$, presented in figure\COM{s} \ref{fig:nvd_0pt2}\COM{ and \ref{fig:nvd_0pt4}}, are very similar to those obtained in figures \ref{fig:MuRhoZmax0pt2} and \ref{fig:MuRhoZmax0pt4}\COM{. Indeed, these fits differ mostly at very small redshifts ($z < 0.03$ for $\zmax = 0.4$)}, and the conclusions made in section~\ref{sec:NOvcYESvd} regarding \Keenan and 2M++ still hold.

\begin{table}[ht!]
\centering
\footnotesize
\begin{tabular}{|c|cccccc|}
\hline
$\zmax$		&Fig.			&$\chi_R^2$&{\it Threshold} (\%)& PRESS	& \modelparam	& Removal	\\
\hline
\multirow{8}{*}{0.2}& N/A		& 11.9	& Not Preferred		& 982.9	& $76.22\pm3.87$&			\\
			& N/A				& 1.25	& $25  \sim 100 $	& 166.2	& $(0, 0, 5)$	&			\\
\cline{2-7}
			& N/A				& 2.15	& $97.6\sim 100 $	& 116.1	& $(0, 0, 0)$	& \multirow{3}{*}{NGC 4536}	\\
		&\ref{fig:svd_0pt2_Mu_1}& 1.99	& $95.0\sim 97.5$	& 111.5	& $63.37\pm1.50$&			\\
			& N/A				& 1.88	& $73  \sim 94.9$	& 109.0	& $(0, 0, 1)$	&			\\
\cline{2-7}
			& N/A				& 1.65	& $99.2\sim 100 $	& 87.3	& $(0, 0, 0)$	& \multirow{3}{*}{+ 1999cl}	\\
			& N/A				& 1.30	& $74  \sim 99.1$	& 76.6	& $(0, 0, 1)$	&			\\
		&\ref{fig:svd_0pt2_Mu_2}& 1.29	& Not Preferred		& 83.3	& $(1, 0, 1)$	&			\\
\hline
\multirow{7}{*}{0.4}& N/A		& 11.0	& $97.6\sim 99.0$	& 988.8	& $76.23\pm3.72$&			\\
			& N/A				& 2.46	& $15  \sim 97.5$	& 662.5	& $(0, 0, 5)$	&			\\
\cline{2-7}
			& N/A				& 2.08	& $97.7\sim 100 $	& 122.8	& $(0, 0, 0)$	& \multirow{3}{*}{NGC 4536}	\\
			& N/A				& 1.94	& $69  \sim 97.6$	& 118.5	& $63.40\pm1.47$&			\\
		&\ref{fig:svd_0pt4_Mu_1}& 1.97	& Not Preferred		& 126.8	& $(1, 0, 1)$	&			\\
\cline{2-7}
			& N/A				& 1.47	& $88  \sim 99.1$	& 89.0	& $63.32\pm1.28$& \multirow{2}{*}{+ 1999cl}	\\
			& N/A				& 1.43	& $61  \sim 87  $	& 92.1	& $(0, 0, 1)$	&			\\
\hline
\end{tabular}
\caption{\label{tab:nvd_Chi2}
Distance modulus best fit model parameters with progressive removal of the outliers for F3, without peculiar velocity correction, with $v_c = 40$, and with 250 \ks velocity dispersion for SNe. The {\it Threshold} column shows the F-test threshold of the model. 
}
\end{table}

\newpage


\section{Union2.1 + Riess et al. 2016 Cepheids dataset}
\label{app:U2p1plusRiessdata}

We present here tables containing the data described in section~\ref{sec:data} and on which our methodology of section~\ref{sec:method} is applied. In table~\ref{Tab:Cepheids} are presented the Cepheid calibrators of \cite{Riess2016} with their associated SNe. Several quantities such as the angular position and redshifts are extracted from external sources: \UTPO, \Ned and \Simbad. Among the 3 fields, there are 8 SNe related to a Cepheid host-galaxy, among which 3 of them have redshift and magnitude recorded in \UTPO (2007af, 1994ae and 1995al, indicated by $\ast$ in tables \ref{Tab:SneIaF2} and \ref{Tab:SneIaF3}) and 5 others for which redshift and magnitude are only given from the Cepheids (2009ig, 1981B, 1990N, 2012cg and 2012ht, indicated by $\dagger$ in table~\ref{Tab:Cepheids}), i.e. SNe which are not in \UTPO.
We already explained in section~\ref{sec:sneia} how the distance moduli are obtained. The ICRS sky positions are written in degrees and the magnitude $\mu$, plus its error $\Delta \mu$ \HWC{before adding the calibration error}, are presented. For types, the letters denote the following: `o' indicates the so-called outliers in \UTPO,  `z' is for redshifts $< 0.015$ and `h' for redshifts $>0.2$ and $\leq 0.4$; $\ast$ and $\dagger$ were explained above.
Finally, we present in tables \ref{Tab:SneIaF1Part1}-\ref{Tab:SneIaF1Part2}, \ref{Tab:SneIaF2} and \ref{Tab:SneIaF3} the respective \UTPO SNe of \cite{Union2.1} and used in fields F1, F2 and F3. Angular positions are extracted from \Simbad, and given with 3 digits of precision. redshifts $z$ are from \UTPO and shown with 5 digits (when available). $(\mu,\Delta\mu)$ are also from \UTPO and $\Delta\mu^{250}$ is computed with the additional 250 \ks velocity dispersion. They are shown with only 4 digits of precision.

\begin{table}[ht!]
\begin{center}
\caption{\label{Tab:Cepheids} Cepheids-hosting galaxies of \cite{Riess2016} with additional data. The Cepheid-hosting galaxy name and the name of the hosted SN are from \cite{Riess2016}. Angular positions of host galaxies are from \Simbad, in ICRS decimal format. Values of $(\mu,\Delta \mu)$ come from \cite{Riess2016}, while $\Delta\mu^{250}$ refers to the modified distance modulus error after considering an additional velocity dispersion of 250 \ks to the hosted SN redshift $z_{\rm SN}$\HWC{, but before adding the calibration error}. Values $z$(NED) are the host redshifts taken from \Ned and $z_{\rm SN}$ are from \UTPO, \Simbad, or Extragalactic Distance Database (\Edd) 
from \cite{Tully:2009ir} depending on availability. For some cases (denoted by ``/") none of these 3 sources have redshifts. NGC 4258 does not host a SN. Angles are truncated to 3 digits, similarly for $\Delta\mu^{250}$ and $\Delta\mu$ (except when \UTPO does not reach this precision).}
\vspace{0.3cm}
\scriptsize
\begin{tabular}{|ccccccccccc|}
\hline
Cepheid	&	Co-hosted	&	R.A.	&	Dec.	&	$\mu$	&	$\Delta\mu$	&	$\Delta\mu^{250}$	&	$z$(\Ned)	&	$z_{\rm SN}$	&	Origin($z_{\rm SN}$)	& Field \\
Host	&	SN	&	(deg)	&	(deg)	&	(mag)	&	(mag)	&	(mag)	&	Host	&		&		& \\
\hline
 M 101	&	SN 2011fe	&	210.802	&	54.349	&	29.135	&	0.045	&	1.499	&	0.001207	&	0.001208	&	\Simbad	& \\
 NGC 1015	&	SN 2009ig	&	39.548	&	-1.319	&	32.497	&	0.081	&	0.081	&	0.00801	&	/\	&	/\	& F1$^\dagger$ \\
 NGC 1309	&	SN 2002fk	&	50.527	&	-15.400	&	32.523	&	0.055	&	0.280	&	0.006618	&	0.0066	&	\UTPO	& \\
 NGC 1365	&	SN 2012fr	&	53.402	&	-36.141	&	31.307	&	0.057	&	0.057	&	0.005133	&	/\	&	/\	& \\
 NGC 1448	&	SN 2001el	&	56.133	&	-44.645	&	31.311	&	0.045	&	0.505	&	0.003676	&	0.0036	&	\Edd	& \\
 NGC 2442	&	SN 2015F	&	114.099	&	-69.531	&	31.511	&	0.053	&	0.053	&	0.005284	&	/\	&	/\	& \\
 NGC 3021	&	SN 1995al	&	147.738	&	33.553	&	32.498	&	0.09	&	0.373	&	0.006017	&	0.005	&	\UTPO	& F3 \\
 NGC 3370	&	SN 1994ae	&	161.767	&	17.274	&	32.072	&	0.049	&	0.424	&	0.005387	&	0.0043	&	\UTPO	& F3 \\
 NGC 3447	&	SN 2012ht	&	163.350	&	16.772	&	31.908	&	0.043	&	0.043	&	0.004688	&	/\	&	/\	& F3$^\dagger$ \\
 NGC 3972	&	SN 2011by	&	178.938	&	55.321	&	31.587	&	0.07	&	0.536	&	0.003406	&	0.003402	&	\Simbad	& \\
 NGC 3982	&	SN 1998aq	&	179.117	&	55.125	&	31.737	&	0.069	&	0.417	&	0.004265	&	0.0044	&	\Edd	& \\
 NGC 4038	&	SN 2007sr	&	180.471	&	-18.868	&	31.29	&	0.112	&	0.292	&	0.006661	&	0.0067	&	\UTPO	& \\
 NGC 4424	&	SN 2012cg	&	186.798	&	9.421	&	31.08	&	0.292	&	0.292	&	0.002581	&	/\	&	/\	& F3$^\dagger$ \\
 NGC 4536	&	SN 1981B	&	188.613	&	2.188	&	30.906	&	0.053	&	0.259	&	0.007174	&	0.00715	&	\Edd	& F3$^\dagger$ \\
 NGC 4639	&	SN 1990N	&	190.718	&	13.258	&	31.532	&	0.071	&	0.427	&	0.004463	&	0.0043	&	\Edd	& F3$^\dagger$ \\
 NGC 5584	&	SN 2007af	&	215.599	&	-0.387	&	31.786	&	0.046	&	0.291	&	0.006293	&	0.0063	&	\UTPO	& F2 \\
 NGC 5917	&	SN 2005cf	&	230.386	&	-7.377	&	32.263	&	0.102	&	0.278	&	0.006935	&	0.007	&	\UTPO	& \\
 NGC 7250	&	SN 2013dy	&	334.574	&	40.562	&	31.499	&	0.078	&	0.078	&	0.002884	&	/\	&	/\	& \\
 UGC 9391	&	SN 2003du	&	218.654	&	59.338	&	32.919	&	0.063	&	0.277	&	0.006652	&	0.0067	&	\UTPO	& \\
NGC 4258	&	/\	&	184.740	&	47.304	&	29.387	&	0.0568	&	0.057	&	0.002181	&	/\	&	/\	& \\
\hline
\end{tabular}
\end{center}
\end{table}

\begin{table}[ht!]
\begin{center}
\vspace{-.7cm}
\caption{\label{Tab:SneIaF1Part1} \UTPO SNe of F1 -- Part I.
}
\tiny
\begin{tabular}{|P{.45cm}P{.55cm}P{.55cm}P{.6cm}P{.6cm}P{.6cm}P{.6cm}P{.4cm}|P{.45cm}P{.55cm}P{.55cm}P{.6cm}P{.6cm}P{.6cm}P{.6cm}P{.4cm}|}
\hline
Name	&	R.A.	&	Dec.	&	z	&	$\mu$	&	$\Delta\mu$	&	$\Delta\mu^{250}$	&	Type	& Name	&	R.A.	&	Dec.	&	z	&	$\mu$	&	$\Delta\mu$	&	$\Delta\mu^{250}$	&	Type	\\
SN	&	(deg)	&	(deg)	&		&	(mag)	&	(mag)	&	(mag)	&   &	SN	&	(deg)	&	(deg)	&		&	(mag)	&	(mag)	&	(mag)	&		\\
\hline
1995ak	&	41.453	&	3.231	&	0.02198	&	34.7547	&	0.1875	&	0.2048	&	F1	&	2005gu	&	12.238	&	-0.906	&	0.3305	&	41.1317	&	0.16	&	0.1601	&	F1h	\\
1995ao	&	44.378	&	-1.689	&	0.24	&	40.6417	&	0.4	&	0.4001	&	F1h	&	2005gv	&	38.476	&	0.281	&	0.3619	&	41.2817	&	0.17	&	0.1701	&	F1h	\\
1995aw	&	36.231	&	0.885	&	0.4	&	42.2117	&	0.48	&	0.48	&	F1h	&	2005gw	&	354.498	&	0.642	&	0.2754	&	40.6417	&	0.14	&	0.1402	&	F1h	\\
1999dr	&	345.073	&	-0.087	&	0.178	&	39.3565	&	0.2357	&	0.236	&	F1	&	2005gx	&	359.884	&	0.737	&	0.14462	&	39.1926	&	0.1127	&	0.1134	&	F1	\\
1999du	&	16.775	&	-0.132	&	0.26	&	40.7217	&	0.2	&	0.2001	&	F1h	&	2005gy	&	21.528	&	0.677	&	0.3306	&	40.9517	&	0.15	&	0.1501	&	F1h	\\
1999dv	&	17.246	&	0.007	&	0.186	&	39.6139	&	0.189	&	0.1893	&	F1	&	2005hc	&	29.200	&	-0.214	&	0.04498	&	36.4521	&	0.1083	&	0.1156	&	F1	\\
1999dx	&	23.498	&	0.071	&	0.269	&	40.6817	&	0.26	&	0.2601	&	F1h	&	2005hj	&	21.702	&	-1.238	&	0.0576	&	36.982	&	0.1619	&	0.1649	&	F1	\\
1999dy	&	23.956	&	0.144	&	0.215	&	40.2817	&	0.19	&	0.1902	&	F1h	&	2005hn	&	329.267	&	-0.223	&	0.10671	&	38.5384	&	0.1176	&	0.1188	&	F1	\\
1999fw	&	352.971	&	0.159	&	0.278	&	40.4717	&	0.2	&	0.2001	&	F1h	&	2005ho	&	14.850	&	0.003	&	0.06184	&	37.0327	&	0.1157	&	0.1193	&	F1	\\
2002ha	&	311.827	&	0.313	&	0.0131	&	33.7317	&	0.23	&	0.2683	&	F1z	&	2005hp	&	307.219	&	-0.779	&	0.17391	&	39.4337	&	0.129	&	0.1294	&	F1	\\
2004ey	&	327.283	&	0.444	&	0.0147	&	33.9617	&	0.18	&	0.2181	&	F1z	&	2005hq	&	312.582	&	-0.825	&	0.3996	&	41.6317	&	0.21	&	0.21	&	F1h	\\
2005a	&	37.680	&	-2.939	&	0.01832	&	34.2735	&	0.1605	&	0.1884	&	F1	&	2005hr	&	49.641	&	0.123	&	0.11635	&	38.6479	&	0.1123	&	0.1134	&	F1	\\
2005ed	&	0.706	&	0.751	&	0.0857	&	37.8955	&	0.118	&	0.1198	&	F1	&	2005hs	&	52.342	&	-1.095	&	0.3003	&	40.7517	&	0.15	&	0.1501	&	F1h	\\
2005eg	&	15.535	&	-0.879	&	0.18971	&	39.8413	&	0.1197	&	0.1201	&	F1	&	2005ht	&	312.603	&	-0.168	&	0.18581	&	39.7258	&	0.1207	&	0.1211	&	F1	\\
2005ei	&	329.199	&	0.318	&	0.12669	&	38.817	&	0.1239	&	0.1248	&	F1	&	2005hu	&	328.670	&	0.413	&	0.2186	&	40.1017	&	0.12	&	0.1203	&	F1h	\\
2005ex	&	25.464	&	-0.877	&	0.09294	&	38.0476	&	0.119	&	0.1205	&	F1	&	2005hv	&	333.183	&	-0.035	&	0.1776	&	39.9724	&	0.1248	&	0.1252	&	F1	\\
2005ey	&	34.273	&	0.281	&	0.14703	&	39.2053	&	0.1117	&	0.1123	&	F1	&	2005hx	&	3.251	&	0.248	&	0.11967	&	38.6573	&	0.1107	&	0.1118	&	F1	\\
2005ez	&	46.796	&	1.120	&	0.12928	&	38.8197	&	0.1344	&	0.1351	&	F1	&	2005hy	&	3.598	&	0.333	&	0.15463	&	39.2257	&	0.112	&	0.1127	&	F1	\\
2005fa	&	24.900	&	-0.758	&	0.16086	&	39.2209	&	0.1186	&	0.1192	&	F1	&	2005hz	&	11.634	&	0.838	&	0.12873	&	38.766	&	0.1095	&	0.1104	&	F1	\\
2005fc	&	320.414	&	0.895	&	0.2956	&	41.0417	&	0.22	&	0.2201	&	F1h	&	2005ia	&	17.896	&	-0.006	&	0.2507	&	40.6517	&	0.12	&	0.1202	&	F1h	\\
2005fd	&	323.799	&	0.163	&	0.2606	&	40.4217	&	0.14	&	0.1402	&	F1h	&	2005ic	&	327.786	&	-0.843	&	0.3095	&	40.9217	&	0.13	&	0.1301	&	F1h	\\
2005fe	&	334.864	&	0.494	&	0.2155	&	40.2017	&	0.14	&	0.1403	&	F1h	&	2005id	&	349.139	&	-0.663	&	0.18255	&	39.6858	&	0.1141	&	0.1145	&	F1	\\
2005ff	&	337.671	&	-0.776	&	0.08569	&	37.8992	&	0.1164	&	0.1183	&	F1	&	2005ie	&	34.761	&	-0.273	&	0.2789	&	40.6717	&	0.13	&	0.1302	&	F1h	\\
2005fh	&	349.374	&	0.430	&	0.11763	&	38.4812	&	0.1136	&	0.1146	&	F1	&	2005if	&	52.554	&	-0.974	&	0.06644	&	37.2739	&	0.1176	&	0.1207	&	F1	\\
2005fi	&	1.995	&	0.638	&	0.2635	&	40.6717	&	0.13	&	0.1302	&	F1h	&	2005ig	&	337.631	&	-0.503	&	0.2795	&	40.4017	&	0.14	&	0.1401	&	F1h	\\
2005fj	&	317.837	&	-0.445	&	0.14179	&	39.1756	&	0.1195	&	0.1202	&	F1	&	2005ih	&	1.807	&	0.349	&	0.2575	&	40.4917	&	0.13	&	0.1302	&	F1h	\\
2005fl	&	311.842	&	-1.253	&	0.2328	&	40.1017	&	0.15	&	0.1502	&	F1h	&	2005ii	&	3.266	&	-0.620	&	0.2925	&	40.8217	&	0.14	&	0.1401	&	F1h	\\
2005fm	&	312.043	&	-1.171	&	0.15186	&	39.0575	&	0.1138	&	0.1145	&	F1	&	2005ij	&	46.089	&	-1.063	&	0.12427	&	38.6217	&	0.1111	&	0.1121	&	F1	\\
2005fn	&	312.221	&	0.191	&	0.09391	&	38.0746	&	0.118	&	0.1195	&	F1	&	2005ik	&	322.815	&	-1.057	&	0.3095	&	41.0917	&	0.15	&	0.1501	&	F1h	\\
2005fo	&	328.943	&	0.594	&	0.2605	&	40.7117	&	0.15	&	0.1502	&	F1h	&	2005ir	&	19.182	&	0.795	&	0.07535	&	37.4818	&	0.1023	&	0.1051	&	F1	\\
2005fp	&	6.807	&	1.121	&	0.2116	&	40.4617	&	0.16	&	0.1602	&	F1h	&	2005is	&	5.437	&	-0.325	&	0.17063	&	39.3766	&	0.1164	&	0.1169	&	F1	\\
2005fr	&	17.092	&	-0.096	&	0.2866	&	40.9417	&	0.13	&	0.1302	&	F1h	&	2005it	&	16.190	&	0.514	&	0.3086	&	40.8117	&	0.18	&	0.1801	&	F1h	\\
2005fs	&	31.221	&	-0.326	&	0.3388	&	41.2117	&	0.17	&	0.1701	&	F1h	&	2005iu	&	305.065	&	0.217	&	0.08902	&	37.7302	&	0.1174	&	0.1192	&	F1	\\
2005ft	&	40.521	&	-0.541	&	0.18012	&	39.5403	&	0.1156	&	0.116	&	F1	&	2005iv	&	307.936	&	0.245	&	0.2988	&	40.9317	&	0.15	&	0.1501	&	F1h	\\
2005fu	&	42.634	&	0.807	&	0.19215	&	39.941	&	0.1211	&	0.1215	&	F1	&	2005ix	&	310.483	&	1.092	&	0.2658	&	40.4017	&	0.12	&	0.1202	&	F1h	\\
\hline
\end{tabular}
\end{center}
\end{table}


\begin{table}[ht!]
\begin{center}
\caption{\label{Tab:SneIaF1Part2} \UTPO SNe of F1 -- Part II.}
\tiny
\begin{tabular}{|P{.45cm}P{.55cm}P{.55cm}P{.6cm}P{.6cm}P{.6cm}P{.6cm}P{.4cm}|P{.45cm}P{.55cm}P{.55cm}P{.6cm}P{.6cm}P{.6cm}P{.6cm}P{.4cm}|}
\hline
Name	&	R.A.	&	Dec.	&	z	&	$\mu$	&	$\Delta\mu$	&	$\Delta\mu^{250}$	&	Type	& Name	&	R.A.	&	Dec.	&	z	&	$\mu$	&	$\Delta\mu$	&	$\Delta\mu^{250}$	&	Type	\\
\hline
2005fv	&	46.343	&	0.858	&	0.11728	&	38.6477	&	0.1137	&	0.1147	&	F1	&	2005iz	&	328.069	&	0.267	&	0.2006	&	39.7517	&	0.13	&	0.1303	&	F1h	\\
2005fw	&	52.704	&	-1.238	&	0.1424	&	39.0182	&	0.1137	&	0.1144	&	F1	&	2005ja	&	358.969	&	0.877	&	0.3264	&	40.8717	&	0.14	&	0.1401	&	F1h	\\
2005fx	&	344.201	&	0.401	&	0.2884	&	40.7417	&	0.15	&	0.1501	&	F1h	&	2005jb	&	339.013	&	-0.368	&	0.2565	&	40.5117	&	0.16	&	0.1602	&	F1h	\\
2005fy	&	50.090	&	-0.886	&	0.19432	&	39.8633	&	0.1264	&	0.1267	&	F1	&	2005jc	&	11.352	&	1.076	&	0.2116	&	39.8817	&	0.12	&	0.1203	&	F1h	\\
2005fz	&	315.922	&	0.570	&	0.12283	&	38.7015	&	0.1211	&	0.122	&	F1	&	2005jd	&	34.276	&	0.535	&	0.3129	&	40.9117	&	0.14	&	0.1401	&	F1h	\\
2005ga	&	16.932	&	-1.040	&	0.17274	&	39.4047	&	0.1135	&	0.1139	&	F1	&	2005je	&	38.861	&	1.075	&	0.09315	&	38.1827	&	0.1145	&	0.1161	&	F1	\\
2005gb	&	19.053	&	0.792	&	0.08585	&	37.8542	&	0.1126	&	0.1145	&	F1	&	2005jg	&	345.262	&	-0.207	&	0.3024	&	40.9317	&	0.13	&	0.1301	&	F1h	\\
2005gc	&	20.407	&	-0.977	&	0.1638	&	39.2963	&	0.1131	&	0.1137	&	F1	&	2005jh	&	350.019	&	-0.056	&	0.10864	&	38.5522	&	0.1195	&	0.1206	&	F1	\\
2005gd	&	26.963	&	0.641	&	0.15989	&	39.2523	&	0.1147	&	0.1153	&	F1	&	2005ji	&	4.326	&	-0.257	&	0.2146	&	40.1017	&	0.12	&	0.1203	&	F1h	\\
2005ge	&	34.561	&	0.797	&	0.205	&	39.9217	&	0.12	&	0.1203	&	F1h	&	2005jj	&	314.186	&	0.408	&	0.3666	&	41.7617	&	0.2	&	0.2001	&	F1h	\\
2005gf	&	334.069	&	0.708	&	0.2485	&	40.1717	&	0.13	&	0.1302	&	F1h	&	2005jk	&	26.499	&	1.196	&	0.18885	&	39.6938	&	0.1197	&	0.1201	&	F1	\\
2005gg	&	334.672	&	0.639	&	0.2285	&	40.1517	&	0.12	&	0.1203	&	F1h	&	2005jl	&	323.234	&	-0.700	&	0.17969	&	39.6227	&	0.117	&	0.1174	&	F1	\\
2005gh	&	312.651	&	-0.354	&	0.2577	&	40.5517	&	0.14	&	0.1402	&	F1h	&	2005jm	&	328.079	&	0.472	&	0.2026	&	39.8717	&	0.13	&	0.1303	&	F1h	\\
2005gj	&	45.300	&	-0.554	&	0.18222	&	39.4956	&	0.1193	&	0.1197	&	F1	&	2005jn	&	4.754	&	-0.281	&	0.3204	&	41.0017	&	0.13	&	0.1301	&	F1h	\\
2005go	&	17.705	&	1.008	&	0.2636	&	40.4517	&	0.14	&	0.1402	&	F1h	&	2005jo	&	52.090	&	-0.326	&	0.2183	&	40.1217	&	0.12	&	0.1203	&	F1h	\\
2005gp	&	55.497	&	-0.783	&	0.12647	&	38.6122	&	0.1192	&	0.1201	&	F1	&	2005jp	&	32.460	&	-0.062	&	0.2109	&	39.8817	&	0.12	&	0.1203	&	F1h	\\
2005gq	&	53.454	&	0.709	&	0.3893	&	41.5017	&	0.2	&	0.2001	&	F1h	&	2005jt	&	42.667	&	-0.066	&	0.36	&	41.0617	&	0.18	&	0.1801	&	F1h	\\
2005gr	&	54.156	&	1.079	&	0.2444	&	40.1017	&	0.12	&	0.1202	&	F1h	&	2005ju	&	39.117	&	0.511	&	0.258	&	40.5417	&	0.14	&	0.1402	&	F1h	\\
2005gs	&	333.293	&	1.051	&	0.2495	&	40.6817	&	0.13	&	0.1302	&	F1h	&	2005jw	&	310.080	&	-0.007	&	0.3797	&	41.3317	&	0.15	&	0.1501	&	F1h	\\
2005gt	&	31.016	&	-0.366	&	0.2779	&	40.7417	&	0.17	&	0.1701	&	F1h	&	2005jy	&	348.465	&	1.257	&	0.2704	&	40.5717	&	0.15	&	0.1501	&	F1h	\\
2005jz	&	22.863	&	-0.632	&	0.2517	&	40.4417	&	0.12	&	0.1202	&	F1h	&	2005ln	&	6.751	&	-0.587	&	0.14567	&	38.9585	&	0.1258	&	0.1264	&	F1	\\
2005ka	&	333.483	&	1.087	&	0.3164	&	41.3017	&	0.19	&	0.1901	&	F1h	&	2005lo	&	9.299	&	-1.203	&	0.2975	&	40.6717	&	0.21	&	0.2101	&	F1h	\\
2005kn	&	318.885	&	-0.355	&	0.19672	&	39.8262	&	0.135	&	0.1353	&	F1	&	2005lp	&	26.928	&	0.207	&	0.3018	&	41.3917	&	0.25	&	0.2501	&	F1h	\\
2005ko	&	357.521	&	-0.921	&	0.18357	&	39.6125	&	0.1238	&	0.1242	&	F1	&	2005lq	&	40.400	&	0.205	&	0.379	&	41.4817	&	0.2	&	0.2001	&	F1h	\\
2005kp	&	7.721	&	-0.719	&	0.11471	&	38.5708	&	0.1134	&	0.1145	&	F1	&	2005mh	&	41.236	&	0.204	&	0.394	&	41.6817	&	0.15	&	0.1501	&	F1h	\\
2005kq	&	347.837	&	-0.609	&	0.3873	&	41.7917	&	0.19	&	0.1901	&	F1h	&	2005mi	&	335.261	&	-0.748	&	0.2125	&	39.9917	&	0.13	&	0.1303	&	F1h	\\
2005kt	&	17.742	&	0.276	&	0.06386	&	37.2177	&	0.1196	&	0.1229	&	F1	&	2005ml	&	33.518	&	-0.239	&	0.11304	&	38.4532	&	0.1192	&	0.1203	&	F1	\\
2005ku	&	344.928	&	-0.014	&	0.04372	&	36.2883	&	0.1272	&	0.1338	&	F1	&	2005mm	&	3.290	&	1.146	&	0.3804	&	41.5617	&	0.22	&	0.2201	&	F1h	\\
2005ld	&	325.002	&	-0.008	&	0.14371	&	39.0875	&	0.1163	&	0.117	&	F1	&	2005mo	&	57.554	&	-0.240	&	0.2735	&	40.6317	&	0.16	&	0.1601	&	F1h	\\
2005le	&	337.885	&	-0.494	&	0.2525	&	40.4217	&	0.15	&	0.1502	&	F1h	&	2005mq	&	350.091	&	-0.350	&	0.3483	&	41.2117	&	0.22	&	0.2201	&	F1h	\\
2005lf	&	349.675	&	-1.205	&	0.2984	&	40.9617	&	0.22	&	0.2201	&	F1h	&	2006cm	&	320.073	&	-1.684	&	0.0153	&	34.6089	&	0.2132	&	0.2438	&	F1	\\
2005lg	&	19.084	&	-0.808	&	0.3486	&	41.2017	&	0.18	&	0.1801	&	F1h	&	2006eq	&	322.155	&	1.228	&	0.04839	&	36.5876	&	0.1192	&	0.1249	&	F1	\\
2005lh	&	328.951	&	1.181	&	0.2166	&	40.2617	&	0.15	&	0.1502	&	F1h	&	2006gj	&	49.399	&	-1.692	&	0.0277	&	35.6068	&	0.1833	&	0.1946	&	F1	\\
2005li	&	335.814	&	0.253	&	0.2555	&	40.3917	&	0.15	&	0.1502	&	F1h	&	2006oa	&	320.929	&	-0.843	&	0.0589	&	37.0134	&	0.1657	&	0.1685	&	F1	\\
2005lj	&	29.430	&	-0.180	&	0.077	&	38.2317	&	0.14	&	0.142	&	F1o	&	2006ob	&	27.952	&	0.264	&	0.0583	&	36.96	&	0.1681	&	0.1709	&	F1	\\
2005lk	&	329.956	&	-1.194	&	0.10272	&	38.389	&	0.1199	&	0.1212	&	F1	&	2006on	&	328.994	&	-1.070	&	0.0688	&	37.3884	&	0.198	&	0.1998	&	F1	\\
2005ll	&	337.029	&	-1.128	&	0.2425	&	40.0317	&	0.15	&	0.1502	&	F1h	&	2006py	&	340.425	&	-0.137	&	0.05668	&	36.8654	&	0.1101	&	0.1146	&	F1	\\
\hline
\end{tabular}
\end{center}
\end{table}

\begin{table}[ht!]
\begin{center}
\vspace{-.7cm}
\caption{\label{Tab:SneIaF2} \UTPO SNe of F2.}
\tiny
\begin{tabular}{|P{.45cm}P{.55cm}P{.55cm}P{.6cm}P{.6cm}P{.6cm}P{.6cm}P{.4cm}|P{.45cm}P{.55cm}P{.55cm}P{.6cm}P{.6cm}P{.6cm}P{.6cm}P{.4cm}|}
\hline
Name	&	R.A.	&	Dec.	&	z	&	$\mu$	&	$\Delta\mu$	&	$\Delta\mu^{250}$	&	Type	& Name	&	R.A.	&	Dec.	&	z	&	$\mu$	&	$\Delta\mu$	&	$\Delta\mu^{250}$	&	Type	\\
\hline
1994m	&	187.786	&	0.605	&	0.02431	&	35.0112	&	0.1817	&	0.1963	&	F2	&	2002ck	&	236.753	&	-0.990	&	0.0303	&	35.53	&	0.1767	&	0.1866	&	F2	\\
1994t	&	200.378	&	-2.149	&	0.03572	&	35.8623	&	0.1712	&	0.1785	&	F2	&	2007af	&	215.588	&	-0.394	&	0.0063	&	31.9817	&	0.38	&	0.4763	&	F2z$^\ast$	\\
1999ar	&	140.067	&	0.561	&	0.1561	&	39.131	&	0.0841	&	0.0849	&	F2	& & & & & & & & \\
\hline
\end{tabular}
\end{center}
\end{table}


\begin{table}[ht!]
\begin{center}
\vspace{-.7cm}
\caption{\label{Tab:SneIaF3} \UTPO SNe of F3.}
\tiny
\begin{tabular}{|P{.45cm}P{.55cm}P{.55cm}P{.6cm}P{.6cm}P{.6cm}P{.6cm}P{.4cm}|P{.45cm}P{.55cm}P{.55cm}P{.6cm}P{.6cm}P{.6cm}P{.6cm}P{.4cm}|}
\hline
Name	&	R.A.	&	Dec.	&	z	&	$\mu$	&	$\Delta\mu$	&	$\Delta\mu^{250}$	&	Type	& Name	&	R.A.	&	Dec.	&	z	&	$\mu$	&	$\Delta\mu$	&	$\Delta\mu^{250}$	&	Type	\\
\hline
1992p	&	190.704	&	10.360	&	0.02649	&	35.3824	&	0.1913	&	0.2031	&	F3	&	2004bg	&	170.256	&	21.340	&	0.0221	&	34.8239	&	0.189	&	0.206	&	F3	\\
1994ae	&	161.758	&	17.275	&	0.0043	&	32.0317	&	0.53	&	0.6768	&	F3z	&	2004gs	&	129.600	&	17.631	&	0.02757	&	35.3772	&	0.1281	&	0.144	&	F3$^\ast$	\\
1994s	&	187.841	&	29.134	&	0.01517	&	34.0034	&	0.2152	&	0.2461	&	F3	&	2004gu	&	191.603	&	11.949	&	0.04697	&	36.3962	&	0.1103	&	0.1168	&	F3	\\
1995al	&	147.733	&	33.553	&	0.005	&	32.2217	&	0.46	&	0.5853	&	F3z	&	2004l	&	156.767	&	16.019	&	0.0334	&	35.7817	&	0.1758	&	0.1839	&	F3$^\ast$	\\
1995ba	&	124.777	&	7.723	&	0.388	&	42.1117	&	0.47	&	0.47	&	F3h	&	2005ag	&	224.180	&	9.327	&	0.08005	&	37.5875	&	0.102	&	0.1044	&	F3	\\
1995d	&	145.228	&	5.141	&	0.0065	&	32.5117	&	0.37	&	0.463	&	F3z	&	2005bg	&	184.322	&	16.372	&	0.02419	&	34.9541	&	0.1334	&	0.1529	&	F3	\\
1996ab	&	230.285	&	27.927	&	0.1244	&	38.9465	&	0.1643	&	0.1649	&	F3	&	2005ki	&	160.118	&	9.183	&	0.02037	&	34.5582	&	0.1456	&	0.1706	&	F3	\\
1997ac	&	126.022	&	4.190	&	0.32	&	41.1417	&	0.42	&	0.42	&	F3h	&	2005m	&	144.387	&	23.203	&	0.02297	&	35.0356	&	0.1356	&	0.1569	&	F3	\\
1997n	&	125.958	&	3.481	&	0.18	&	40.0913	&	0.4454	&	0.4456	&	F3	&	2005mc	&	126.777	&	21.646	&	0.026	&	35.283	&	0.1839	&	0.1967	&	F3	\\
1997o	&	126.010	&	4.126	&	0.374	&	43.0917	&	0.92	&	0.92	&	F3h	&	2006ac	&	190.436	&	35.062	&	0.0239	&	34.9367	&	0.1814	&	0.1966	&	F3	\\
1999aa	&	126.925	&	21.487	&	0.015	&	34.0653	&	0.1615	&	0.2015	&	F3	&	2006al	&	159.868	&	5.183	&	0.069	&	37.4678	&	0.1762	&	0.1781	&	F3	\\
1999ac	&	241.813	&	7.972	&	0.0095	&	33.0817	&	0.24	&	0.3064	&	F3z	&	2006an	&	183.661	&	12.230	&	0.0651	&	37.2084	&	0.1628	&	0.1652	&	F3	\\
1999cl	&	187.983	&	14.426	&	0.0087	&	31.0417	&	0.26	&	0.333	&	F3z	&	2006br	&	202.508	&	13.416	&	0.0255	&	35.6219	&	0.1962	&	0.2086	&	F3	\\
1999gd	&	129.603	&	25.759	&	0.01926	&	34.8544	&	0.2406	&	0.2583	&	F3	&	2006bt	&	239.141	&	20.052	&	0.0325	&	35.7148	&	0.1704	&	0.1793	&	F3	\\
2001ay	&	216.571	&	26.249	&	0.0309	&	35.8312	&	0.1837	&	0.1928	&	F3	&	2006bu	&	208.199	&	5.314	&	0.0843	&	37.9536	&	0.2003	&	0.2014	&	F3	\\
2001fe	&	144.488	&	25.495	&	0.0145	&	33.8217	&	0.22	&	0.2529	&	F3z	&	2006bw	&	218.486	&	3.799	&	0.0308	&	35.5307	&	0.1784	&	0.1878	&	F3	\\
2001iv	&	117.556	&	10.286	&	0.3965	&	41.3917	&	0.2	&	0.2001	&	F3h	&	2006cj	&	194.852	&	28.348	&	0.0684	&	37.6329	&	0.1706	&	0.1726	&	F3	\\
2001iw	&	117.664	&	10.339	&	0.3396	&	40.9817	&	0.23	&	0.2301	&	F3h	&	2006cp	&	184.812	&	22.427	&	0.0233	&	34.7839	&	0.1834	&	0.1992	&	F3	\\
2001n	&	159.992	&	24.091	&	0.0221	&	34.8133	&	0.19	&	0.2069	&	F3	&	2006cq	&	201.105	&	30.956	&	0.0491	&	36.6319	&	0.1758	&	0.1797	&	F3	\\
2001v	&	179.354	&	25.203	&	0.016	&	33.7264	&	0.208	&	0.2367	&	F3	&	2006s	&	191.413	&	35.087	&	0.0329	&	35.8435	&	0.1704	&	0.1791	&	F3	\\
2002bo	&	154.527	&	21.828	&	0.0053	&	31.9417	&	0.44	&	0.5569	&	F3z	&	2006x	&	185.729	&	15.820	&	0.0063	&	30.9317	&	0.36	&	0.4605	&	F3z	\\
2002de	&	244.127	&	35.708	&	0.0283	&	35.422	&	0.1801	&	0.1911	&	F3	&	2007bc	&	169.819	&	20.813	&	0.0219	&	34.7507	&	0.1884	&	0.2057	&	F3	\\
2002g	&	196.980	&	34.085	&	0.0345	&	35.88	&	0.2116	&	0.218	&	F3	&	2007bz	&	194.224	&	22.373	&	0.0227	&	35.5517	&	0.19	&	0.206	&	F3o	\\
2003cg	&	153.567	&	3.467	&	0.0053	&	31.9317	&	0.44	&	0.5569	&	F3z	&	2007ci	&	176.441	&	19.771	&	0.0192	&	34.3933	&	0.1971	&	0.2185	&	F3	\\
2003kc	&	146.643	&	30.655	&	0.0341	&	35.7437	&	0.1752	&	0.1831	&	F3	&	2007s	&	150.130	&	4.407	&	0.015	&	34.0132	&	0.2134	&	0.2451	&	F3	\\
2003w	&	146.706	&	16.044	&	0.0211	&	34.5625	&	0.1884	&	0.207	&	F3	&	2008af	&	224.869	&	16.653	&	0.0341	&	35.6722	&	0.1776	&	0.1853	&	F3	\\
2004as	&	171.413	&	22.830	&	0.0321	&	35.7977	&	0.1738	&	0.1828	&	F3	&	2008bf	&	181.012	&	20.245	&	0.0251	&	34.85	&	0.1792	&	0.1932	&	F3	\\
\hline
\end{tabular}
\end{center}
\end{table}

\end{appendices}

~
\newpage
~
\newpage

\bibliographystyle{JHEP-3}

\bibliography{bibliography}

\end{document}